\pdfoutput=1 % if your are submitting a pdflatex (i.e. if you have
\documentclass[11pt,a4paper]{article}

\usepackage{jheppub} % for details on the use of the package, please
\usepackage[T1]{fontenc} % if needed
 
\usepackage{amsmath,amsthm,amssymb}
\usepackage{multicol}
\usepackage{blkarray}
	 
\numberwithin{equation}{section}
 
\usepackage[utf8]{inputenc}
\usepackage[english]{babel}
\usepackage{textcomp}
\usepackage{amsmath}
\usepackage{mathtools}
\usepackage{gensymb}
\usepackage{multirow}
\usepackage{multicol}
\usepackage{amsbsy}
\usepackage{amssymb}
\usepackage{capt-of}%%To get the captionx
\usepackage[T1]{fontenc} % Use 8-bit encoding that has 256 glyphs
\usepackage{physics}
\usepackage{multicol} % Used for the two-column layout of the document
\usepackage{multirow}
\usepackage[table]{xcolor}
\usepackage{lscape}
\usepackage[hang, small,labelfont=bf,up]{caption} \usepackage[title]{appendix}
\usepackage[format=plain,
            labelfont={bf}]{caption}

\usepackage{booktabs} % Horizontal rules in tables
\usepackage{float} % Required for tables and figures in the multi-column environment - they need to 
\usepackage{hyperref} % For hyperlinks in the PDF
\usepackage{graphicx} % Para imagens
\usepackage{epstopdf} %para imagens .eps
\usepackage{paralist} % Used for the compactitem environment which makes bullet points with less 
\usepackage{subcaption}
\usepackage[framemethod=TikZ]{mdframed}
\def\be{\begin{equation}}
\def\ee{\end{equation}}
\def\bea{\begin{eqnarray}}
\def\eea{\end{eqnarray}}

\newcommand{\Z}{\mathbb{Z}}

\newcommand{\C}{\mathbb{C}}

\newcommand{\V}{\mathcal{V}}
\newcommand{\antiD}{\overline{\text{D3}}}
\newcommand{\ed}{\text{d}}

\newcommand{\Ricci}{\mathcal{R}}
\newcommand{\Donder}{\mathcal{G}}
\newcommand{\M}{\mathcal{M}}
\newcommand{\Op}{\mathcal{O}}
\newcommand{\K}{\mathcal{K}}

\newcommand{\uPhi}{\mathrm{\Phi}}

\newcommand{\Lag}{\mathcal{L}}
\newcommand{\T}{\mathcal{T}}

\usepackage{cancel}
  
\RequirePackage{color}
\usepackage{colortbl}
\definecolor{green2}{cmyk}{0, 1, 0.5, 0.3}
\definecolor{green3}{cmyk}{1, 0.75, 1.0, 0.0}
\definecolor{darkred}{cmyk}{0.2, 0.9, 1.0, 0.2}

\definecolor{lightgreen}{cmyk}{0.2, 0, 0.2, 0.2}
\definecolor{lightgray}{cmyk}{0.1,0.2,0,0.1}
\definecolor{lightgray2}{cmyk}{0.4,0.4,0,0.8}
\definecolor{black}{cmyk}{1.0,1.0,1.0,1.0} 
  \hypersetup{
    colorlinks=true,
    linkbordercolor={white},
    linkcolor={darkred},
    citecolor={darkred},      
    urlcolor={darkred},
}

\title{\huge Gravity at the Tip of the Throat}

\author[a]{Bruno Valeixo Bento,}
\author[b]{Dibya Chakraborty,}
\author[a]{Susha Parameswaran,}
\author[c]{Ivonne Zavala}

\affiliation[a]{Department of Mathematical Sciences, University of Liverpool, Liverpool L69 7ZL}
\affiliation[b]{Facultad de Ciencias Físico-Matemáticas,
Benemérita Universidad Autónoma de Puebla,
C.P. 72570, Puebla, Mexico}
\affiliation[c]{Physics Department, Swansea University, SA2 8PP, UK}

\emailAdd{Bruno.Bento@liv.ac.uk}
\emailAdd{dibyac@fisica.ugto.mx}
\emailAdd{susha@liv.ac.uk}
\emailAdd{e.i.zavalacarrasco@swansea.ac.uk}

\abstract{We study the gravitational signatures that arise from compactifying Type IIB supergravity on a compact space containing a Klebanov-Strassler warped throat. 
After reviewing the dimensional reduction of the 10d graviton and explicitly obtaining the equations of motion for the 4d tensor $h_{\mu\nu}$, vector $h_{\mu n}$ and scalar $h_{mn}$ modes, we find the masses and wavefunctions of the Kaluza-Klein tower of spin-2 states. We explore how the masses and wavefunctions depend on the balance between the strength of the warping and the size of the bulk, and how these relate to the range and strength of the interactions which correct the Newtonian gravitational potential.  By computing the modified Newtonian potential for sources on a brane somewhere along the throat, and applying consistency constraints on the Klebanov-Strassler parameters, we obtain predictions for the phenomenological parameter space.  In the case of a fully warped throat, and depending on where the brane is along the throat, these predictions are narrow in range and consistent with current observational and experimental constraints.  We also begin an exploration of gravitational wave signatures of KK gravitons in warped throats, finding that strong warping can bring the corresponding frequencies down to the windows of current and proposed experiments.}

\begin{document}
\maketitle

\section{Introduction}
\label{S:Introduction}

The recent success of the LIGO and Virgo collaboration in directly observing gravitational waves (GW) from the merger of two black holes \cite{LIGO2016} kick-started the era of GW astronomy. Since then several signals were detected, not only originating from black hole mergers \cite{LIGOScientific:2016sjg,LIGOScientific:2017bnn,Marion:2017enj,LIGOScientific:2017ycc}, but also from black hole-neutron star \cite{LIGOScientific:2021qlt} and binary neutron star \cite{LIGOScientific:2017vwq} mergers, with associated electromagnetic signals which can be used to extract more information from these events. In particular, the study of GW signals can be used to test General Relativity (GR) in an unprecedented way \cite{Berti:2015itd,LIGOScientific:2016lio,LIGOScientific:2018dkp,LIGOScientific:2019fpa,Johnson-McDaniel:2019zkl,LIGOScientific:2020tif,LIGOScientific:2021sio}, constraining deviations from GR and therefore alternative theories of gravity and quantum gravity completions, such as string theory. With several ground and space-based experiments, such as the Einstein Telescope (ET) \cite{Punturo:2010zz} and LISA \cite{eLISA:2013xep,LISA:2017pwj}, planned for the near future, and interest in GW searches at ultra-high frequencies (UHF) in the range MHz--GHz \cite{Aggarwal:2020olq} not covered by these experiments, the gravitational signals of modifications to GR will have the potential to test any theory (such as a UV completion) in which they arise.

This exciting progress in gravitational wave detection is complemented by a host of other diverse experiments and observations. From torsion table-top experiments, astronomical tests \cite{Adelberger:2003zx} and atom interferometry \cite{Dimopoulos:2006nk}, to the Event Horizon telescope \cite{EventHorizonTelescope:2019dse,Psaltis:2018xkc} and collider searches \cite{Murata:2014nra}, GR is being tested in all possible regimes, with strong and weak field tests. One should ultimately combine all these results, looking for how they match, differ or complement each other, in order to know what kind of deviations of GR are still possible and which are excluded \cite{Baker:2014zba}. A useful way to combine some of these tests is by choosing a common parameterisation, e.g. expressing the results in terms of a correction to the Newtonian potential \cite{Murata:2014nra} --- we will see that one can compare a specific string theory compactification setup with several experimental and observational results by looking at such corrections in the form of a single Yukawa interaction. 

A characteristic feature of string theory is the presence of extra dimensions and the need to obtain a 4d effective field theory (EFT) compatible with all available observations. One usually considers the compactification of the extra dimensions onto an internal compact space, which results in a 4d EFT in which each higher-dimensional field gives rise to an infinite tower of massive modes, known as the Kaluza-Klein (KK) tower\footnote{The existence of such a discrete tower of states relies on the compactness of the internal space --- if the extra dimensions are not compact, the spectrum will be a continuum of states (e.g. the spectrum of \cite{Randall:1999vf} is continuous whereas the very similar setup in \cite{Randall:1999ee} gives a discrete spectrum because the extra dimension is now compact). This has a direct impact on the form of the corrections to the Newtonian potential that arise from these extra-dimensional models, which will take a Yukawa-type form for compact cases such as \cite{Randall:1999ee} but a power-law form for non-compact cases such as \cite{Randall:1999vf}.}. Since in string theory gravity is described by the ten-dimensional graviton, the KK towers include a tower of massive graviton KK modes which might have direct effects on gravitational waves and other gravitational effects \cite{Seahra:2004fg,Clarkson:2005eb,Clarkson:2006pq,Chakraborty:2017qve,Andriot:2017oaz,Andriot:2019hay,Andriot:2021gwv,Du:2020rlx} . These massive graviton states can be integrated out if their masses are much higher than the energy scale of interest, such that the low-energy 4d theory gives simply GR. However, for high enough energies, i.e. for energies close to the masses of these states, the first KK modes will start contributing with corrections to the effective theory and, in particular with corrections to the Newtonian potential.  

On the other hand, within string theory, our Universe could be confined to a (3+1)-dimensional brane (or stack of branes), since the states giving rise to the Standard Model can come from open strings which end on different types of branes --- these states are then confined to live on the brane and cannot directly probe the extra dimensions. The brane itself could be located at the tip of a warped throat in the internal compact space --- the warped throat allows the natural high scale of the higher-dimensional theory, typically the string scale, to be supressed on the brane, helping to bridge the gap between the UV scales considered in string theory and the observed IR scales of the 4d theory. Warped throats are also useful in de Sitter constructions, such as KKLT \cite{KKLT} and LVS \cite{originalLVS}, being responsible for the suppression of the naturally high scale of an $\antiD$-brane responsible for uplifting an AdS minimum into dS. An explicit solution of Type IIB supergravity describing a warped throat, known as the Klebanov-Strassler (KS) solution, uses a warped deformed conifold as the internal space \cite{KS2000supergravity} --- this is a non-compact solution, but one usually considers smoothly gluing a finite portion of this solution to a compact Calabi-Yau 3-fold (CY$_3$), such that the internal space is compact.

In this work we consider a flux compactification of Type IIB supergravity in the presence of a warped throat and study its effects on the gravitational sector of the 4d EFT.  Our main focus is the corrections to the Newtonian potential which can be compared to observations across diverse scales \cite{Murata:2014nra}. The dimensional reduction of a D-dimensional gravitational theory down to 4d was performed in \cite{Andriot:2017oaz}, with warping taken into account, in order to study the effects of extra dimensions in gravitational wave signals. The effects of the warping are further explored in \cite{Andriot:2019hay,Andriot:2021gwv} in the context of warped toroidal backgrounds. Focusing specifically on Type IIB supergravity, we dimensionally reduce the 10d action down to 4d by considering the warped background to be described by a compact CY$_3$ with a warped throat described by the KS solution. The effects of this warped geometry on the tower of KK states were previously considered in \cite{Tye:2005qs}, where the mass spectrum of graviton KK modes was obtained. We reproduce the results found in \cite{Tye:2005qs}, paying moreover careful attention to the normalisation of the graviton KK mode wavefunctions which provide the couplings to other modes in the theory. The importance of this normalisation was already emphasised in \cite{Shiu:2007tn}, where it was noted that higher KK modes have stronger couplings when considering a KS warped throat rather than the Randall-Sundrum model (RSI) \cite{Randall:1999ee}, the latter giving a good approximation only away from the tip of the throat. Finally, we consider a braneworld model within this warped Type IIB setup and study the corrections to the Newtonian potential between masses living on the brane due to the presence of the KK tower (this was done in the context of RSI in \cite{Callin:2004detail,Callin:2004short}).  We identify the exclusion region in the parameter space for a Yukawa-type correction to the Newtonian potential arising from a Type IIB brane model in which the Standard Model hierarchy is achieved by placing the brane somewhere along a KS warped throat.  We moreover begin a study on the implications of warped throats for gravitational wave experiments, identifying points in the parameters space of the KS solution which bring gravitational wave frequencies down to observable scales.

The paper is organised as follows. In Section \ref{sec:4dEFT} we obtain the 4d wave equations describing the tensor, vector and scalar modes that descend from the 10d graviton and depend on the geometry of the compact space, which we take to include a cut-off KS warped deformed conifold. 
%The dimensional reduction was previously done in \cite{Andriot:2017oaz}, considering a generic D-dimensional theory. 
In Section \ref{sec:conifoldKKtower} we find the corresponding KK tower of tensor modes, which describes an infinite set of massive spin-2 fields in 4d, by computing the masses and wavefunctions of the tensor modes, with their respective normalisations. 
%We find that the zero mode has a constant wavefunction on the compact internal space, whereas the massive modes are localised at the tip of the throat, in agreement with \cite{Tye:2005qs}, where the KK spectrum on a warped deformed conifold was also computed. 
In Section \ref{sec:NewtPotCorrections} we use these results to compute the corrections to the Newtonian gravitational potential due to the massive tower and compare these predicitions with current experimental and observational constraints.  In Section \ref{sec:GW} we begin an exploration of gravitational wave signatures of a warped throat.  Finally, in Section \ref{S:Conclusions}, we provide a summary of our results and an outlook to future research.

\section{Dimensional Reduction of the 10d Graviton}
\label{sec:4dEFT}

In this section we study the dimensional reduction of the gravitational sector in 10d Type IIB supergravity. We are interested in the equations for the 4d modes arising from the 10d graviton, which not only provide their masses in terms of the characteristics of the compact internal space, but also determine how the different modes couple to one another and to modes descending from other 10d fields. In Section \ref{sec:4dEFT_waveEq} we obtain the 10d wave equation for fluctuations $h_{MN}$ of the 10d metric $G_{MN}$ (\ref{eq:10dWaveEq}) on a fully general background. In Section \ref{sec:4dEFT_compactification} we discuss the deformed conifold and the Klebanov-Strassler solution, which will constitute part of our 6d compact space together with a compact Calabi-Yau 3-fold. Finally, in Section \ref{sec:dimensionalReduction} we dimensionally reduce the wave equation (\ref{eq:10dWaveEq}) using the Klebanov-Strassler background and obtain the 4d equations describing the tensor, vector and scalar modes that arise from $h_{MN}$.

%%%%%%%%%%%%%%%%%%%%%%%%%%%%%%%%%%%%%%%%%%%%%%%%%%%%%%%%%%%%%%%%%%%%%%%%%%%%
%	2.1 PERTURBED EINSTEIN EQUATIONS IN TYPE IIB
%%%%%%%%%%%%%%%%%%%%%%%%%%%%%%%%%%%%%%%%%%%%%%%%%%%%%%%%%%%%%%%%%%%%%%%%%%%%

\subsection{Perturbed Einstein equations in Type IIB}
\label{sec:4dEFT_waveEq}

We start with the 10d action for Type IIB supergravity, which is the low-energy EFT of Type IIB string theory, valid at energies $E\ll M_s\sim 1/\sqrt{\alpha'}$,
\begin{align}
    S_{IIB}^{E} 
    =& \frac{1}{2\kappa^2}\int d^{10}x\sqrt{-G}
    \left(R - \frac{1}{2}(\partial_M\uPhi)(\partial^M\uPhi) - \frac{g_s}{2}e^{-\uPhi} |H_3|^2\right) \nonumber \\ 
    &-\frac{1}{2\kappa^2}\int d^{10}x\sqrt{-G} \left(\frac{e^{2\uPhi}}{2}|F_1|^2 + \frac{g_s}{2}e^{\uPhi}|\Tilde{F}_3|^2 + \frac{g_s^2}{4}|\Tilde{F}_5|^2\right) \nonumber \\
    &- \frac{g_s^2}{4\kappa^2}\int C_4\wedge H_3\wedge F
    _3 \,,
    \label{eq:TypeIIB_action_EinsteinFrame}
\end{align}
where $R$ is the Ricci scalar, $\uPhi$ is the dilaton, $H_3$ is the field-strength of the NS 2-form $B_2$ and $F_p$ is the field-strength of the RR $(p-1)$-form $C_{p-1}$, 
\begin{align}
    & H_3 = dB_2  \,,
    && \Tilde{F}_3 = F_3 - C_0 H_3 \,,\\
    & F_p = dC_{p-1} \,,
    && \Tilde{F}_5 = F_5 - \frac{1}{2}C_2\wedge H_3 + \frac{1}{2}B_2\wedge F_3  \,,\\
    & |F_p|^2 = \frac{1}{p!}F_{M_1...M_p}F^{M_1...M_p}\,.
\end{align}
The action (\ref{eq:TypeIIB_action_EinsteinFrame}) is written in the 10d Einstein frame, with $2\kappa^2 = (2\pi)^7g_s^2\alpha'^4$, and must be supplemented with the self-duality condition $\Tilde{F}_5=\star \Tilde{F}_5$.
We use capital Latin letters $M,N,P,Q$ for 10d indices, while 4d indices are represented by lower-case Greek letters $\mu,\nu,\rho,\sigma$ and 6d ones by lower-case Latin letters $m,n,p,q$. 
The (trace-reversed) Einstein equations for Type IIB take the form\footnote{We usually write 
\begin{equation}
	R_{MN}
	= \T_{MN} - \frac{1}{8}(\T + \Lag)G_{MN} 
	= T_{MN} - \frac{1}{8} G_{MN} T\,,
\end{equation}
in terms of the energy-momentum tensor $T_{MN}\equiv -\frac{2}{\sqrt{-G}}\frac{\delta(\sqrt{-G}\Lag)}{\delta G^{MN}} = \T_{MN} + \frac{1}{2}G_{MN}\Lag$ and the trace $T=G^{MN}T_{MN}$. However it is useful here to have $\T$ and $\Lag$ separately.
}
\begin{equation}
    R_{MN} + \frac{1}{8}(\T + \Lag)G_{MN} - \T_{MN} = 0\,,
    \label{eq:TypeIIB_EinsteinEq}
\end{equation}
where we define 
\begin{align}
	\T_{MN} \equiv -\frac{\delta\Lag}{\delta G^{MN}} 
	=& \frac{1}{2}(\partial_M\uPhi)(\partial_N\uPhi) 
	+ \frac{e^{2\uPhi}}{2}(\partial_M C_0)(\partial_N C_0)
	+ \frac{g_s}{4}e^{-\uPhi}(H_3)_{MPQ}(H_3)_N^{\phantom{M}PQ} \nonumber \\
	\hspace{10mm}
	&+ \frac{g_s}{4}e^{\uPhi}(\tilde{F}_3)_{MPQ}(\tilde{F}_3)_N^{\phantom{N}PQ} 
	+\frac{g_s^2}{4\times 4!}(\tilde{F}_5)_{MPQRS}(\tilde{F}_5)_N^{\phantom{N}PQRS} \, 
	\label{eq:tau_MN} 
\end{align}
so that $\T = G^{MN}\T_{MN}$, and $\Lag$ is the matter Lagrangian that couples to the metric (i.e. does not include the topological Chern-Simons term),
\begin{equation}
	\Lag 
	= - \frac{1}{2}(\partial_M\uPhi)(\partial^M\uPhi) - \frac{e^{2\uPhi}}{2}|F_1|^2 - \frac{g_s}{2}e^{-\uPhi} |H_3|^2 - \frac{g_s}{2}e^{\uPhi}|\Tilde{F}_3|^2 - \frac{g_s^2}{4}|\Tilde{F}_5|^2\,.
	\label{eq:L}
\end{equation}

We want to study the perturbations $h_{MN}$ to a background 10d metric $g_{MN}$. An equation for $h_{MN}$ is obtained from (\ref{eq:TypeIIB_EinsteinEq}) by perturbing $G_{MN}\to g_{MN} + h_{MN}$ to first order in $h_{MN}$, 
\begin{align}
	0 =& ~R_{MN}^{(0)} 
	+\frac{1}{8}(\T^{(0)} + \Lag^{(0)})g_{MN}
	- \T_{MN}^{(0)} \nonumber \\
	&+ R_{MN}^{(1)}
	+\frac{1}{8}(\T^{(1)} + \Lag^{(1)})g_{MN} 
	+\frac{1}{8}(\T^{(0)} + \Lag^{(0)})h_{MN} 
	- \T_{MN}^{(1)} \,,
    \label{eq:TypeIIB_EinsteinEq_perturbed}
\end{align}
where $R_{MN}$, $\T_{MN}$, $\Lag$ have been expanded to linear order in $h_{MN}$, so that $R_{MN}^{(0)}$, $\T_{MN}^{(0)}$, $\Lag^{(0)}$ only depend on the background $g_{MN}$ and $R_{MN}^{(1)}$, $\T_{MN}^{(1)}$, $\Lag^{(1)}$ depend linearly on $h_{MN}$. In particular, $\T^{(1)} = g^{MN}\T_{MN}^{(1)} - g^{MP}h_{PQ}g^{QN}\T_{MN}^{(0)}$. Imposing that the background metric satisfies (\ref{eq:TypeIIB_EinsteinEq}), we obtain
\begin{align}
    & R_{MN}^{(0)} + \frac{1}{8}(\T^{(0)} + \Lag^{(0)})g_{MN} - \T_{MN}^{(0)} = 0 \,,
    \label{eq:Ddim_EinsteinEq_TraceReversed_Background} \\
    & R_{MN}^{(1)}
	+\frac{1}{8}(\T^{(1)} + \Lag^{(1)})g_{MN} 
	+\frac{1}{8}(\T^{(0)} + \Lag^{(0)})h_{MN} 
	- \T_{MN}^{(1)} = 0 \,.
    \label{eq:Ddim_EinsteinEq_TraceReversed_Fluctuations}
\end{align}
Expanding $R_{MN}$ to first order in $h_{MN}$ to find $R_{MN}^{(1)}$,
\begin{equation}
    R_{MN} = R_{MN}^{(0)} + 
    \left(-\frac{1}{2}\Box_D h_{MN} 
    + g^{PQ}\nabla_P\nabla_{(M}h_{N)Q}
    - \frac{1}{2}\nabla_M\nabla_N h_D
    \right)\,,
\end{equation}
where the covariant derivatives $\nabla_M$ are with respect to the background metric $g_{MN}$, $\Box_D = g^{PQ}\nabla_P\nabla_Q$ and $h_D = g^{PQ}h_{PQ}$. 
Commuting the covariant derivatives using 
\begin{equation}
    \nabla_P\nabla_M h_{NQ} = \nabla_M\nabla_Ph_{NQ} + g^{AB}R^{(0)}_{PMNA}h_{QB}
    + g^{AB}R^{(0)}_{PMQA}h_{BN}
\end{equation}
and contracting the $PQ$ indices with $g^{PQ}$
\begin{equation}
    g^{PQ}\nabla_P\nabla_M h_{QN} = g^{PQ}\nabla_M\nabla_Ph_{QN} + g^{PQ}R^{(0)A}_{\phantom{(0)A}NMP}h_{QA}
    + g^{AB}R^{(0)}_{MA}h_{BN} \,,
\end{equation}
we can write $R_{MN}^{(1)}$ as
\begin{equation}
    R_{MN}^{(1)} =
    -\frac{1}{2}\Box_D h_{MN}
    + g^{PQ}h_{Q(M}R^{(0)}_{N)P}
    + R^{(0)S}_{\phantom{(0)S}MNP}g^{PQ}h_{QS} + \nabla_{(M}\Donder_{N)} \,,
\end{equation}
where $\Donder_N$ is defined as 
\begin{equation}
    \Donder_N = g^{PQ}\nabla_P h_{QN} - \frac{1}{2}\nabla_N h_D \,.
\end{equation}
We can now use (\ref{eq:Ddim_EinsteinEq_TraceReversed_Background}) to replace $R^{(0)}_{NP}$, so that 
\begin{equation}
    g^{PQ}h_{Q(M}R^{(0)}_{N)P} 
    = - \frac{1}{8}(\T^{(0)} + \Lag^{(0)})h_{MN} + g^{PQ}h_{Q(M}\T_{N)P}^{(0)} \,,
\end{equation}
and the first order equation becomes
\begin{align}
	-\frac{1}{2}\Box_D h_{MN} 
	+ R^{(0)S}_{\phantom{(0)S}MNP}g^{PQ}h_{QS} + \nabla_{(M}\Donder_{N)}
	+\frac{1}{8}(\T^{(1)} + \Lag^{(1)})g_{MN} & \nonumber \\ 
	+ g^{PQ}h_{Q(M}\T_{N)P}^{(0)}
	- \T_{MN}^{(1)} &= 0 \,.
	\label{eq:10dWaveEq_preGauge}
\end{align}
Note that, by definition of $\T_{MN}$, $\Lag^{(1)} = \T_{MN}^{(0)}h^{MN}$, so that $\T^{(1)} + \Lag^{(1)} = g^{MN}\T_{MN}^{(1)}$, where $\T_{MN}^{(1)}$ is obtained by expanding $\T_{MN}$ to first order in $h_{MN}$, 
\begin{align}
	\T_{MN}^{(1)} = \Bigg(&
	\frac{g_s}{2}\Big(e^{-\uPhi}(H_3)_M^{\phantom{M}RS}(H_3)_{NRP} 
	+ e^{\uPhi}(\tilde{F}_3)_M^{\phantom{M}RS}(\tilde{F}_3)_{NRP}\Big) \nonumber \\
	& +\frac{g_s^2}{4!}(\tilde{F}_5)_M^{\phantom{M}SIJR}(\tilde{F}_5)_{NPIJR}
	\Bigg) g^{PQ}h_{SQ}\,.
	\label{eq:tau_MN_1}
\end{align}
Finally, choosing the gauge such that $\Donder_N = 0$, known as de Donder (or harmonic) gauge, equation (\ref{eq:10dWaveEq_preGauge}) becomes
\begin{equation}
	\Box_D h_{MN} 
	- 2  R^{(0)S}_{\phantom{(0)S}MNP}g^{PQ}h_{QS}
	- \frac{1}{4}\big(g^{PQ}\T_{PQ}^{(1)}\big)g_{MN}  
	- 2 g^{PQ}h_{Q(M}\T_{N)P}^{(0)}
	+ 2 \T_{MN}^{(1)} = 0 \,.
	\label{eq:10dWaveEq}
\end{equation}
This is a wave equation for the 10d metric fluctuations $h_{MN}$, sourced by $\T_{MN}$. We see that (\ref{eq:10dWaveEq}) depends on the background geometry $g_{MN}$, as well as on the background values of the other Type IIB fields ($\Phi, B_2, C_0, C_2, C_4$). Upon compactifying the Type IIB action from 10d to 4d, (\ref{eq:10dWaveEq}) gives rise to three equations describing tensor, vector and scalar modes in the 4d EFT, and the way these equations are coupled depends on the background geometry (which depends itself on the background solution for the scalars and form fields). In the next subsection we introduce the deformed conifold which, together with a compact CY$_3$, describes our internal space.

%%%%%%%%%%%%%%%%%%%%%%%%%%%%%%%%%%%%%%%%%%%%%%%%%%%%%%%%%%%%%
%  2.2 BACKGROUND SOLUTION - THE WARPED DEFORMED CONIFOLD
%%%%%%%%%%%%%%%%%%%%%%%%%%%%%%%%%%%%%%%%%%%%%%%%%%%%%%%%%%%%%

\subsection{Background solution -- the warped deformed conifold}
\label{sec:4dEFT_compactification}

\def\rs{r_{\ll}}
\def\rl{r_{\gg}}

In order to obtain the 4d EFT, we consider a Type IIB flux compactification with an internal space composed of a compact CY$_3$, which constitutes the bulk of the compact space, together with a warped throat, which can be described by the Klebanov-Strassler solution \cite{KS2000supergravity}, corresponding to a warped deformed conifold\footnote{The Klebanov-Strassel solution is a special case of a broader class of solutions for the deformation of the conifold metric with fluxes that preserve the $SU(2)\times SU(2)\times\Z_2$ symmetry of the geometry \cite{PTconifold}.}, smoothly glued to the bulk. Considering surfaces in $\C^4$ given by $\sum_{i=1}^4 w_i^2 = \epsilon^2$, with $\epsilon$ a non-zero complex parameter (if $\epsilon=0$ the conifold is singular), gives a deformed conifold with metric \cite{candelas1990conifolds,Minasian:1999tt,Aganagic:1999fe,KS2000supergravity} 
\begin{align}
    ds_{con}^2 = \frac{\epsilon^{4/3}}{2}\mathcal{K}(\tau) 
    \Bigg( \frac{1}{3\mathcal{K}^3(\tau)} \left(d\tau^2 + (g^5)^2 \right) 
    &+ \sinh^2(\tau/2) \left( (g^1)^2 + (g^2)^2 \right) \nonumber \\
    &+ \cosh^2(\tau/2) \left( (g^3)^2 + (g^4)^2 \right)
    \Bigg) \,,
    \label{eq:conifold:deformed_conifold_metric}
\end{align}
where $\mathcal{K}(\tau) = \frac{(\sinh(2\tau) - 2\tau)^{1/3}}{2^{1/3}\sinh(\tau)}$ and $g^i$ are a basis of one-forms \cite{candelas1990conifolds}. In this basis, the metric only depends on $\tau$ which is related to the radial direction along the cone. 
For large $\tau$ the metric approaches the singular conifold metric
\begin{align}
    ds_{con}^2 
    &= d\rl^2 + \rl^2\left(\frac{1}{9}(g^5)^2 +\frac{1}{6}\sum_{i=1}^{4}(g^i)^2 \right),
    \quad \text{ as } \tau\to\infty \,,
    \label{eq:metric_conifold_largetau}
\end{align}
where the coordinate $\rl$ is defined as $\rl^2 = \frac{3}{2^{5/3}}\epsilon^{4/3}e^{2\tau/3}$, while for small $\tau$
\begin{align}
    ds_{con}^2 
    &= d\rs^2 + \frac{\rs^2}{8}d\mathrm{\Omega}_2^2 + R_\epsilon^2 d\mathrm{\Omega}_3^2,
    \quad \text{ as } \tau\to 0 \,,
    \label{eq:metric_conifold_smalltau}
\end{align}
where $\rs^2 = \frac{\epsilon^{4/3}}{4}\left(\frac{2}{3}\right)^{1/3}{\tau}^2$, $d\mathrm{\Omega}_2^2=((g^1)^2+(g^2)^2)$ is the metric of an $S^2$ and $d\mathrm{\Omega}_3^2=((g^3)^2+(g^4)^2+\frac{1}{2}(g^5)^2)$ is the metric of an $S^3$. We see that $\rs\rightarrow 0$ as $\tau\rightarrow 0$, and we are left with the metric for an $S^3$ whose size $R_\epsilon$ is controlled by the deformation parameter $\epsilon$,
\begin{align}
    R_\epsilon^2 = \frac{\epsilon^{4/3}}{2}  \left(\frac{2}{3}\right)^{1/3} \,.
\end{align}

The Klebanov-Strassler solution is obtained in \cite{KS2000supergravity}, using the deformed conifold as an ansatz for the internal space of a warped solution of Type IIB supergravity. The 10d metric can be written as
\begin{align}
    ds_{10}^2 &= e^{2A} g_{\mu\nu}dx^{\mu}dx^{\nu} + e^{-2A} g_{mn}dy^mdy^n \,.
    \label{eq:metric}
\end{align}
Notice that this results in a non-compact solution, which is not suitable for phenomenological applications. Therefore, in practice only a finite portion of the Klebanov-Strassler solution is used and glued to a compact CY$_3$ (or more generally, a compact 6d \textit{bulk}). Due to the symmetries of the conifold, the warp factor can only depend on $\tau$ and is given by \cite{KS2000supergravity}
\begin{align}
    e^{-4A} &=
    2^{2/3}\frac{(\alpha' g_sM)^2}{\epsilon^{8/3}} I(\tau) \,,
    \quad\quad I(\tau)\equiv \int_{\tau}^{\infty} dx ~ \frac{x\coth(x) - 1}{\sinh^2{x}}(\sinh(2x)-2x)^{1/3} \,. 
    \label{eq:deformedConifold_warpfactor}
\end{align}

For the conifold in the small $\tau$ limit the metric takes the form
\begin{align}
    ds_{10}^2 
    = e^{2A} g_{\mu\nu}dx^{\mu}dx^{\nu} + e^{-2A} \left(d\rs^2 + \frac{\rs^2}{8}d\mathrm{\Omega}_2^2 + R_\epsilon^2 d\mathrm{\Omega}_3^2 \right).
\end{align}

\noindent The physical size of the $S^3$ is $R^2_{S^3}=e^{-2A}R_\epsilon^2$, which does not depend on the deformation parameter $\epsilon$
\begin{align}
    R_{S^3}^2 &= e^{-2A} R_\epsilon^2
    = \left(\frac{2^{2/3}}{3^{1/3}} I^{1/2}(\tau)\right)(g_sM)\alpha' 
	\overset{\tau = 0}{\approx} (g_s M)\alpha' \,,
\end{align}
from which follows the condition $g_sM\gg 1$ for a well under control supergravity (i.e. such that the $\alpha'$-expansion with which the Type IIB low-energy supergravity used in \cite{KS2000supergravity} is derived is under control).

Far away from the tip, the warp factor becomes \cite{KT}
\begin{align}
	e^{-4A} \approx \frac{L^4}{\rl^4}\left[1 + \frac{3 g_sM}{8\pi K} + \frac{3g_s M}{2\pi K}\log\Big(\frac{\rl}{r_{UV}}\Big) \right] \,,
	\label{E:mouthwarp}
\end{align}
where we define 
\begin{align}
	L^4 = \frac{27\pi}{4}\frac{g_sMK}{(2\pi)^4}~l_s^4 \,,
\end{align}
with $K$ being the flux number through the non-compact cycle\footnote{More precisely, the cycle which is non-compact in the non-compact Klebanov-Strassler solution, but becomes compact when only a finite portion is glued to the compact bulk.} and $r_{UV}$ the value of $\rl$ where we cut the throat and glue it to the bulk

\begin{equation}
	r_{UV} \equiv \frac{3^{1/2}}{2^{5/6}}\epsilon^{2/3}e^{\tau_{UV}/3} \,.
	\label{eq:rUV}
\end{equation}

The background warp factor, $e^{-4A}$, that solves the 10d Einstein equations in the presence of fluxes, is only fixed up to a constant shift in $y^m$, motivating the following form \cite{frey2009universal}
\begin{equation}
    e^{-4A(y)} = e^{-4A_0(y)} + c(x).
\end{equation}
On the other hand, $g_{mn}\rightarrow\lambda g_{mn}$ together with $e^{2A}\rightarrow \lambda e^{2A}$ is a gauge redundancy of the 6d part of the metric \cite{Giddings:2005ff,Aparicio:2015psl}, and we can choose $\lambda=c(x)^{1/2}$ to rewrite (\ref{eq:metric}) as
\begin{equation}
    ds^2 
    = \left[1 + \frac{e^{-4A_0(y)}}{c(x)}\right]^{-1/2} e^{2\mathrm{\Omega}(x)} g_{\mu\nu}dx^{\mu}dx^{\nu}  + \left[1 + \frac{e^{-4A_0(y)}}{c(x)}\right]^{1/2}  c(x)^{1/2} g_{mn}dy^mdy^n \,,
    \label{eq:10dmetric_ourconventions}
\end{equation}
and define the warp factor as 
\be\label{WF}
H=1+\frac{e^{-4A_0(y)}}{c(x)} \,.
\ee
In the limit $c(x)\rightarrow\infty$, $H\to 1$ and we naturally recover the unwarped case. Approaching that limit we identify $c(x)=\mathcal{V}^{2/3}$, where $\mathcal{V}\,l_s^6$ is the unwarped volume of the compact space, as we would expect for a volume modulus once coordinates are chosen such that $V_{6} = \int d^6y\sqrt{g_6} = l_s^6$. The factor $e^{2\mathrm{\Omega}(x)}$ anticipates the Weyl-rescaling necessary to go to the Einstein frame in 4d. 

Importantly, the question of whether a region is warped, through $H(\tau)$, depends not only on the solution $e^{-4A_0(\tau)}$, but also on the size of the bulk to which the throat is glued. The warping will dominate when $e^{-4A_0(\tau)}\gg c$ and be negligible when $e^{-4A_0(\tau)}\ll c$, so that effectively the throat ends at $e^{-4A_0(\tau)}\sim c$. This interplay between $e^{-4A_0(\tau)}$ and $c$ gives an interesting intermediate regime of weak warping which was explored in \cite{Bento:2021nbb}. For concreteness, we can define the gluing point $\tau_c$ such that $e^{-4A_0(\tau_c)} = c$, keeping in mind that the gluing must involve a smooth transition between the warped throat and the bulk around $\tau_c$. 
Therefore $\tau_c$ can be defined implicitly in terms of the conifold parameters
\begin{equation}
	\frac{e^{-4A_0(\tau_c)}}{c} = 1 \implies 2^{2/3}\frac{(\alpha' g_s M)^2}{c~\epsilon^{8/3}} = \frac{1}{I(\tau_c)}\,, 
	\label{eq:tauc_parameters}
\end{equation}
which allows us to rewrite the warp factor as
\begin{equation}
	H_{\tau_c}(\tau) = 1 + \frac{I(\tau)}{I(\tau_c)} \,.
	\label{eq:warp_factor_tauc}
\end{equation}

\begin{figure}[t]
     \centering
	\begin{subfigure}[b]{0.48\textwidth}
        	\centering
         	\includegraphics[width=\textwidth]{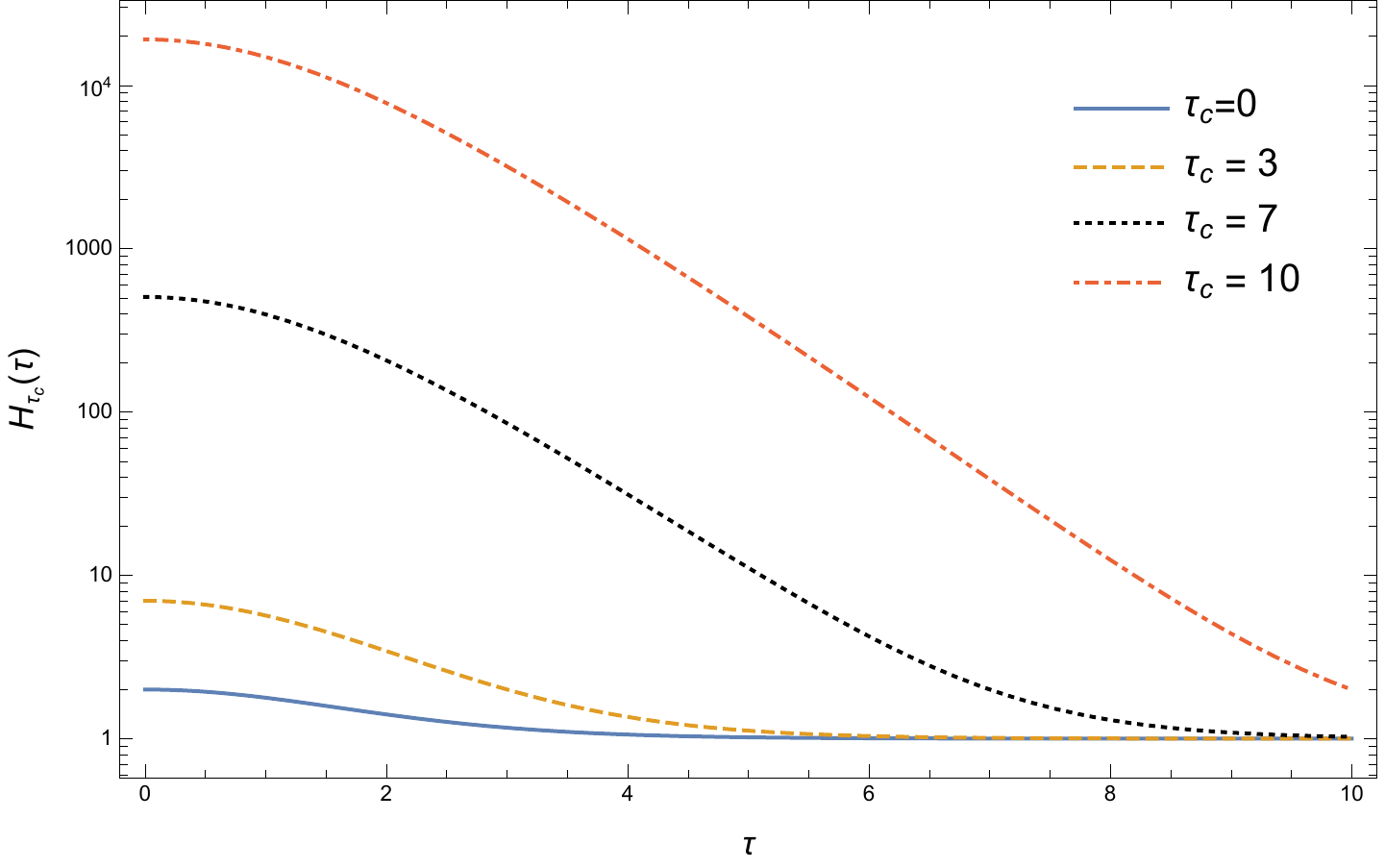}
     	\end{subfigure}
     \hfill
	\begin{subfigure}[b]{0.48\textwidth}
        	\centering
         	\includegraphics[width=\textwidth]{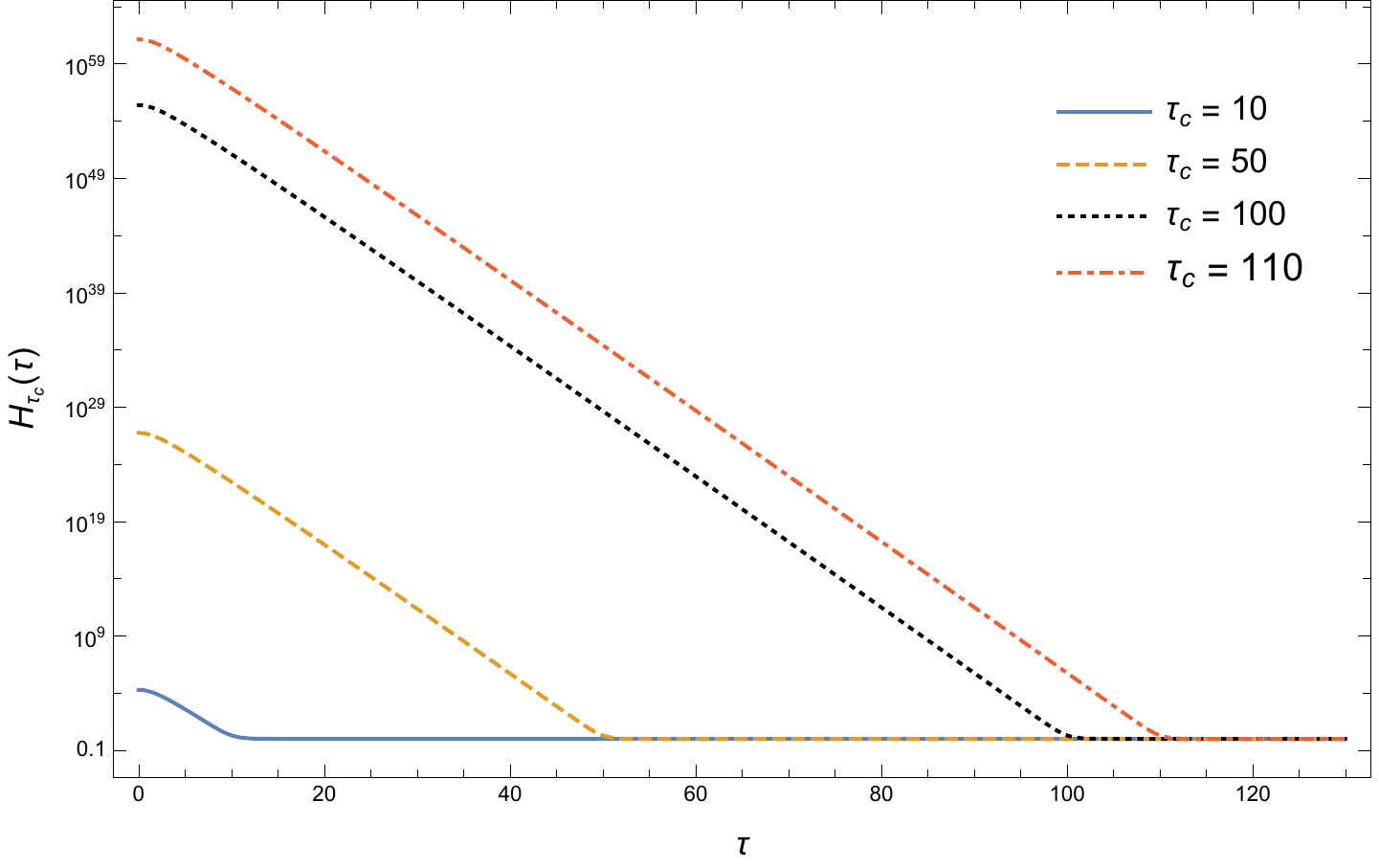}
     	\end{subfigure}
    \caption{Warp factor $H(\tau)$ (in log scale) for different choices of $\tau_c$. We can see how the choice of $\tau_c$ is directly related to the strength of the warping and the size of the warped throat (defined as the region with non-trivial warping).}
    \label{fig:warp_factor}
\end{figure}

\noindent This highlights the fact that the warp factor depends on one parameter only, $\tau_c$. Even though it is implicitly determined by a specific combination of the more familiar parameters through (\ref{eq:tauc_parameters}), it is a convenient parametrisation since a choice of $\tau_c$ has a clear physical interpretation. In Fig.~\ref{fig:warp_factor} we show how the warp-factor $H(\tau)$ behaves for different choices of $\tau_c$.  

Dimensionally reducing the 10d Einstein-Hilbert term in (\ref{eq:TypeIIB_action_EinsteinFrame}) down to 4d using the metric (\ref{eq:10dmetric_ourconventions}), we obtain
\begin{equation}
    S  = \frac{V_w^0}{2\kappa^2}\int d^{4}x \sqrt{-g_4}  R_4 + ... \,,
    \label{eq:dimRed_kineticTerms}
\end{equation}
where the Weyl factor $e^{2\mathrm{\Omega}(x)}$ is defined as 
\begin{equation}
    	e^{2\mathrm{\Omega}(x)} = \frac{V_w^0}{c(x)^{3/2}\int d^6y \sqrt{g_6} ~H} = \frac{V_w^0}{V_w} \,,
    	\quad V_w = ~c(x)^{3/2}\int d^6y \sqrt{g_6} ~H\,.
	\label{eq:Vw_definition}
\end{equation}
The choice of constant $V_w^0$, which ensures that the Weyl factor is dimensionless, is arbitrary\footnote{Note that all the mass-scales in units of $M_{p}$ will be independent of the normalisation of $e^{2\mathrm{\Omega}(x)}$.}. The most convenient choice is $V_w^0 \equiv \langle V_w \rangle$, such that $\langle e^{2\mathrm{\Omega}(x)}\rangle =1$, i.e. the two frames are the same at the vev and we can relate the string scale with the 4d Planck scale using frame independent volumes.  With this choice, the definition of $M_P$ is
\begin{equation}
    \frac{M_P^2}{2} = \frac{1}{2\kappa_4^2} = \frac{V_w^0}{2\kappa^2}
	\implies
	\frac{m_s}{M_P} = \frac{g_s}{\sqrt{4\pi\V_w^0}} \,.
	\label{eq:msMp}
\end{equation}
Note that the relation between the string scale\footnote{In our conventions, $(2\pi)^2\alpha' = l_s^2$ and $m_s=l_s^{-1}$, which differs from $M_s=1/\sqrt{\alpha'} = 2\pi m_s$.} and the 4d Planck scale is only determined once the volume modulus is fixed (analogously to what happens with the dilaton in 10d, where the physical gravitational coupling can only be related to the string frame coupling once the dilaton is fixed). In what follows we assume that all moduli are stabilised, including in particular the volume modulus, and work at the resulting vacuum. Therefore, we use the metric 
\begin{equation}
    ds^2 
    = H^{-1/2} g_{\mu\nu}dx^{\mu}dx^{\nu}  + H^{1/2}  c^{1/2} g_{mn}dy^mdy^n \,,
    \label{eq:10dmetric}
\end{equation}
by setting $e^{2\mathrm{\Omega}(x)}$ to its value at the vacuum and letting $c=\langle c(x)\rangle$ and $V_w = V_w^0$ from now on. 

For the simple case we are considering, with one bulk and one throat, $V_w$ is the sum of the bulk volume $V_\mathcal{B}$ and the throat volume $V_{th}$, 
\be\label{eq:Vw_total}
	V_w = V_{\mathcal{B}} + V_{th} \,,
\ee
where
\begin{align}
	V_{\mathcal{B}} &\approx c^{3/2}\int_{bulk} d^6 y \sqrt{g_6} = c^{3/2} l_s^6 \,,
	\label{eq:V_Bulk}
\end{align}

\noindent whereas for $V_{th}$ we find
\begin{align}
	V_{th} &\approx c^{1/2}\int_{throat} d^6y \sqrt{g_6} e^{-4A_0(y)} \nonumber \\
	&= 2^{5/3}\frac{\pi^3}{3}\frac{(\alpha'g_sM)^2}{\epsilon^{8/3}}\epsilon^4c^{1/2} \int_0^{\tau_c} d\tau \sinh^2(\tau) I(\tau) \nonumber \\
	&\approx\frac{9\pi^3}{16\cdot 2^{1/3}}\frac{(\alpha'g_sM)^2}{\epsilon^{8/3}}\epsilon^4c^{1/2}  \left\{1 + e^{2\tau_c/3}\Big(\frac{2}{3}\tau_c -1\Big) \right\} \,,
\end{align}
where we used the large $\tau$ behaviour of the integrand to perform the integral. Expressing this in terms of $r_{UV}$ (\ref{eq:rUV}), using the flux quantisation condition that defines the flux number $K$ and the fact that strong warping requires $\frac{2\pi K}{g_sM}\gg 1$, we find
\be\label{eq:V_throat}
	V_{th} \approx 2\pi^4 g_s M K (\alpha')^2 c^{1/2} r_{UV}^2 = \frac{g_s MK}{8}\Big(\frac{R_{throat}}{l_s}\Big)^2 ~ l_s^6 \,,
\ee
where we defined the physical size of the throat, $R_{throat} = c^{1/4} r_{UV}$.
This can be compared with the volume of the unwarped deformed conifold of physical size $R_{con}$, 
\be\label{eq:V_uwarped_conifold}
	V_{con} = \frac{(2\pi)^3}{81}\left(\frac{R_{con}}{l_s}\right)^6 ~ l_s^6\,.
\ee

\iffalse
Putting these together, we find
\be\label{eq:Vw_total_explicit}
	\V_w \approx c^{3/2}\left(1 + \frac{g_s MK}{8c^{3/2}}\Big(\frac{r_{UV}}{l_s}\Big)^2 \right) \,. 
\ee
\fi

Our goal is to find the towers of KK modes that follow from the 10d metric after compactifying the Type IIB action on the background geometry (\ref{eq:10dmetric}) and identify their effects on 4d gravitational interactions. In the next subsection we decompose the wave equation (\ref{eq:10dWaveEq}) using the background (\ref{eq:10dmetric}) and find the equations for infinite towers of 4d tensor, vector and scalar modes. 

%%%%%%%%%%%%%%%%%%%%%%%%%%%%%%%%%%%%%%%%%%%%%%%%%%%%%%%%%%%%%%%%%%%%%%%%
%   2.3 WAVE EQUATIONS IN THE WARPED DEFORMED CONIFOLD
%%%%%%%%%%%%%%%%%%%%%%%%%%%%%%%%%%%%%%%%%%%%%%%%%%%%%%%%%%%%%%%%%%%%%%%%

\subsection{Wave equations in the warped deformed conifold}
\label{sec:dimensionalReduction}

Using the background metric (\ref{eq:10dmetric}) we can  rewrite the wave equation (\ref{eq:10dWaveEq}) in terms of the 4d and 6d operators, which gives equations for $h_{\mu\nu}$, $h_{\mu n}$ and $h_{mn}$. 

On the other hand, the background solution for the metric (\ref{eq:10dmetric}) together with the requirement that the 4d spacetime preserves Poincaré invariance implies that the background solutions for $H_3,F_3,F_5$ take the form
\begin{align}
	H_3 &= \frac{1}{3!} H_{mnp} ~\ed y^m\wedge \ed y^n\wedge \ed y^p \,, \\
	F_3 &= \frac{1}{3!} F_{mnp} ~\ed y^m\wedge \ed y^n\wedge \ed y^p \,, \\
	F_5 &= \frac{1}{5!} (1 + \star_{10}) \sqrt{g_4}~\ed\alpha \wedge \ed x^\mu\wedge \ed x^{\nu}\wedge \ed x^\rho\wedge \ed x^\sigma \,,
\end{align}
where we also use the self-duality condition that must be imposed on $F_5$, $\star_{10}$ is the 10d Hodge star operator and $\alpha=H^{-1}$ \cite{GKP}. Recalling that $\alpha=\alpha(y)$, we can write these in components as
\begin{align}
	(H_3)_{MNP} &= \delta_{M}^{m} \delta_{N}^{n} \delta_{P}^{p} H_{mnp} \\
	(F_3)_{MNP} &= \delta_{M}^{m} \delta_{N}^{n} \delta_{P}^{p} F_{mnp} \\
	(F_5)_{MPQRS} &= \delta_{M}^{m}\delta_{P}^{\pi}\delta_{Q}^{\eta}\delta_{R}^{\rho}\delta_{S}^{\sigma} \sqrt{g_4} (\partial_m\alpha)\epsilon_{\pi\eta\rho\sigma} \nonumber \\
	&+  \delta_{M}^{a}\delta_{P}^{p}\delta_{Q}^{q}\delta_{R}^{r}\delta_{S}^{s} \frac{\sqrt{-G}}{5!}\sqrt{g_4}(\partial_m\alpha)\epsilon_{\pi\eta\rho\sigma} \epsilon^{m\pi\eta\rho\sigma}_{\phantom{m\nu\pi\eta\rho}apqrs} \,.
\end{align}

Similarly, the background values of $\partial_M\Phi$, $\partial_M C_0$ are only non-zero for internal components. We can therefore compute the different components of $\T_{MN}^{(0)}$ (\ref{eq:tau_MN}) and $\T_{MN}^{(1)}$ (\ref{eq:tau_MN_1}),
\begin{align}
	\T_{\mu\nu}^{(0)} &= \frac{g_s^2}{4c^{1/2}H^5}(\partial H)^2 g_{\mu\nu}\,, \\
	\T_{\mu n}^{(0)} &= 0 \,, \\
	\T_{mn}^{(0)} &=  \frac{1}{2}(\partial_m\uPhi)(\partial_n\uPhi) 
	+ \frac{e^{2\uPhi}}{2}(\partial_m C_0)(\partial_n C_0)
	+ \frac{g_s}{4cH}e^{-\uPhi}(H_3)_{mpq}(H_3)_n^{\phantom{n}pq} \nonumber \\
	\hspace{10mm}
	&+ \frac{g_s}{4cH}e^{\uPhi}(\tilde{F}_3)_{mpq}(\tilde{F}_3)_n^{\phantom{n}pq} 
	- \frac{g_s^2}{4H^4}(\partial_m H)(\partial_n H) \,,
\end{align}
\begin{align}
	\T_{\mu\nu}^{(1)} &= \frac{g_s^2}{4c^{1/2}H^{9/2}}(\partial H)^2(h_{\mu\nu} - h^\rho_{\phantom{\rho}\rho} g_{\mu\nu}) 
	- \frac{g_s^2}{4cH^{11/2}}\{g^{mp}g^{nq}(\partial_p H)(\partial_q H)h_{mn} \}g_{\mu\nu} \,, \\
	\T_{\mu n}^{(1)} &= - \frac{g_s^2}{4c^{1/2}H^{9/2}}h_{\mu m}g^{mp}(\partial_p H)(\partial_n H) \, \\
	\T_{mn}^{(1)} &= \frac{g_s}{2c^{3/2}H^{3/2}}(e^{-\Phi}H_{m}^{\phantom{m}rs}H_{nrp} + e^{\Phi}F_{m}^{\phantom{m}rs}F_{nrp})g^{pq}h_{sq}
	-\frac{g_s^2}{4H^{7/2}}h^\rho_{\phantom{\rho}\rho}(\partial_m H)(\partial_n H) \,,
\end{align}
where $(\partial H)^2\equiv g^{pq}(\partial_p H)(\partial_qH)$, from which follows the trace
\begin{align}
	g^{PQ}\T_{PQ}^{(1)} &= -\frac{g_s^2}{c^{1/2}H^{4}}h^\rho_{\phantom{\rho}\rho}(\partial H)^2 - \frac{g_s^2}{cH^5}g^{mp}g^{nq}(\partial_p H)(\partial_q H)h_{mn} \nonumber \\ 
	&+\frac{g_s}{2c^2H^2}(e^{-\Phi}H_{mrs}H^{mrq} + e^{\Phi}F_{mrs}F^{mrq})g^{sp}h_{pq} \,.
\end{align}

Finally, we can decompose the wave equation (\ref{eq:10dWaveEq}) into the three equations describing the 4d dynamics of the tensor, vector and scalar modes. In doing so, we are only taking into account fluctuations of $h_{MN}$, whereas all other field fluctuations are set to zero for simplicity. One should ultimately check that this is consistent, e.g. because the fields are heavier than the first KK  modes of the graviton, or take into account such fluctuations.

\begingroup
\allowdisplaybreaks
\begin{align}
	%
	% TENSOR EQUATION
	%
	\mu\nu:\quad  
	&H^{1/2}\Box_4 h_{\mu\nu} 
	+ \frac{\Delta_\M h_{\mu\nu}}{c^{1/2}H^{1/2}} 
	+ \frac{h_{\mu\nu}\Delta_M H}{2c^{1/2}H^{3/2}} 
	- 2H h_{\rho\alpha}g^{\alpha\sigma}R^{\rho}_{\phantom{\rho}\mu\nu\sigma} 
	-\frac{g^{pq}\nabla_{(\mu} h_{\nu) q}\partial_q H}{2c^{1/2}H^{3/2}} 
	\label{eq:eom_tensormodes}  \\
	&+\frac{g^{pq}\partial_p h_{\mu\nu}\partial_q H}{c^{1/2}H^{3/2}}
	-\frac{1}{4c^{1/2}H^{5/2}}(\partial H)^2 h_{\mu\nu}
	-\frac{h^{\rho}_{\phantom{\rho}\rho}}{8c^{1/2}H^{5/2}} h_{\mu\nu}(\partial H)^2\nonumber \\
	&
	+g_{\mu\nu}\frac{g^{pq}g^{rs}}{2c H^{7/2}}
	\Big(h_{pr}(H\nabla_q\partial_s H 
	- \partial_q H\partial_s H) 
	+\frac{1}{4}h_{pq}\partial_r H\partial_s H \Big) \nonumber \\
	&-\frac{g_s}{8c^2H^{5/2}}g_{\mu\nu}(e^{-\Phi}H_{mrs}H^{mrq} + e^{\Phi}F_{mrs}F^{mrq})g^{sp}h_{pq}
	= 0 \,,
	\nonumber \\
	% 
	% VECTOR EQUATION
	%
	\mu n:\quad  
	&H^{1/2}\Box_4 h_{\mu n} 
	+ \frac{\Delta_\M h_{\mu n}}{c^{1/2}H^{1/2}}
	-\frac{g^{pq}h_{\mu p}\nabla_p\partial_n H}{2c^{1/2}H^{3/2}}
	- \frac{1}{c^{1/2}H^{5/2}}(\partial H)^2 h_{\mu n} \label{eq:eom_vectormodes} \\
	&-\frac{g^{pq}\nabla_\mu h_{pn}\partial_q H}{2c^{1/2}H^{3/2}}
	+ \frac{1}{2}H^{-1/2}g^{\rho\sigma}\nabla_\rho h_{\mu\sigma}\partial_n H
	+\frac{g^{pq}\partial_{[q} H~\nabla_{n]} h_{\mu p}}{c^{1/2}H^{3/2}} 
	\nonumber \\
	&-\frac{1}{c^{1/2}H^{1/2}}g^{pq}h_{\mu q}\tau_{np}^{(0)}
	-\frac{g_s^2}{4c^{1/2}H^{9/2}}(\partial H)^2 h_{\mu n}
	- \frac{g_s^2}{2c^{1/2}H^{9/2}}h_{\mu m}g^{mp}(\partial_p H)(\partial_n H)
	= 0 \,, \nonumber \\
	%
	% SCALAR EQUATION
	%
	mn:\quad
	&H^{1/2}\Box_4 h_{mn} 
	+ \frac{\Delta_\M h_{mn}}{c^{1/2}H^{1/2}}
	- \frac{h_{mn}\Delta_\M H}{2c^{1/2}H^{3/2}}
	- \frac{2g^{rs}h_{pr}R^{p}_{\phantom{p}mns}}{c^{1/2}H^{1/2}} \label{eq:eom_scalarmodes} \\
	&\frac{g^{\rho\sigma}\nabla_\rho h_{\sigma (m}\partial_{n)}H}{H^{1/2}} 
	- \frac{g^{pq}\nabla_p h_{q(m}\partial_{n)}H}{c^{1/2}H^{3/2}}
	+ \frac{g^{pq}\nabla_{(n}h_{m)p}\partial_q H}{c^{1/2}H^{3/2}} \nonumber \\
	&-\frac{g^{pq}\nabla_p h_{mn}\partial_q H}{c^{1/2}H^{3/2}}
	+\frac{1}{c^{1/2}H^{5/2}}(\partial H)^2 h_{mn}
	- \frac{5g^{pq}h_{p(n}\partial_{m)}H\partial_q H}{4c^{1/2}H^{5/2}} \nonumber \\
	&+\frac{ h^{\rho}_{\phantom{\rho}\rho}}{2H^{3/2}}
	\Big(H\nabla_m\partial_n H 
	- \partial_m H\partial_n H 
	+ \frac{1}{4}g_{mn}g^{pq}\partial_p H\partial_q H
	\Big) \nonumber \\
	&+\frac{h_{rs}\partial_p H\partial_q H}{8c^{1/2}H^{5/2}}\Big(
	\delta^{p}_{\phantom{p}m}\delta^{q}_{\phantom{q}n}g^{rs}
	+ g_{mn}(g^{pr}g^{qs} - g^{rs}g^{pq})
	\Big)  \nonumber \\
	& \Big( 
	\frac{g_s^2}{4H^{7/2}}h^\rho_{\phantom{\rho}\rho}(\partial H)^2 
	+ \frac{g_s^2}{4c^{1/2}H^{9/2}}g^{mp}g^{nq}(\partial_p H)(\partial_q H)h_{mn} \nonumber \\
	&-\frac{g_s}{2c^{3/2}H^{3/2}}(e^{-\Phi}H_{mrs}H^{mrq} + e^{\Phi}F_{mrs}F^{mrq})g^{sp}h_{pq}
	\Big)g_{mn} 
	- \frac{2}{c^{1/2}H^{1/2}}g^{pq}h_{q(m}\tau_{n)p}^{(0)} \nonumber \\
	& + \frac{g_s}{c^{3/2}H^{3/2}}(e^{-\Phi}H_{m}^{\phantom{m}rs}H_{nrp} + e^{\Phi}F_{m}^{\phantom{m}rs}F_{nrp})g^{pq}h_{sq}
	-\frac{g_s^2}{2H^{7/2}}h^\rho_{\phantom{\rho}\rho}(\partial_m H)(\partial_n H)
	= 0 \,, \nonumber 
\end{align}
\endgroup
where $\Box_4 = g^{\mu\nu}\nabla_\mu\nabla_\nu$, $\Delta_\M = g^{pq}\nabla_p\nabla_q$ and the covariant derivative $\nabla_\rho$ ($\nabla_p$) is with respect to the 4d metric $g_{\mu\nu}$ (6d metric $g_{pq}$).

%%%%
\iffalse
\begin{align}
    \mu\nu:\quad
    &\Box_4 h_{\mu\nu} 
    + \frac{\Delta_\M h_{\mu\nu}}{c^{1/2}H} 
    %+\frac{D-10}{4}\frac{g^{pq}\partial_p h_{\mu\nu}\partial_q H}{c^{1/2}H^2}
    - 2h_{\rho\alpha}g^{\alpha\sigma}R^{\rho}_{\phantom{\rho}\mu\nu\sigma}  \nonumber \\
    %
    &-\frac{g^{pq}\nabla_{(\mu} h_{\nu) q}\partial_q H}{2c^{1/2}H^{3/2}}
    -\frac{h_{\mu\nu}h^{\rho}_{\phantom{\rho}\rho}}{8c^{1/2}H^{5/2}}(g^{pq}\partial_p H\partial_q H) \nonumber \\
    %
    &+g_{\mu\nu}\frac{g^{pq}g^{rs}}{2c H^{7/2}}
    \Big(h_{pr}(H\nabla_q\partial_s H 
    - \partial_q H\partial_s H) 
    +\frac{1}{4}h_{pq}\partial_r H\partial_s H \Big)
    = 0
\end{align}
\fi
%%%%	
The resulting equations (\ref{eq:eom_tensormodes}-\ref{eq:eom_scalarmodes}) couple $h_{\mu\nu}$, $h_{\mu n}$ and $h_{mn}$, which makes it hard to find a general solution. Since we are most interested in the 4d gravitational modes $h_{\mu\nu}$, we may consider the simpler case where $h_{\mu n}=h_{mn}=0$, i.e. a solution for which the vector and scalar modes vanish.\footnote{Note that by choosing this simple solution, we are missing some possibly interesting effects, e.g. the breathing mode identified in \cite{Andriot:2017oaz} will not be present. We should also recall the relation between $h_{mn}$ and the complex structure and K\"ahler moduli.} We also make the field redefinition $h_{\mu\nu}\to H^{-1/2}h_{\mu\nu}$, so that (\ref{eq:eom_tensormodes}) describes fluctuations of the 4d (unwarped) metric $g_{\mu\nu}$. The equations become
\begin{align}
    \mu\nu:\quad
    &\Box_4 h_{\mu\nu} 
    + \frac{\Delta_\M h_{\mu\nu}}{c^{1/2}H} 
    %+\frac{D-10}{2}\frac{g^{pq}\partial_p h_{\mu\nu}\partial_q H}{\sqrt{c}H^3}
    - 2h_{\rho\alpha}g^{\alpha\sigma}R^{\rho}_{\phantom{\rho}\mu\nu\sigma} 
    -\frac{h_{\mu\nu}h^{\rho}_{\phantom{\rho}\rho}}{8c^{1/2}H^{5/2}}(g^{pq}\partial_p H\partial_q H)
    = 0 \,, \\
    \mu n:\quad  
    &
    \partial_n H (g^{\rho\sigma}\nabla_\rho h_{\mu\sigma})
    = 0 \,, \\
    mn:\quad
    & h^{\rho}_{\phantom{\rho}\rho}
    \Big(H\nabla_m\partial_n H 
    - \partial_m H\partial_n H 
    + \frac{1}{4}g_{mn}g^{pq}\partial_p H\partial_q H
    \Big)  \nonumber \\
    &+\frac{g_s^2}{2H^2}h^\rho_{\phantom{\rho}\rho} \big( (\partial H)^2 g_{mn} 
    -2(\partial_m H)(\partial_n H)\big)
    = 0 \,. 
\end{align}
We need to check all the equations to ensure that $h_{\mu n}=h_{mn}=0$ is a solution. For a non-constant warp factor, which is the case we want to study, the vector equation implies that $g^{\rho\sigma}\nabla_\rho h_{\mu\sigma}$ vanishes. Tracing the scalar equation we find
\begin{equation}
	h^{\rho}_{\phantom{\rho}\rho} (H\Delta_\M H + 2g_s^2(\partial\ln H)^2) = 0 \,,
\end{equation}
%since $\Delta_\M H$ on a compact space vanishes only if $H$ is constant, 
which gives $h^{\rho}_{\phantom{\rho}\rho}=0$. Therefore, the vector and scalar equations impose conditions on $h_{\mu\nu}$ which correspond to the 4d transverse-traceless gauge,
\begin{equation}
    g^{\rho\sigma}\nabla_\rho h_{\mu\sigma} = 0 \,,
    \quad\quad 
    h^{\rho}_{\phantom{\rho}\rho} = 0\,.
\end{equation}
The equation for $h_{\mu\nu}$ is then 
\begin{equation}
    \Box_4 h_{\mu\nu} 
    + \frac{\Delta_\M h_{\mu\nu}}{c^{1/2}H} 
    %+\frac{D-10}{2}\frac{g^{pq}\partial_p h_{\mu\nu}\partial_q H}{\sqrt{c}H^3}
    - 2h_{\rho\alpha}g^{\alpha\sigma}R^{\rho}_{\phantom{\rho}\mu\nu\sigma}
    = 0 \,.
    \label{eq:4d_wave_equation_ourconventions_general}
\end{equation}
On the other hand, if we insist that the 4d spacetime is maximally symmetric, we have the following relation
\begin{equation}
    R^\rho_{\phantom{\rho}\mu\nu\sigma} = \frac{\mathrm{\Lambda}_4}{3}(\delta^\rho_\nu g_{\mu\sigma} - \delta^\rho_\sigma g_{\mu\nu})\,,
\end{equation}
where $\mathrm{\Lambda}_4 = \Ricci_4/4$, which gives
\begin{equation}
    R^{\rho}_{\phantom{\rho}\mu\nu\sigma} g^{\sigma\alpha}h_{\alpha\rho} 
    = \frac{\mathrm{\Lambda}_4}{3}(h_{\mu\nu} - g_{\mu\nu}h^{\rho}_{\phantom{\rho}\rho}) \,,
\end{equation}
and using $h^{\rho}_{\phantom{\rho}\rho}=0$, the wave equation becomes
\begin{equation}
    \Box_4 h_{\mu\nu} 
    + \frac{\Delta_\M h_{\mu\nu}}{c^{1/2}H} 
    - \frac{2}{3}\mathrm{\Lambda}_4 h_{\mu\nu}
    = 0 \,.
    \label{eq:4d_wave_equation}
\end{equation}
This equation describes the 4d tensor components of $h_{MN}$, but we should remember that $h_{\mu\nu}(x^\mu,y^m)$ is still a function of both external and internal coordinates. In the 4d EFT, we must express it in terms of modes which are only functions of external coordinates $h_{\mu\nu}(x^{\mu})$ --- more precisely, equation (\ref{eq:4d_wave_equation}) corresponds to an infinite tower of 4d spin-2 modes $h_{\mu\nu}^k(x^{\mu})$, where $k$ labels each mode in the tower. In the next section we study this tower, including the masses of each mode, their wavefunctions in the extra dimensions and the respective normalisations.
\section{Graviton KK Tower in a Deformed Conifold}
\label{sec:conifoldKKtower}

In this section we obtain the Kaluza-Klein tower of graviton modes arising from equation (\ref{eq:4d_wave_equation}). In Section \ref{sec:KKtower} we identify the operator $\Op_\M$ whose eigenvalues determine the masses of each mode $h_{\mu\nu}^k(x^{\mu})$ in the tower and whose eigenfunctions correspond to their wavefunctions in the extra dimensions --- these form a complete basis with which we can decompose $h_{\mu\nu}$ in order to have fully 4-dimensional equations and are solutions of the Schr\"odinger-like equation (\ref{eq:Bk_schrodinger}). We discuss in particular how the masses and wavefunctions of these modes depend on the warping of the deformed conifold. We solve the equations numerically (Section \ref{sec:numericalSolutions}) and compare the results with analytical approximations (Section \ref{sec:analyticalApproximations}) in the limits $\tau_c\to T$ (fully warped conifold) and $\tau_c\to 0$ (unwarped conifold).  

%%%%%%%%%%%%%%%%%%%%%%%%%%%%%%%%%%%%%%%%%%%%%%%%%%%%%%%%%%%%%%%%%%%%%%%%
%   3.1 KK WAVE EQUATIONS AND BOUNDARY CONDITIONS
%%%%%%%%%%%%%%%%%%%%%%%%%%%%%%%%%%%%%%%%%%%%%%%%%%%%%%%%%%%%%%%%%%%%%%%%

\subsection{KK wave equations and boundary conditions}
\label{sec:KKtower}

We may rewrite (\ref{eq:4d_wave_equation}) as
\begin{equation}
    \Box_4 h_{\mu\nu} + \Op_\M h_{\mu\nu} = 0\,,
    \label{eq:generalWaveEqmunu_nosource}
\end{equation}
with $\Op_\M \equiv \frac{\Delta_\M}{c^{1/2}H} - \frac{2}{3}\mathrm{\Lambda}_4$. 
In the 4d EFT, we must express $h_{\mu\nu}(x^\mu,y^p)$ in terms of modes which are only functions of external coordinates $h_{\mu\nu}^k(x^{\mu})$, where $k$ labels each mode in an infinite tower, and a complete basis $\uPhi^{k}$ of eigenmodes of $\Op_\M$,
\begin{equation}
    h_{\mu\nu}(x^\mu,y^p) = \sum_k h_{\mu\nu}^{k}(x^\mu)\mathrm{\uPhi}_{k}(y^p) \,.
    \label{eq:decomposition_munu}
\end{equation}

\noindent The decomposition is such that   
\begin{equation}
    \Op_\M h_{\mu\nu} = \Op_\M (h_{\mu\nu}^k(x^\mu)\uPhi_k(y^p)) = h^k_{\mu\nu} \Op_\M\uPhi_k(y^p) = -m_k^2 h^k_{\mu\nu}\uPhi_k(y^p)
    \label{eq:eigenmodeEq_definition}
\end{equation}
where there is an implicit sum in $k$, following from the eigenvalue equation $\Op_\M \uPhi_k(y^p) = - m_k^2 \uPhi(y^p)$.

Since $\uPhi_k$ is a scalar,
\begin{equation}
    \Delta_\M \uPhi_k = \frac{1}{\sqrt{g}}\partial_p(\sqrt{g}g^{pq}\partial_q\uPhi_k)\,,
\end{equation}
and hence the equation we must solve can be written as
\begin{equation}
    \frac{1}{\sqrt{g}}\partial_p(\sqrt{g}g^{pq}\partial_q\uPhi_k) + c^{1/2}H\Big(m_k^2 - \frac{2}{3}\Lambda_4\Big)\uPhi_k = 0 \,.
    \label{eq:eigenmodeEq}
\end{equation}

We should remember that the compact space contains two pieces glued together --- a warped throat described by the Klebanov-Strassler solution and a compact bulk, which is usually chosen to be a Calabi-Yau whose metric we do not explicitly know.  Therefore the 6d metric $g_{pq}$ which appears in (\ref{eq:eigenmodeEq}) will be different inside the throat and in the bulk --- it corresponds to the metric of the warped deformed conifold in the throat region and (usually) to the unknown metric of a compact CY$_3$. However, we can still solve the equation in the region where the metric is unkown if $\uPhi_k=0$ in the CY$_3$ bulk, which for consistency implies that the modes must vanish at the point where these two regions meet. We will motivate further this boundary condition below.

\iffalse
We therefore have two regions in which to solve the above equation and the solutions should match at the gluing point. In particular, we have
\begin{align}
    &c^{1/2}H \to c^{-1/2}e^{-4A_0(y)}, && \text{in the throat} \\
    &c^{1/2}H \to c^{1/2}, && \text{in the bulk} \,,
\end{align}
so that we have two different equations,
\begin{align}
    \frac{1}{\sqrt{g}}\partial_p(\sqrt{g}g^{pq}\partial_q\uPhi_k) + c^{-1/2}e^{-4A_0}(m_k^2 -\lambda^2)\uPhi_k = 0 &&& \text{in the throat} \label{eq:warped_throat_equation}\,,\\
    \frac{1}{\sqrt{g}}\partial_p(\sqrt{g}g^{pq}\partial_q\uPhi_k) + c^{1/2}(m_k^2 - \lambda^2)\uPhi_k = 0 &&& \text{in the bulk.}
\end{align}
\fi

\begin{figure}[t]
     \centering
         \includegraphics[width=0.7\textwidth]{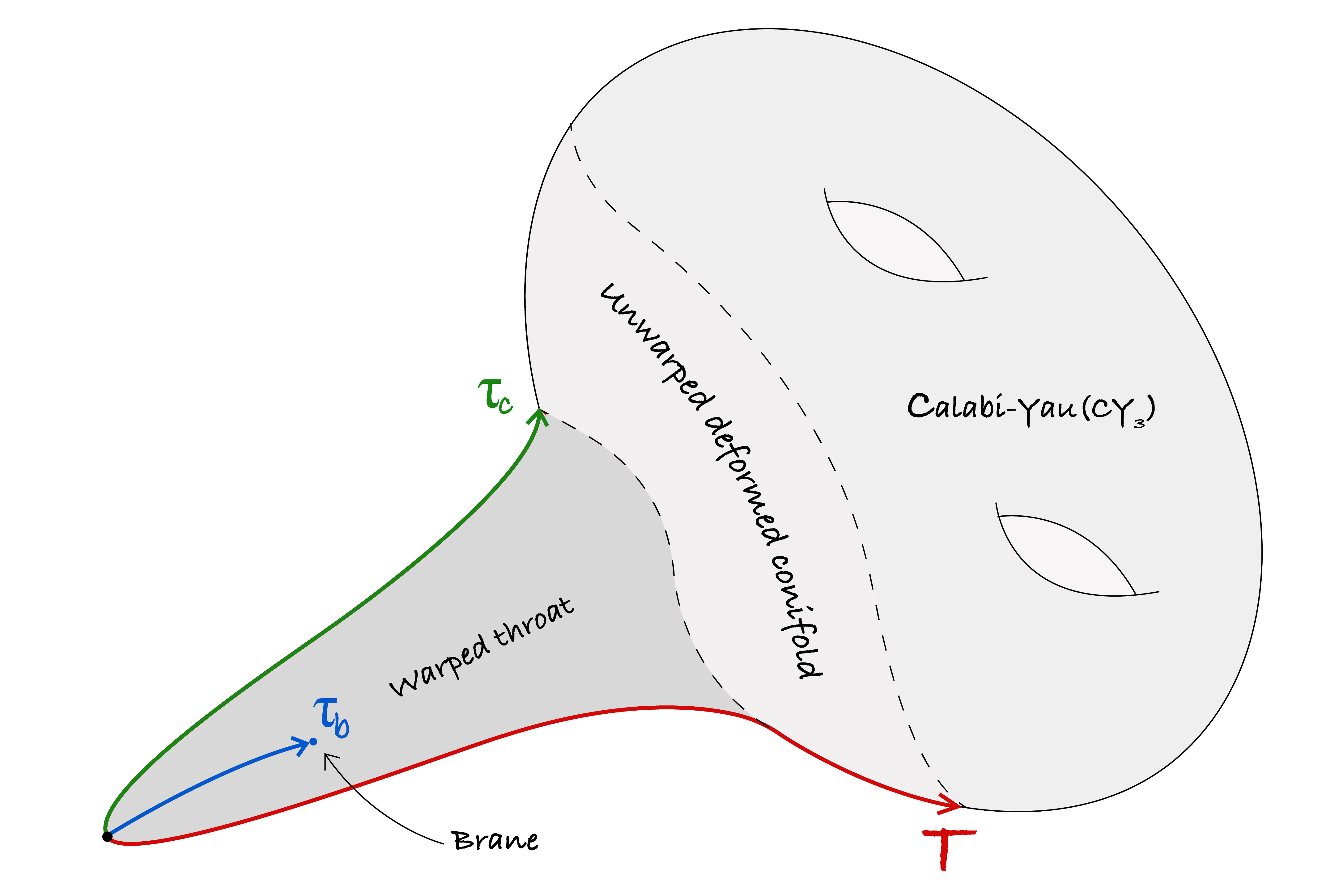}
     \caption{The internal space consists of a 6d compact manifold with a warped region described by the Klebanov-Strassler solution (i.e. a warped deformed conifold). We split the bulk into two different pieces: one piece is a generic CY$_3$ and the other takes the metric of an \textit{unwarped} deformed conifold and serves as a transition between the warped throat and the CY$_3$ (with $\tau_c<\tau< T$, where $\tau$ is the radial coordinate in the deformed conifold metric (\ref{eq:conifold:deformed_conifold_metric})).}
    \label{fig:compactspace}
\end{figure}

This means, however, that we cannot take the interesting limit where there is no warping and the whole internal space is the CY$_3$ whose metric we do not know, since the wavefunctions will be identically zero. In order to consider this regime, we may split the bulk region into two different pieces: one piece is the generic CY$_3$ and the other takes the metric of an \textit{unwarped} deformed conifold and serves as a transition (with $\tau_c<\tau< T$) between the warped throat and the CY$_3$ (see Fig.~\ref{fig:compactspace}). While we still solve the equation in the CY$_3$ with $\uPhi_k=0$, we can now have a non-vanishing wavefunction in the piece of the bulk described by the unwarped deformed conifold. While $\tau_c$ determines the size of the warped throat, $T$ determines the portion of the bulk in which the wavefunctions do not vanish (more precisely, $T-\tau_c$ determines the extension of the wavefunction into the bulk). Notice that a fully warped conifold corresponds to the limit $\tau_c\to T$ and an unwarped conifold corresponds, roughly, to $\tau_c\to 0$.\footnote{Strictly speaking, the unwarped limit corresponds to $H(\tau)=1$ for all $\tau$, whereas when $\tau_c=0$, $H(0)=2$. At $\tau_c=0$, the second term in (\ref{WF}) becomes of the same order as the first, marking the boundary between a warped and an unwarped regime.}

Therefore we will write (\ref{eq:eigenmodeEq}) explicitly using the metric (\ref{eq:conifold:deformed_conifold_metric}) in terms of the warp factor $H_{\tau_c}(\tau)$. We consider splitting the 6d coordinates $y^m$ into a radial coordinate $\tau$ and angular coordinates $\theta^a$, $a=1,...,5$ (these are related to the 1-forms $g^i$ in the conifold metric). This will split the Laplacian into two pieces, one for $\tau$ and one along the angular coordinates. With this in mind, we can decompose the functions $\uPhi_k(\tau,\theta^a)$ as (see \cite{Tye:2005qs})
\begin{equation}
    \uPhi_k(\tau,\theta^a) = G(\tau)^{-1/2}B_{k}(\tau)\varphi_k(\theta^a)\,, 
	\label{eq:wavefunction}
\end{equation}
such that (\ref{eq:eigenmodeEq}) becomes a Schr\"odinger equation for $B_k(\tau)$
\begin{equation}
	B_k'' - V_{eff} B_k = 0\,,
	\label{eq:Bk_schrodinger}
\end{equation}
with an effective potential given by 
\begin{equation}
    V_{eff} = -g_{\tau\tau}\left(c^{1/2}H\Big(m_k^2 - \frac{2}{3}\Lambda_4\Big) 
    + \frac{\Op_{ang}\uPhi_k}{\uPhi_k}\right)
    + \frac{(G^{1/2})''}{G^{1/2}}\,,
    \label{eq:Bk_effPotential}
\end{equation}
where we factorise $(g_{con})_{mn} = K_{mn}(\tau)\gamma_{mn}(\theta)$ (with no summation) and define
\begin{align}
	G(\tau) &= \frac{2^4}{\epsilon^{8/3}} g^{\tau\tau}\sqrt{g_{con}} = \K(\tau)^2 \sinh^2(\tau) \,, \\
	\Op_{ang}\varphi_k &\equiv K^{ab}(\tau)\frac{1}{\sqrt{\gamma}}\partial_{\theta^a}\big(\sqrt{\gamma}\gamma^{ab}(\theta)\partial_{\theta^b}\varphi_k \big) \,,
\end{align}
with $g_{\tau\tau} = \frac{\epsilon^{4/3}}{6\K(\tau)^2}$ following directly from the metric (\ref{eq:conifold:deformed_conifold_metric}), the operator $\Op_{ang}$ containing the angular information of the metric and the sum in $a,b$ implied. Since the contribution from the angular coordinates is more complicated, we look at modes with $\Op_{ang}\varphi_k = 0$ (usually known as the s-orbital). In this case, we have $\varphi_k(\theta^a)=$ const. and we can absorb it into the overall normalisation of the wavefunction. Notice that trying a constant wavefunction (i.e. with no dependence on the internal coordinates) requires $B_k(\tau)=G(\tau)^{1/2}$, in which case (\ref{eq:Bk_schrodinger}) reduces to
\begin{equation}
    g_{\tau\tau}\Big(m_k^2 - \frac{2}{3}\Lambda_4\Big) = 0\,,
\end{equation}
from which it follows that $m_k^2 = \frac{2}{3}\Lambda_4$. We see that for a flat Minkowski background, i.e. $\Lambda_4=0$, the constant wavefunction $\uPhi_0$ corresponds to a massless 4d graviton, which no longer exists if $\Lambda_4\neq 0$.\footnote{Interestingly, if we choose a de Sitter background the zero mode mass matches the Higuchi bound \cite{Higuchi,Higuchi:1989gz}. Note that this mass is consistent with current constraints on the graviton mass \cite{deRham:2016nuf}.} In what follows we assume a Minkowski background and set $\Lambda_4=0$.

\begin{figure}[t]
     \centering
	\begin{subfigure}[b]{0.48\textwidth}
        	\centering
         	\includegraphics[width=\textwidth]{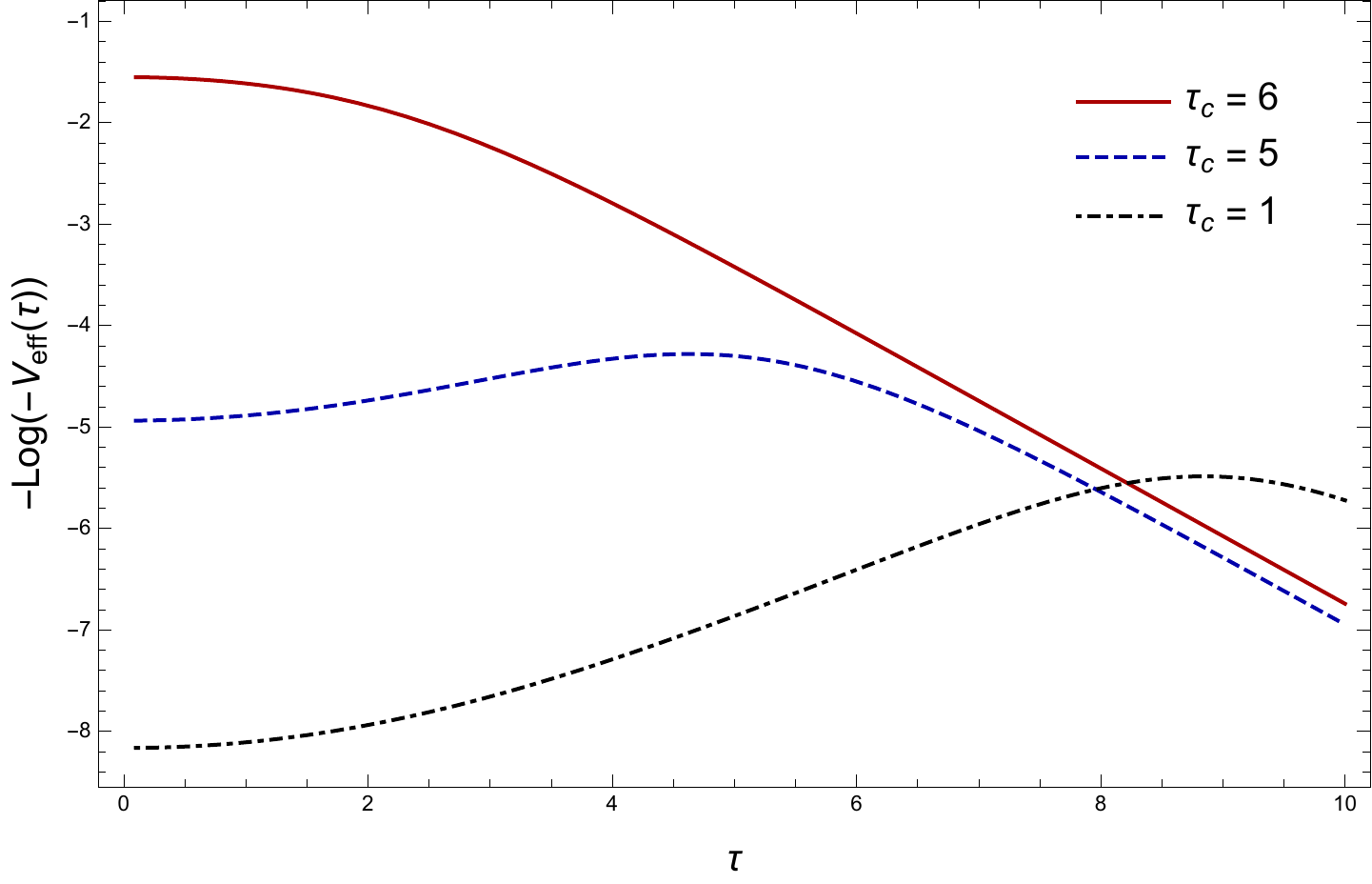}
     	\end{subfigure}
     \hfill
	\begin{subfigure}[b]{0.48\textwidth}
        	\centering
         	\includegraphics[width=\textwidth]{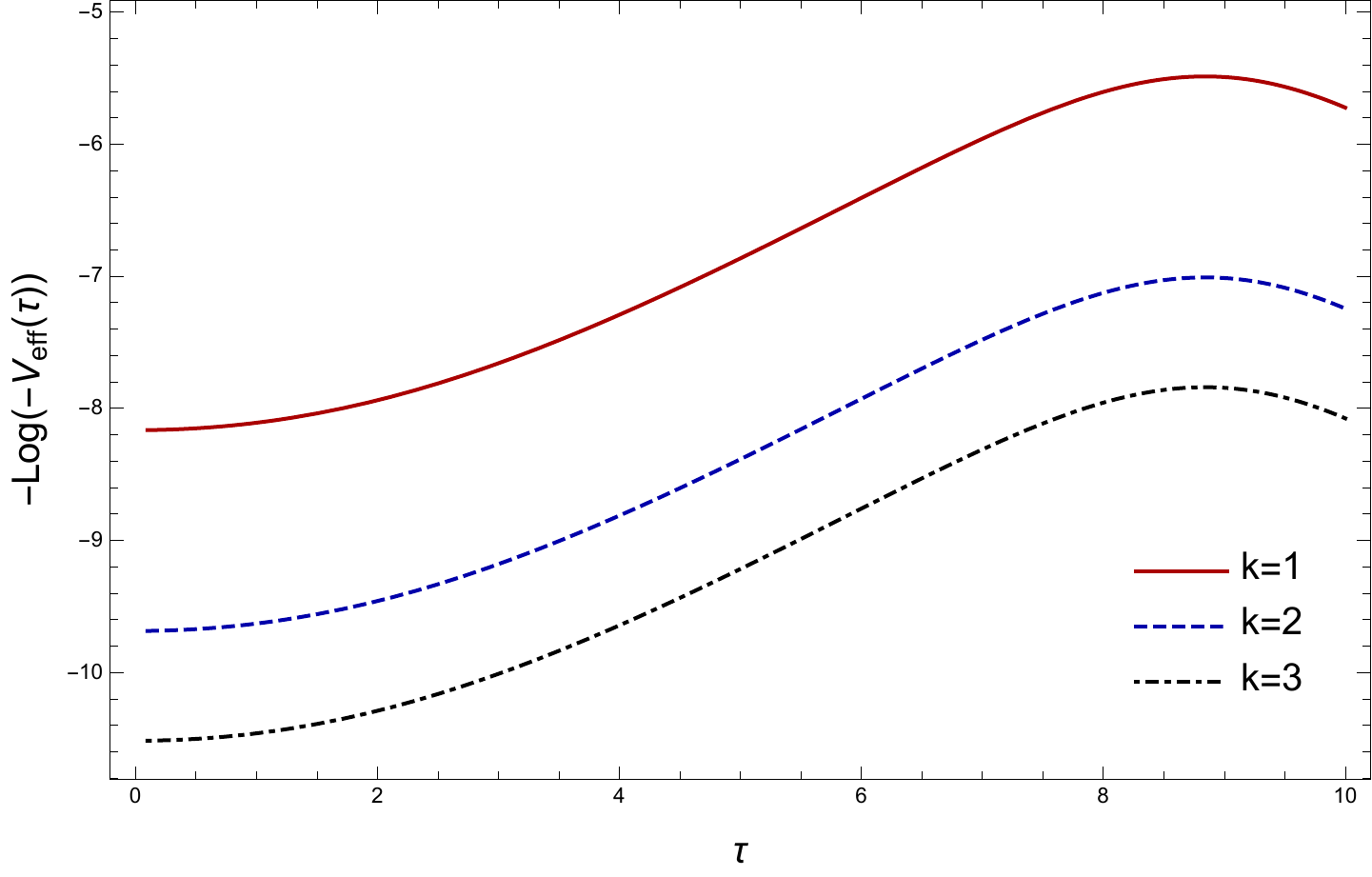}
     	\end{subfigure}     
     \caption{Effective potential $V_{eff}$ corresponding to the first mode for different choices of $\tau_c$ (left) and for the first 3 modes when $\tau_c=9$ (right), both with $T=10$.}
     \label{fig:Veff_tauc}
\end{figure}

Defining also $\hat{E}_k = \epsilon^{2/3} c^{1/4} m_k$ the effective potential can be written as 
\begin{equation}
	V_{eff} = - \frac{H_{\tau_c}(\tau)}{6K(\tau)^2}\hat{E}_k^2 + \frac{(G^{1/2})''}{G^{1/2}} \,,
	\label{eq:Veff}
\end{equation}
which we can think of as a family of potentials, each member of which is determined by a choice of $\tau_c$ and is only a function of $\tau$ (Fig.~\ref{fig:Veff_tauc}).  There is no analytical solution for the corresponding Schr\"odinger equations --- it can either be solved numerically (Section \ref{sec:numericalSolutions}) or we can consider approximations to this potential (Section \ref{sec:analyticalApproximations}).

In order to obtain the wavefunctions, we must also impose boundary conditions, which we choose as follows.
\begin{enumerate}
	\item The wavefunction is finite at $\tau=0$, i.e. 
	\begin{equation*} 
		\lim_{\tau\to 0}\uPhi_k(\tau) < \infty \implies \lim_{\tau\to 0}B_k(\tau) = 0 \,;
	\end{equation*} 
	\item The wavefunctions vanish as they approach the CY$_3$ region, i.e.
	\begin{equation*}
		\uPhi_k(T) = 0 \implies B_k(T) = 0 \,. 
	\end{equation*}
\end{enumerate} 
The boundary condition at $\tau=T$ has two main motivations. Firstly, it connects to the infinite throat limit ($\tau_c\to T\to\infty$), for which the wavefunction must decay towards zero in order to be normalisable. Secondly, it is consistent with the localisation of massive KK modes at the tip of the throat, which is what one finds in a Randall-Sundrum scenario \cite{Randall:1999ee,Randall:1999vf}. In addition, as we discussed below (\ref{eq:eigenmodeEq}), it is also useful if one wants to remain agnostic with respect to the geometry of the bulk beyond the conifold, since vanishing modes will solve (\ref{eq:eigenmodeEq}) for a generic metric.

%%%%%%%%%%%%%%%%%%%%%%%%%%%%%%%%%%%%%%%%%%%%%%%%%%%%%%%%%%%%%%%%%%%%%%%%%%%
% NORMALISATION
%%%%%%%%%%%%%%%%%%%%%%%%%%%%%%%%%%%%%%%%%%%%%%%%%%%%%%%%%%%%%%%%%%%%%%%%%%%

Finally, we fix the normalisation of the eigenfunctions by looking at the kinetic terms for $h_{\mu\nu}$, 
%(only schematically, the correct terms would involve different contractions of the indices involved --- see \cite{Rattazzi} for an explicit discussion)
\begin{align}
    S &= \frac{1}{2\kappa^2}\int d^D x \sqrt{-G} R \nonumber \\
    &=\frac{1}{2\kappa^2}\int d^4x \sqrt{-g_4} ~c^{3/2}\int ~d^6y \sqrt{g_{6}} H \Ricci_4 + ... \nonumber \\
    %&= \frac{1}{2\kappa^2}\int d^4 x \sqrt{-g_4}\left\{c^{3/2} \int d^6y\sqrt{g_6}H \Big( \Ricci_4^{(0)} + \nabla_\rho h_{\mu\nu}\nabla^\rho h^{\mu\nu} + ... \Big)\right\} + ... \nonumber \\
    &= S^{(0)} + \frac{1}{2\kappa^2}\int d^4 x \sqrt{-g_4}\left\{c^{3/2}\int d^6 y\sqrt{g_6}H ~\nabla_\rho h_{\mu\nu}\nabla^\rho h^{\mu\nu}\right\} + ... \\
    &= S^{(0)} + \int d^4 x \sqrt{-g_4} \left\{\frac{1}{2}\nabla_\rho h_{\mu\nu}^k\nabla^\rho h^{k',\mu\nu}\Big(c^{3/2}\int d^6y\sqrt{g_6}~H~\uPhi_k(y)\uPhi_{k'}(y)\Big) + ... \right\}\nonumber
    \label{eq:graviton_kinetic_terms}
\end{align}
In the last line, we make the usual field redefinition $h_{\mu\nu}\to \kappa ~ h_{\mu\nu}$, which gives the standard mass dimension of 4 for a 10d bosonic field $h_{\mu\nu}$ (notice that this means $\delta g_{\mu\nu} = \kappa ~ h_{\mu\nu} = \frac{\sqrt{V_w}}{M_P}h_{\mu\nu}$), and substitute the decomposition (\ref{eq:decomposition_munu}), with an implicit sum in the indices $k$ and $k'$. The modes $\uPhi_k\equiv\uPhi_k(y^p)$ are the eigenmodes of the operator $\Op_\M$ and form an orthogonal basis\footnote{One can show this by noting that (\ref{eq:Bk_schrodinger}) is a regular Sturm-Liouville problem, which guarantees (i) real eigenvalues, (ii) a unique (up to normalisation) eigenfunction for each eigenvalue and (iii) normalised eigenfunctions which form an orthonormal basis under the inner product $\int \sqrt{g_6}H \Phi_k(y) \Phi_{k'}(y)$. This is equivalent to verifying that the Hamiltonian for the corresponding Schr\"odinger equation is hermitian.} under the inner product weighted by $H$, 
\begin{equation}
    \int d^{6}y\sqrt{g_6}~H~\uPhi_k(y)\uPhi_{k'}(y) = \delta_{kk'} \,.
    \label{eq:orthogonal_modes}
\end{equation}
In order to have canonical kinetic terms for each spin-2 mode in 4d, $h_{\mu\nu}^k$, we include a normalisation constant in each wavefunction $\uPhi_k = N_{(k)}\tilde{\uPhi}_k$, with $N_{(k)}$ defined as
\begin{equation}
    N_{(k)}^{-2} = c^{3/2}\int d^6y\sqrt{g_6}~H~|\tilde{\uPhi}_k(y)|^2\,,
    \label{eq:eigenmodeNormalisation}
\end{equation}
where $\tilde{\uPhi}_k(y)=G(\tau)^{-1/2}B_k(\tau)$ is the wavefunction obtained by solving (\ref{eq:Bk_schrodinger}). From the canonically normalised action (at this order in perturbations)
\begin{align}
    S &= \int d^4 x \sqrt{-g_4} \left\{\frac{1}{2}\nabla_\rho h_{\mu\nu}^k\nabla^\rho h_{k}^{\mu\nu} - \frac{1}{2}m_k^2 h_{\mu\nu}^k h_{k}^{\mu\nu} \right\} \,,
    \label{eq:canonically_normalised_action}
\end{align}
we find decoupled equations of motion for each $h_{\mu\nu}^k$,

%\footnote{It is common to have $M_P^2/2$ instead of $1/2$, since in GR we consider the metric perturbations to be dimensionless. However, from the field theory perspective, where we want to consider a particle (the graviton) with a canonical kinetic term, $h_{\mu\nu}^k$ must have mass dimension 1. This is important for the Newtonian potential computation later on. Notice that this implies $\uPhi_k(\tau)$ has mass dimension $\frac{D-4}{2}$.} 

\begin{equation}
    \Box_4 h^{k}_{\mu\nu} - m_k^2 h^k_{\mu\nu} = 0\,.
    \label{eq:waveEq_KKmodes}
\end{equation}
This is an infinite set of equations describing spin-2 modes of mass $m_k$ in 4d. 
%A 4d massless graviton requires $m_0=0$ (notice that for $\mathrm{\Lambda}_4=0$ this mode always exists, whereas a very small cosmological constant implies a zero mode with a tiny but non-zero mass). 
For the zero mode, $\tilde{\uPhi}_0(\tau)=$ const, which we can set to one by absorbing the constant in the normalisation $N_{(0)}$,
\begin{equation}
    \uPhi_0(\tau) = N_{(0)}\tilde{\uPhi}_0(\tau) = N_{(0)} \,,
\end{equation}
so that we have
\begin{align}
    N_{(0)}^{-2} = c^{3/2} \int d^6y\sqrt{g_6}~H = V_w \,.
\end{align}
Hence the graviton zero mode wavefunction is simply 
\begin{equation}
    \uPhi_0(\tau) = \frac{1}{\sqrt{V_w}} \,,
    \label{eq:graviton_zeromode_wavefunction}
\end{equation}
constant over the compact dimensions --- the graviton zero mode does not localise.

We are interested in the contributions from higher modes with $m_{k\neq 0}\neq 0$. For these modes we find
\begin{align}
    N_{(k)}^{-2} 
    &= c^{3/2} \int d^{6}y\sqrt{g_{6}}~H|\tilde{\uPhi}_k(y)|^2 \\
    &= \frac{c^{3/2}\epsilon^4}{2^5\cdot 3}\left(\int \prod_i g^i\right) \int_0^T d\tau \frac{\sinh^2(\tau)}{G(\tau)}H_{\tau_c}(\tau)|B_k(\tau)|^2  \\
    &= \frac{2\pi^3}{3}c^{3/2}\epsilon^4 \mathcal{N}_{(k)}^{-2}(\tau_c,T) \,,
    \label{eq:Normalisation_general}
\end{align}

\noindent where we used $\int \prod_i g^i = 64\pi^3$ and the fact that $\tilde{\uPhi}_k = 0$ in the CY$_3$, and with $\tau_c$ given by (\ref{eq:tauc_parameters}). We also define $\mathcal{N}_{(k)}(\tau_c,T)$ as
\begin{equation}
	\mathcal{N}_{(k)}^{-2}(\tau_c,T) \equiv  \int_0^T d\tau \frac{H_{\tau_c}(\tau)}{\K(\tau)^2}|B_k(\tau)|^2  \,,
	\label{eq:Normalisation_general_numerical}
\end{equation}
which only depends on the pair $(\tau_c,T)$, the known functions $I(\tau), G(\tau)$ and the wavefunctions $B_k(\tau)$. Note that one can easily take the limits $\tau_c\to 0$ (unwarped conifold) and $\tau_c\to T$ (fully warped conifold). In particular, 
\begin{equation}
    \mathcal{N}_{(k)}^{-2} 
    \approx \frac{1}{I(\tau_c)} \int_0^{\tau_c} d\tau \frac{I(\tau)}{\K(\tau)^2}|B_k(\tau)|^2 
	+ \int_{\tau_c}^T d\tau \frac{|B_k(\tau)|^2}{\K(\tau)^2} \,.
    \label{eq:Normalisation_limits}
\end{equation}

%%%%%%%%%%%%%%%%%%%%%%%%%%%%%%%%%%%%%%%%%%%%%%%%%%%%%%%%%%%%%%%%%%%%%%%%%%%%
%  3.2 NUMERICAL SOLUTIONS
%%%%%%%%%%%%%%%%%%%%%%%%%%%%%%%%%%%%%%%%%%%%%%%%%%%%%%%%%%%%%%%%%%%%%%%%%%%%

\subsection{Numerical solutions}
\label{sec:numericalSolutions}

We can solve (\ref{eq:Bk_schrodinger}) with the effective potential (\ref{eq:Veff}) and the boundary conditions chosen above, for different values of $(\tau_c,T)$, which may be interpreted as the strength of the warping, given by $\tau_c$, and the proportion of the conifold that is warped, given by $\tau_c/T$. 

\begin{figure}
     \centering
	\begin{subfigure}[b]{0.48\textwidth}
         \centering
         \includegraphics[width=\textwidth]{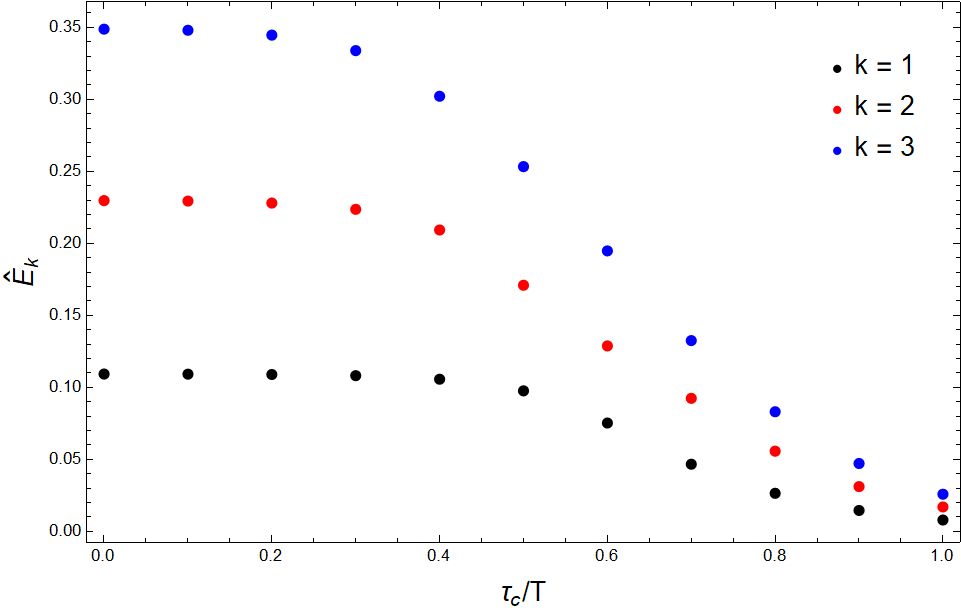}
     \end{subfigure}
     \hfill
	\begin{subfigure}[b]{0.48\textwidth}
         \centering
         \includegraphics[width=\textwidth]{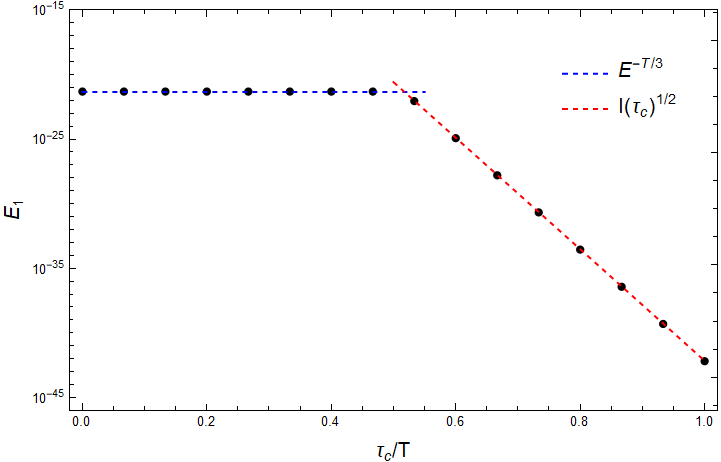}
     \end{subfigure}
      \caption{(Left) Eigenvalues $\hat{E}_k$ for the first 3 modes obtained by solving the Schr\"odinger equation numerically with $V_{eff}$ (\ref{eq:Veff}), for fixed $T=10$ and different values of $\tau_c$; (Right) Eigenvalues $\hat{E}_1$ for the first excited mode obtained by solving the Schr\"odinger equation numerically with $V_{eff}$ (\ref{eq:Veff}), for fixed $T=150$ and different values of $\tau_c$.}
    \label{fig:masses_taucTratio}
\end{figure}

In Fig.~\ref{fig:masses_taucTratio} we plot the eigenvalues $\hat{E}_k$ as a function of the ratio $\tau_c/T$ for $T=10$ and $k=1,2,3$ (left), and for $T=150$ and $k=1$ (right). We can see from the left plot that the effect of the warping only starts influencing the masses once the warped throat dominates over the (unwarped conifold) bulk $\tau_c/T\gtrsim 1/2$, with higher modes starting to feel the effects of warping for smaller values $\tau_c/T$. In the right plot we clearly identify the dominant behaviour of the masses in the different regimes (throat dominated vs bulk dominated) --- suppressed by the warping when the throat dominates and by the volume when the bulk dominates. 

\begin{figure}[t]
	\centering
	\begin{subfigure}[b]{0.48\textwidth}
		\centering
         	\includegraphics[width=\textwidth]{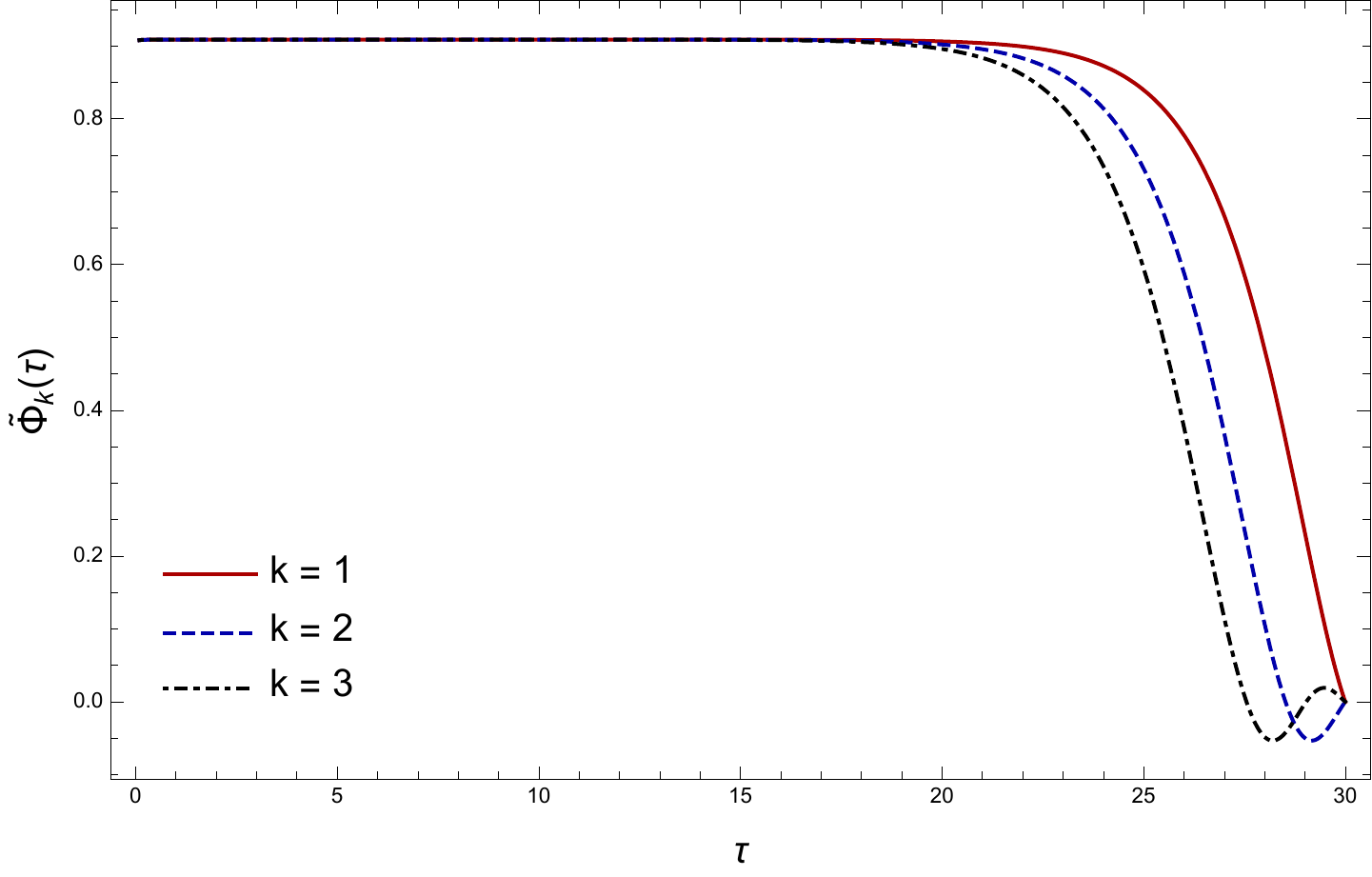}
	\end{subfigure}
     	\hfill
	\begin{subfigure}[b]{0.48\textwidth}
		\centering
         	\includegraphics[width=\textwidth]{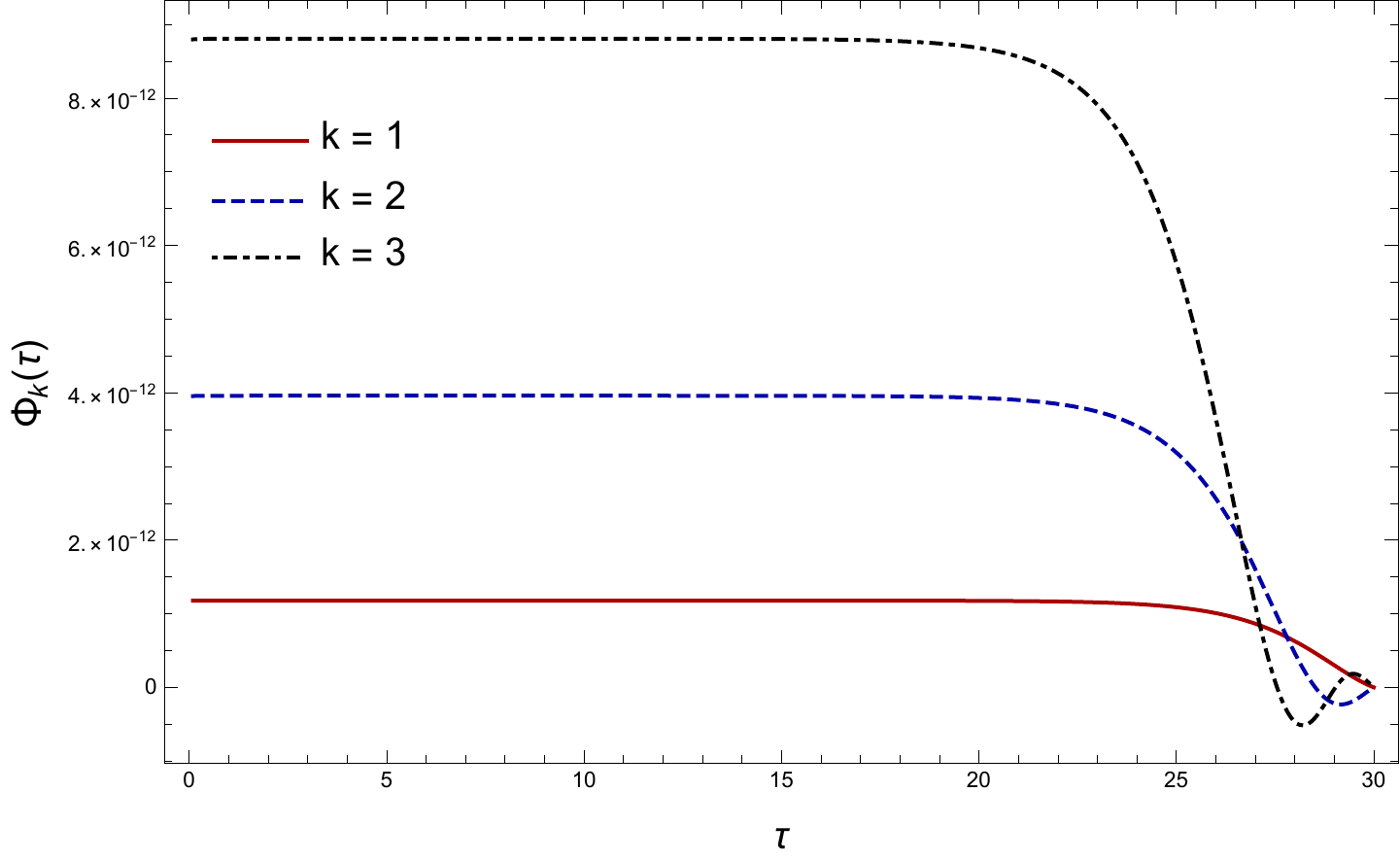}
	\end{subfigure}
	\begin{subfigure}[b]{0.48\textwidth}
		\centering
         	\includegraphics[width=\textwidth]{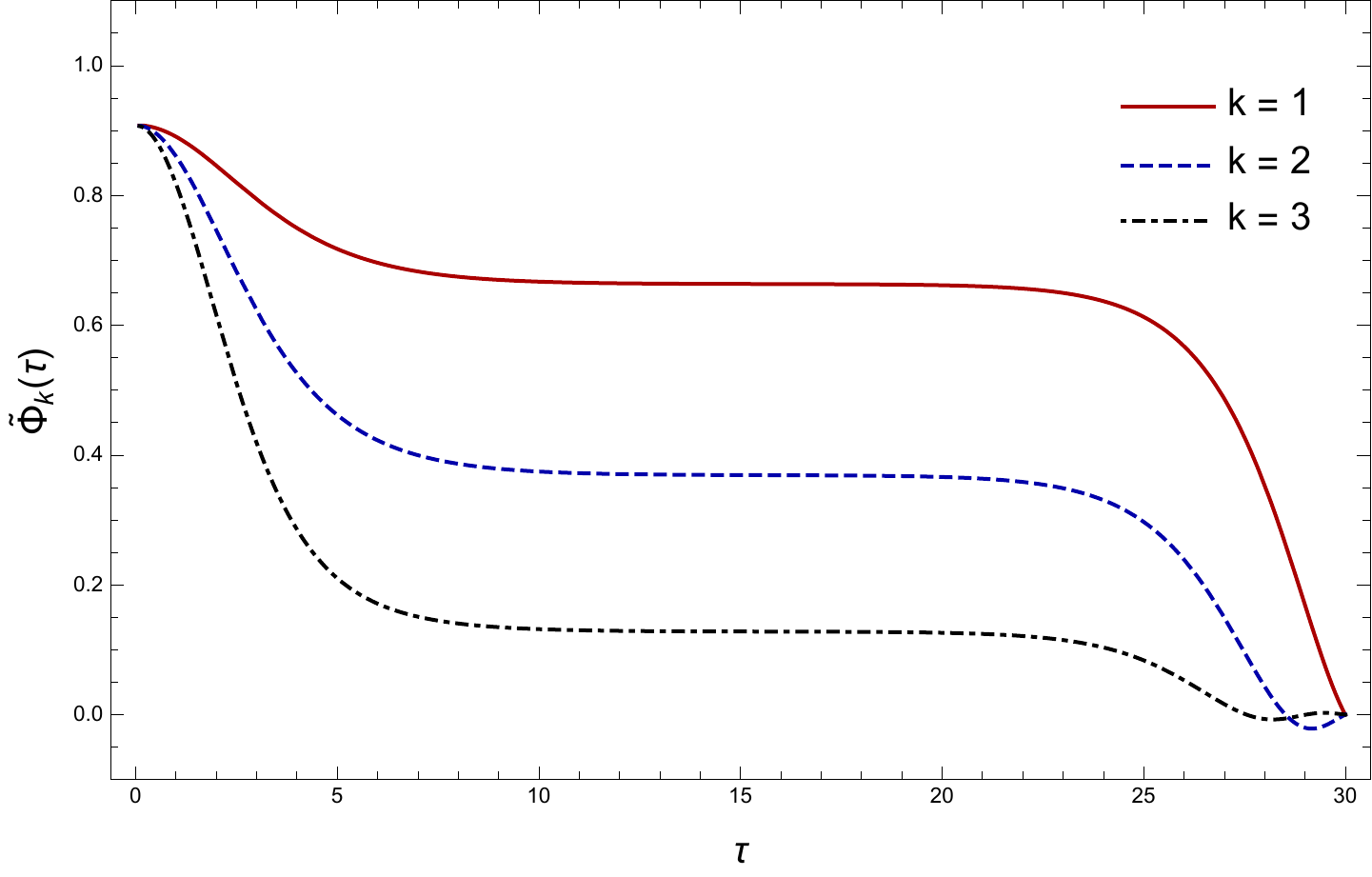}
	\end{subfigure}
     	\hfill
	\begin{subfigure}[b]{0.48\textwidth}
		\centering
         	\includegraphics[width=\textwidth]{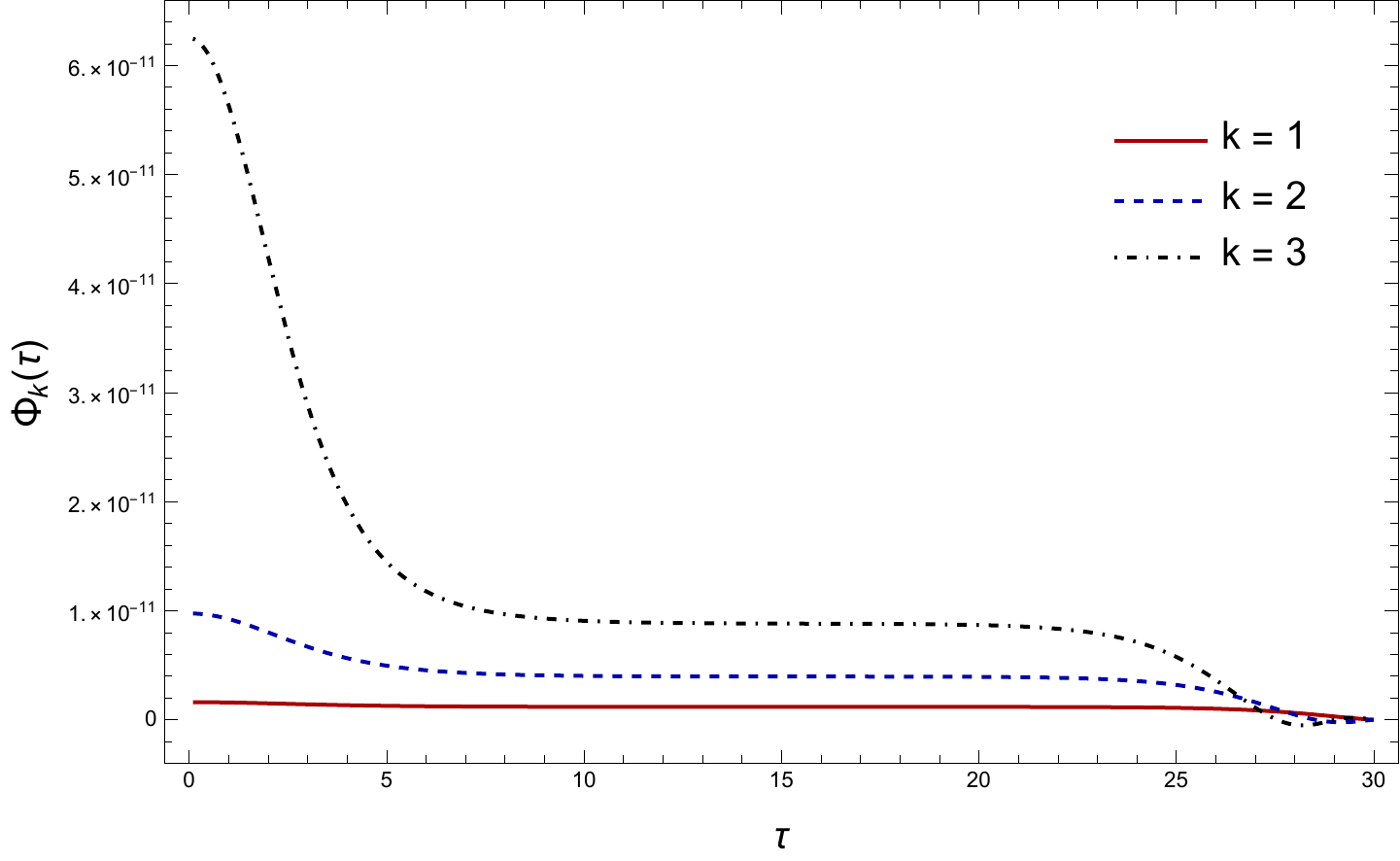}
	\end{subfigure}
	\begin{subfigure}[b]{0.48\textwidth}
		\centering
         	\includegraphics[width=\textwidth]{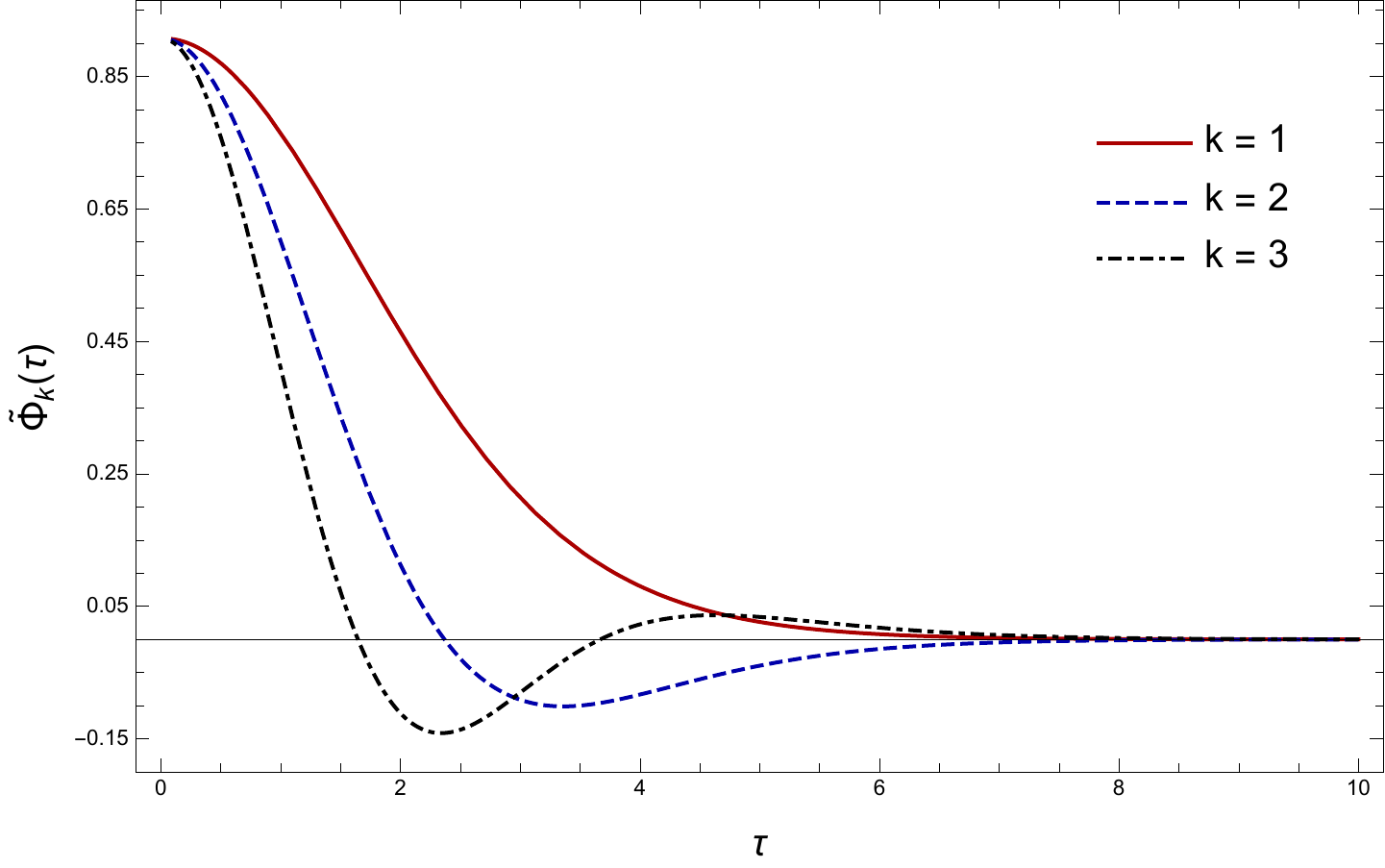}
	\end{subfigure}
     	\hfill
	\begin{subfigure}[b]{0.48\textwidth}
		\centering
         	\includegraphics[width=\textwidth]{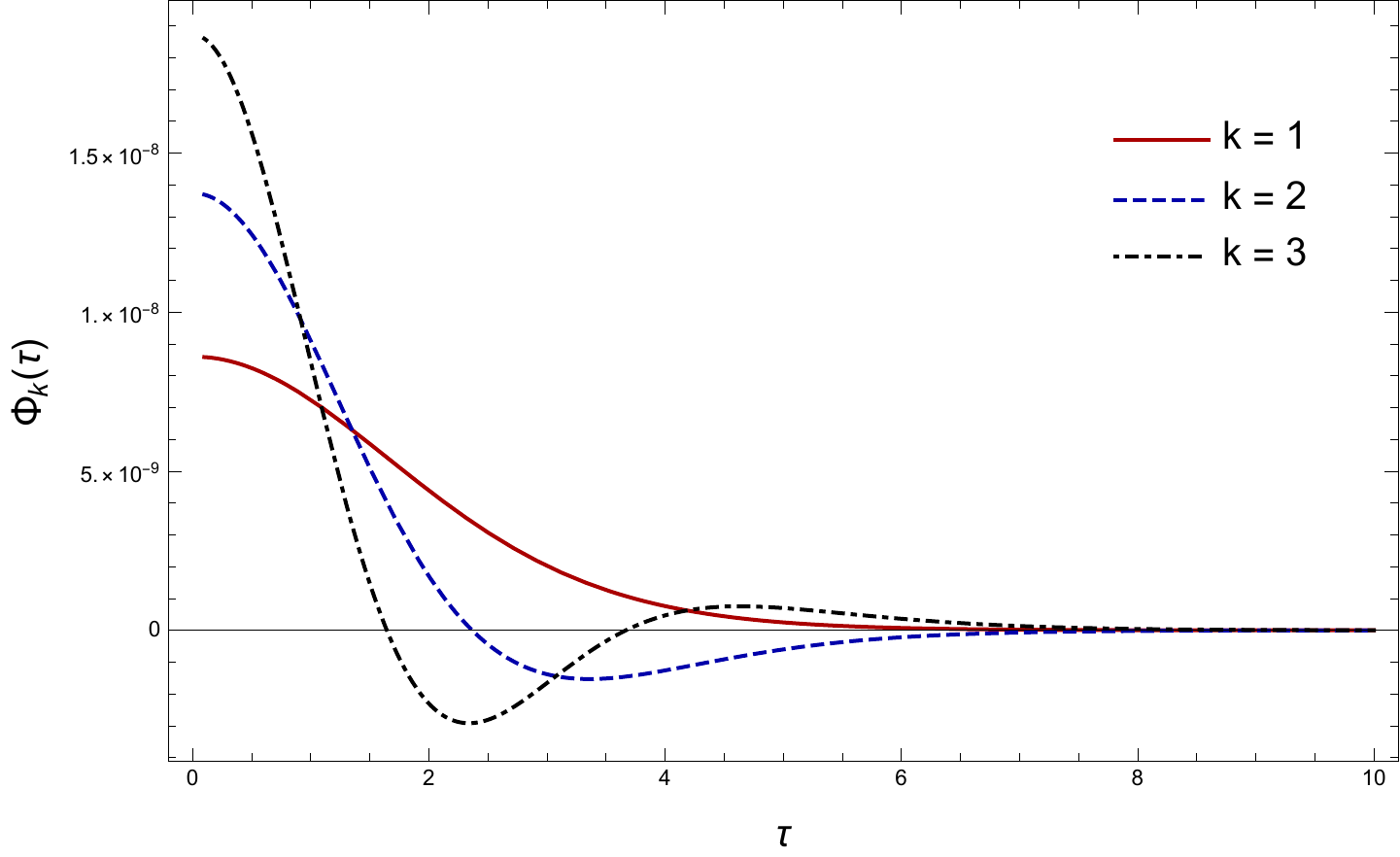}
	\end{subfigure}
	\caption{Plot of $\tilde{\Phi}_k(\tau)$ (left) and $\Phi_k(\tau)$ (right) with $k=1,2,3$, for different values of $\tau_c=0,15,30$ and fixed $T=30$. When $\tau_c=0$ (unwarped case) the wavefunctions spread as much as possible (given our boundary conditions) through the conifold, being constant for most of values of $\tau$, and when $\tau_c=T$ they localise at the tip $\tau=0$. Wavefunctions for modes with higher $k$ always have a larger amplitude than lower modes.}
	\label{fig:NumericalSolutions}
\end{figure}

In Fig.~\ref{fig:NumericalSolutions} we plot the wavefunctions $\tilde{\uPhi}_k(\tau)$ prior to normalisation (left) and the normalised wavefunctions $\uPhi_k(\tau)$ (right) for the first three modes ($k=1,2,3$), when $T=30$ and $\tau_c=0,15,30$, representing the unwarped, partially warped and fully warped regimes respectively. 
Without the warping, $\tau_c = 0$, the wavefunctions will spread throughout the internal space, much like the zero mode, only being forced to go to zero by our choice of boundary conditions. When we increase the warping by setting $\tau_c/T = 1/2$, we see the wavefunction starting to localise near the tip, but reaching a plateau in the bulk --- this is a transition between a warping dominated and a bulk dominated regime. When the warping completely dominates, the wavefunctions localise at the tip and quickly decrease as they approach the bulk. This illustrates how the balance between a strong warping and a large bulk may influence the localisation of the modes, i.e.~the profile of their wavefunctions, which is reflected in the couplings to other modes, as we discuss in Section \ref{sec:NewtPotCorrections}.
Finally, we see how the normalisation affects the wavefunctions, with higher modes having larger amplitudes than lower modes --- this will translate into stronger couplings for higher modes, which was also found in \cite{Shiu:2007tn}. We also see that the wavefunction amplitudes will be smaller for weaker warping and larger volumes, which can be understood by noting that in the absence of warping the wavefunctions will spread through a larger region and therefore have smaller overall amplitudes --- this translates to weaker couplings.

%We see from the first plot that the solutions $B_k(\tau)$ will localise differently depending on the strength of the warping (more precisely, on the portion of the support which is warped), localising near the tip for large $\tau_c/T$ and in the bulk for small $\tau_c/T$. However, looking at the second plot we see that both regimes seem to localise near the tip --- this is not an effect of the warping, but in fact an effect of the conifold geometry, appearing through the factor $G(\tau)^{-1/2}$ in the wavefunction (\ref{eq:wavefunction}). It is clear that strong warping (large $\tau_c/T$) will increse the localisation significantly, with solutions for weaker warping spreading more evenly through the conifold. 

%%%%%%%%%%%%%%%%%%%%%%%%%%%%%%%%%%%%%%%%%%%%%%%%%%%%%%%%%%%%%%%%%%%%%%%
%  3.3 ANALYTICAL APPROXIMATION
%%%%%%%%%%%%%%%%%%%%%%%%%%%%%%%%%%%%%%%%%%%%%%%%%%%%%%%%%%%%%%%%%%%%%%%

\subsection{Analytical approximation}
\label{sec:analyticalApproximations}

In order to find an approximate analytical solution for the Schr\"odinger equation (\ref{eq:Bk_schrodinger}), we split the potential into two pieces, the warped region when $\tau<\tau_c$ and the unwarped region when $\tau>\tau_c$, and approximate each region using the corresponding dominant term in $H_{\tau_c}(\tau)$ (\ref{eq:warp_factor_tauc}). The asymptotic behaviours of the functions appearing in $V_{eff}$ (\ref{eq:Veff}), as $\tau\to\infty$, are given by
\begin{align}
    \frac{I(\tau)}{\K(\tau)^2} &\to \frac{3}{2}\tau e^{-2\tau/3} \,, \\
    \frac{(G^{1/2})''}{G^{1/2}} &\to \frac{4}{9} - \frac{16}{3}\tau e^{-2\tau} 
\end{align}
and $\K(\tau)\to 2^{1/3}e^{-\tau/3}$. Notice that the subleading term in the second line goes as $\tau(e^{-2\tau/3})^3$, which is subdominant compared to $\tau e^{-2\tau/3}$ in the first line. In this limit the effective potential becomes 
\begin{equation}
	V_{eff}\overset{\tau\to\infty}{\approx} 
	V_{asym} \equiv
	\begin{dcases}
		-\frac{1}{4}\frac{\hat{E}_k^2}{I(\tau_c)}~ \tau e^{-2\tau/3} + \frac{4}{9} \,,	& \tau<\tau_c \\
		- \frac{1}{6\cdot 2^{2/3}}\hat{E}_k^2 e^{2\tau/3} + \frac{4}{9} \,			& \tau_c<\tau<T
	\end{dcases}
	\label{eq:Vasym}
\end{equation}

\begin{figure}[t]
     \centering
     \includegraphics[width=0.7\textwidth]{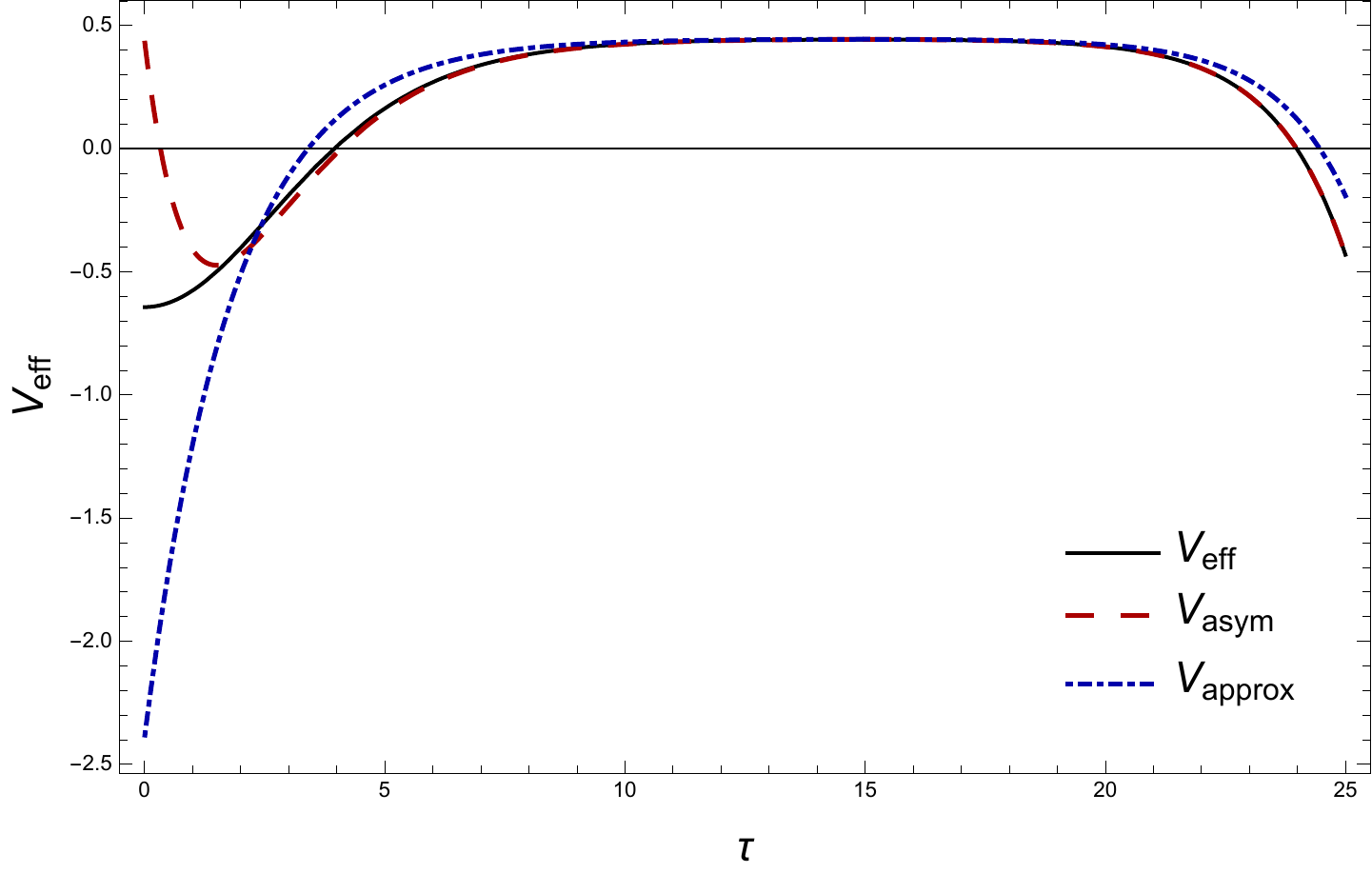}
     \caption{Effective potential  $V_{eff}$ (\ref{eq:Veff}) for the first excited mode, obtained numerically for $\tau_c=15, T=25$, together with $V_{asym}$ (\ref{eq:Vasym}) and $V_{approx}$ (\ref{eq:Vapprox}) for the best fit values $(a,\nu)=(1.96,2.45)$. We see that for small $\tau$, $V_{asym}$ this behaves rather differently from $V_{eff}$. 
Using $V_{approx}$ (\ref{eq:Vapprox}) we can better approximate this regime. Notice that a better fit could be achieved by introducing further terms, but this would not allow for an analytical solution, which is the main goal of this approximation.}
     \label{fig:Veff}
\end{figure}

We plot $V_{asym}$ in (Fig.~\ref{fig:Veff}) together with the exact form of $V_{eff}$. We can see that this is a good approximation for large $\tau$, as expected, but it behaves rather differently at small $\tau$. This can be problematic since we are dealing with an eigenvalue problem, which depends crucially on this behaviour. For small $\tau$, $\tau e^{-\frac{2\tau}{3}}\sim \tau$, so that (\ref{eq:Vasym}) will approach the positive constant $4/9$ as $\tau\to 0$. If instead we have only the exponential behaviour, the negative contribution at small $\tau$ remains, which means we can better approximate this regime (Fig.\ref{fig:Veff}). For larger $\tau$, however, this makes the approximation less accurate. We can modify this approximation by including a couple of free parameters, as suggested in \cite{Tye:2005qs}, to obtain a $V_{approx}$ that looks more like $V_{eff}$ at small $\tau$. 
Using the notation in \cite{Tye:2005qs}, we write the potential as 
\begin{equation}
    V_{approx} \equiv \begin{cases}
	(-\lambda_1^2 e^{-\frac{2}{\nu}\frac{2\tau}{3}} + 4)/9 \,, & \tau<\tau_c \\
	(-\lambda_2^2 e^{\frac{2\tau}{3}} + 4)/9 \,, & \tau_c<\tau<T  
	\end{cases}
    \label{eq:Vapprox}
\end{equation}
where $\lambda_1=a\hat{E}_k/I(\tau_c)^{1/2}$, with free parameters $a$ and $\nu$, and $\lambda_2 = \lambda_1 e^{-\left(1+\frac{2}{\nu}\right)\frac{\tau_c}{3}}$ such that $V_{approx}$ is continuous at $\tau_c$ (in the absence of the throat region, we would have $\lambda_2 = \frac{3^{1/2}}{2^{5/6}}\hat{E}_k$, which is what we use in the limit $\tau_c\to 0$).
%The exact numbers are chosen for convenience (which is only clear in hindsight, once we have the analytical solution). 

These free parameters should be thought of as a compensation for changing the asymptotic $(\tau\to\infty)$ form of the potential.
They should be chosen by comparing the analytical solution obtained with $V_{approx}$ (\ref{eq:Vapprox}) with the numerical result obtained with $V_{eff}$ (\ref{eq:Veff}). 

The wavefunction will have a profile $\uPhi^{(1)}_k(\tau)$ inside the throat ($\tau<\tau_c$), a profile $\uPhi^{(2)}_k$ in the unwarped piece of the conifold ($\tau_c<\tau<T$), and will vanish $\uPhi^{(3)}_k(\tau)=0$ in the CY$_3$ ($\tau>T$). We therefore choose the following boundary conditions
\begin{enumerate}
	\item The wavefunction is finite at $\tau=0$, i.e. 
		\begin{equation*}
			\lim_{\tau\to 0}\uPhi^{(1)}_k(\tau) < \infty \implies \lim_{\tau\to 0}B^{(1)}_k(\tau) = 0 \,; 
		\end{equation*}
	\item The wavefunctions match at $\tau=\tau_c$, i.e. 
		\begin{equation*}
			\uPhi^{(1)}_k(\tau_c) = \uPhi^{(2)}_k(\tau_c) 
			\quad \text{and} 
			\quad \partial_\tau\uPhi^{(1)}_k(\tau_c) = \partial_\tau\uPhi^{(2)}_k(\tau_c) \,;
		\end{equation*}
	\item The wavefunctions vanish as they approach the CY$_3$ region, i.e. $\uPhi^{(2)}_k(T) = 0$.
\end{enumerate} 

\noindent The Schr\"odinger equation in the warped region is
\begin{equation}
        \big(B_k^{(1)}\big)'' + \left[\frac{\lambda_1^2}{9} e^{-4\tau/3\nu} - \frac{4}{9}\right]B_k^{(1)} = 0 \,,
\end{equation}
whose solution is a linear combination of Bessel functions
\begin{equation}
        B_k^{(1)}(\tau) = C_1 J_\nu\left(\frac{\nu\lambda_1}{2}e^{-2\tau/3\nu}\right) + D_1Y_\nu\left(\frac{\nu\lambda_1}{2}e^{-2\tau/3\nu}\right) \,, \quad 0\leq \tau \leq \tau_c \,,
\end{equation}
and the  Schr\"odinger equation in the unwarped region is
\begin{equation}
        \big(B_k^{(2)}\big)'' + \left[\frac{\lambda_2^2}{9} e^{2\tau/3} - \frac{4}{9}\right]B_k^{(2)} = 0 \,,
\end{equation}
with solution 
\begin{equation}
        B_k^{(2)}(\tau) = C_2 J_2\left(\lambda_2e^{\tau/3}\right) + D_2Y_2\left(\lambda_2e^{\tau/3}\right) \,, \quad \tau_c < \tau \leq T \,,
\end{equation}
where $C_{1,2},D_{1,2}$ are integration constants that are fixed using the boundary conditions 1-3 as follows
\begin{enumerate}
	\item The wavefunction is finite at $\tau=0$ (hence we use $B^{(1)}(\tau)$),
		\begin{equation}
			C_1 J_\nu\left(\frac{\nu\lambda_1}{2}\right) + D_1 Y_\nu\left(\frac{\nu\lambda_1}{2}\right) = 0 
			\implies D_1 = -\frac{J_\nu (x_1)}{Y_\nu(x_1)}C_1 \,,
			\label{eq:boundary_condition_1}
		\end{equation}
		with $x_1=\frac{\nu\lambda_1}{2}$.
	\item The wavefunctions match at $\tau=\tau_c$. The first condition $\uPhi^{(1)}_k(\tau_c) = \uPhi^{(2)}_k(\tau_c) \implies B_k^{(1)}(\tau_c) = B_k^{(2)}(\tau_c)$ (cf. (\ref{eq:wavefunction}) with trivial angular dependence), which gives 
		\begin{equation}
			C_1 J_\nu(x_2) + D_1 Y_\nu(x_2) = C_2 J_2(x_2) + D_2 Y_2(x_2)  \,,
			\label{eq:boundary_condition_2a}
		\end{equation}
		with  $x_2 = \frac{\nu\lambda_1}{2}e^{-2\tau_c/3\nu}$.  The second condition $\partial_\tau\uPhi^{(1)}_k(\tau_c) = \partial_\tau\uPhi^{(2)}_k(\tau_c)$, after using
		\begin{equation}
			\partial_\tau \uPhi^{(i)} (\tau) = -\frac{1}{2}\frac{G'(\tau)}{G(\tau)}\uPhi^{(i)}(\tau) + G(\tau)^{-1/2}\partial_\tau B^{(i)} (\tau) \,, \quad i=1,2 \,, 
		\end{equation}
		and $\uPhi^{(1)}_k(\tau_c) = \uPhi^{(2)}_k(\tau_c)$, implies $\partial_\tau B^{(1)} (\tau_c) = \partial_\tau B^{(2)} (\tau_c)$.  Moreover, we can use the relation
		\begin{equation}
			Z'_\alpha (x) = \frac{\alpha}{x}Z_\alpha (x) - Z_{\alpha + 1}(x) \,,
		\end{equation}
		for a Bessel function $Z_\alpha(x)$, to rewrite $\partial_\tau B^{(1)} (\tau_c) = \partial_\tau B^{(2)} (\tau_c)$ as
		\begin{equation}
			\lambda_1 \{C_1 J_{\nu+1}(x_2) + D_1 J_{\nu+1}(x_2)\} = -\lambda_2 \{C_2 J_{3}(x_2) + D_2 Y_{3}(x_2)\} \,.
			\label{eq:boundary_condition_2b}
		\end{equation}
	\item The wavefunctions vanish at $\tau=T$, i.e. $\uPhi^{(2)}_k(T) = 0$, which implies
		\begin{equation}
			C_2 J_2(\lambda_2 e^{T/3}) + D_2 Y_2(\lambda_2 e^{T/3}) = 0 
			\implies D_2 = -\frac{J_2(x_3)}{Y_2(x_3)}C_2 \,, 
			\label{eq:boundary_condition_3}
		\end{equation}
		with $x_3 = \lambda_2 e^{T/3}$.
\end{enumerate} 

\noindent Putting all these together, we find
\begin{align}
	D_1 &= -\frac{J_\nu(x_1)}{Y_\nu(x_1)}C_1 \, \\
	C_2 &= \frac{Y_2(x_3)}{Y_\nu(x_1)}\frac{J_\nu(x_2)Y_\nu(x_1) - J_\nu(x_1)Y_\nu(x_2)}{J_2(x_2)Y_2(x_3) - J_2(x_3)Y_2(x_2)} C_1 \,, \\
	D_2 &= -\frac{J_2(x_3)}{Y_2(x_3)}\frac{Y_2(x_3)}{Y_\nu(x_1)}\frac{J_\nu(x_2)Y_\nu(x_1) - J_\nu(x_1)Y_\nu(x_2)}{J_2(x_2)Y_2(x_3) - J_2(x_3)Y_2(x_2)} C_1 \,,
\end{align}
as well as the quantisation condition (cf. \cite{Tye:2005qs})
\begin{equation}
	e^{\left(1+\frac{2}{\nu}\right)\frac{\tau_c}{3}}\frac{J_2(x_2)Y_2(x_3) - J_2(x_3)Y_2(x_2)}{J_\nu(x_2)Y_\nu(x_1) - J_\nu(x_1)Y_\nu(x_2)} 
	= \frac{J_2(x_3)Y_3(x_2) - J_3(x_2)Y_2(x_3)}{J_{\nu+1}(x_2)Y_\nu(x_1) - J_\nu(x_1)Y_{\nu+1}(x_2)} \,.
\end{equation}
Notice that this is an equation for $\lambda_1 = a\hat{E}_k$ depending on the choices of $\tau_c$ and $T$. Once more, this cannot be solved analytically. Instead, we can look at the different limits $\tau_c\to T$ (fully warped conifold) and $\tau_c\to 0$ (unwarped conifold) for which there are analytical solutions.

\subsubsection{Fully warped conifold \texorpdfstring{$(\tau_c\to T)$}{TEXT}}

In this limit $x_2\to x_3$, so that $J_2(x_2)Y_2(x_3) - J_2(x_3)Y_2(x_2) \to 0$ which, together with the property $J_2(x_3)Y_3(x_3) - J_3(x_3)Y_2(x_3) = -\frac{2}{\pi x_2} \neq 0$, implies 
\begin{equation}
	J_\nu(x_2)Y_\nu(x_1) - J_\nu(x_1)Y_\nu(x_2) \overset{!}{=} 0 \implies J_\nu\left(\frac{\nu\lambda_1}{2}\right) = 0 \,,
\end{equation}
where we assume $x_2\ll 1$. For large $z$, $J_\nu(z)$ can be approximated by
\begin{equation}
    J_\nu(z) \sim \sqrt{\frac{2}{\pi z}}\cos\left(z-\frac{\nu\pi}{2} - \frac{\pi}{4}\right)\,, 
\end{equation}
whose roots are
\begin{equation}
    z = k\pi +\left(\nu - \frac{1}{2}\right)\frac{\pi}{2}.
\end{equation}
Hence, for $z=\frac{\nu\lambda_1}{2}$ and recalling the definitions of $\lambda_1$ and $\hat{E}_k$, we have
\begin{equation}
    \hat{E}_k = I(\tau_c)^{1/2} \Big\{ \frac{2\pi}{a\nu}k + \left(\nu - \frac{1}{2}\right)\frac{\pi}{a\nu} \Big\} \,,
    \label{eq:EnergySpectrum_InfiniteThroat_asymptotic}
\end{equation}
which we can fit to a numerical solution with $\tau_c=T\gg 1$ to find the best values of $a$ and $\nu$ for this analytical approximation\footnote{The difference between our result and the one in \cite{Tye:2005qs} comes from an extra factor of $2^{1/3}$ in our definition of $\hat{E}_k$}, which gives $\nu\sim 2.45$ and $a\sim 1.96$. 
Notice that the energies are suppressed by the warping, as we would expect for a warped throat.
The quantisation condition also implies $D_1 \approx 0$, so that in this limit the wavefunction takes the simpler form
\begin{equation}
	\uPhi_k (\tau) \approx \frac{N_{(k)}}{(\sinh(2\tau) - 2\tau)^{1/3}} J_\nu \left(\frac{\nu\lambda_1}{2}e^{-2\tau/3\nu}\right) \,,
	\label{eq:analytical_solution_warped_wavefunctions}
\end{equation}
which peaks near the tip of the throat and quickly decays towards the bulk (Fig.~\ref{fig:eigenfunctions_analytical}).

\subsubsection{Unwarped conifold \texorpdfstring{$(\tau_c\to 0)$}{TEXT}}

In this limit $x_2\to x_1$, so that $J_\nu(x_2)Y_\nu(x_1) - J_\nu(x_1)Y_\nu(x_2) \to 0$. Together with the property $J_{\nu+1}(x_1)Y_\nu(x_1) - J_\nu(x_1)Y_{\nu+1}(x_1) = \frac{2}{\pi x_1} \neq 0$, this implies 
\begin{equation}
	J_2(x_1)Y_2(x_3) - J_2(x_3)Y_2(x_1) \overset{!}{=} 0 \implies J_2(\lambda_2 e^{T/3}) = 0 \,,
\end{equation}
where we assume $x_1 \ll 1$. Using the same approximation for $J_\nu(z)$, these roots are given approximately by
\begin{equation}
	z = k\pi + \frac{3\pi}{4} \,.
\end{equation}
Hence, for $z = \lambda_2 e^{T/3}$, we have 
\begin{equation}
	E_k = \frac{1}{R_{con}}\Big\{\pi k + \frac{3\pi}{4} \Big\} \,,
\end{equation}
where $R_{con}=c^{1/4}r_T$ is the physical size of the conifold, with $r_T$ the value of the radial coordinate $\rl = \frac{3^{1/2}}{2^{5/6}}\epsilon^{2/3}e^{\tau/3}$ at which the conifold meets the CY$_3$, i.e. at $\tau=T$.
Notice that the energies are suppressed by the size of the conifold, $R_{con}$, which is contributing to a large volume. 
This quantisation condition also implies $D_2\approx 0$, so that in this limit the wavefunctions take the form\footnote{In this limit, it is actually important to keep the $D_2$ term in the wavefunction at small values of $\tau$, otherwise it will not be regular at $\tau=0$.}
\begin{equation}
	\uPhi_k(\tau) \approx \frac{N_{(k)}}{(\sinh(2\tau) - 2\tau)^{1/3}}J_2\left(\lambda_2 e^{\tau/3}\right) \,.
	\label{eq:analytical_solution_unwarped_wavefunctions}
\end{equation}

\begin{figure}
     \centering
     \begin{subfigure}[b]{0.48\textwidth}
         \centering
         \includegraphics[width=\textwidth]{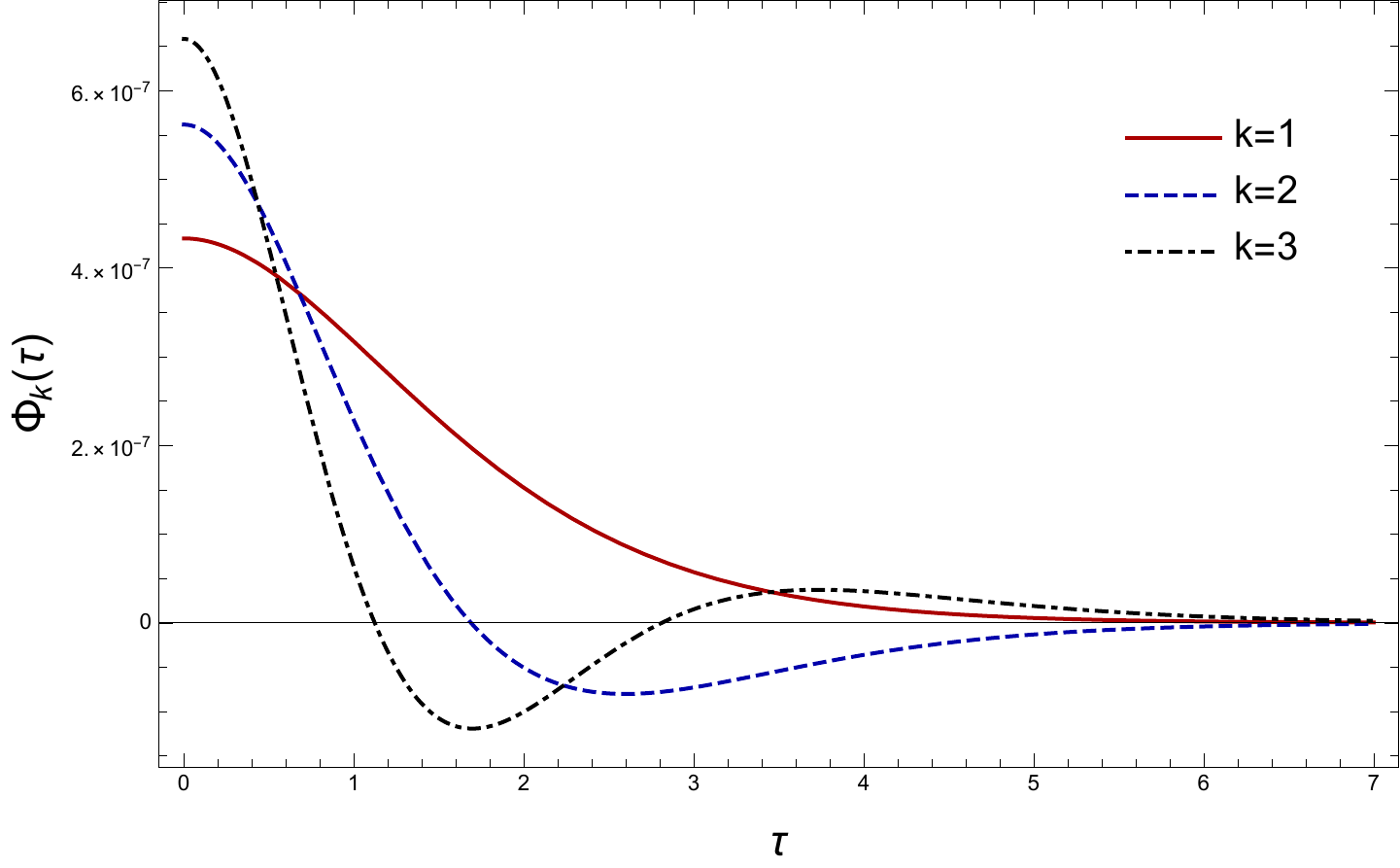}
     \end{subfigure}
     \hfill
     \begin{subfigure}[b]{0.48\textwidth}
         \centering
         \includegraphics[width=\textwidth]{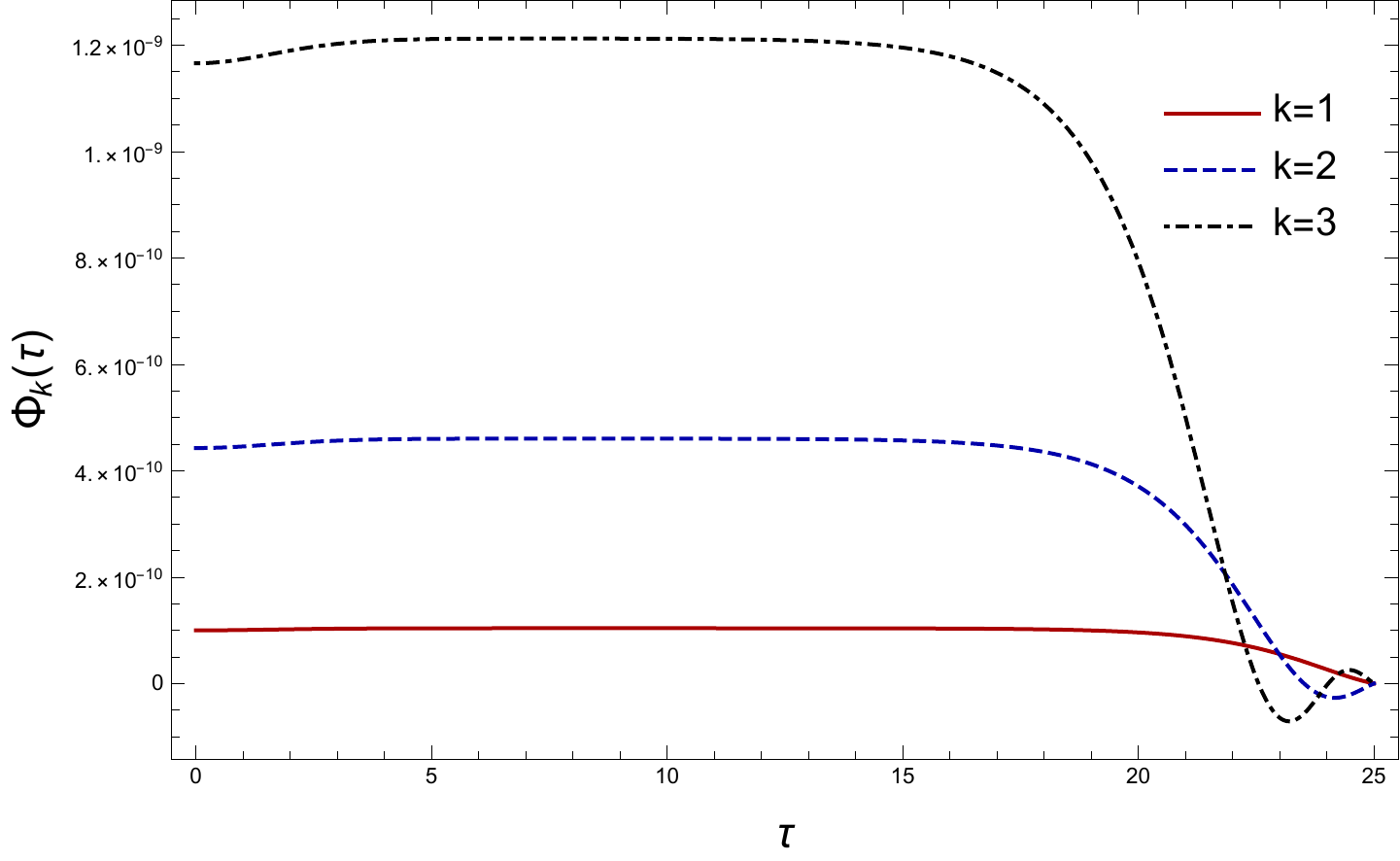}
     \end{subfigure}
    \caption{Wavefunctions obtained analytically in the limits $\tau_c\to T$ (\ref{eq:analytical_solution_warped_wavefunctions}) and $\tau_c\to 0$, with $T=25$. Notice the localisation near the tip in the limit $\tau_c\to T$ and the spreading of the wavefunctions in the limit $\tau_c\to 0$.}
    \label{fig:eigenfunctions_analytical}
\end{figure}

In Fig.~\ref{fig:eigenfunctions_analytical} we plot the approximate wavefunctions in the two limits described above. We can clearly see the localisation near the tip in the limit $\tau_c\to T$ and the spreading of the wavefunctions in the limit $\tau_c\to 0$, as one would expect from the physical interpretation of these limits. These should be compared with the wavefunctions obtained numerically (Fig.~\ref{fig:NumericalSolutions}).

\medskip

Now that we have a full description of the tower of spin-2 KK modes, $h_{\mu\nu}^k(x^{\rho})$, which are the 4d manifestation of the extra dimensions, we may study its effects on 4d gravitational interactions. 
\newpage
\section{Predictions for the Modified Newtonian Potential}
\label{sec:NewtPotCorrections}

In this section we consider the corrections to the Newtonian gravitational potential between two point masses living on a (3+1)-dimensional brane sitting somewhere along the deformed conifold region of the compact space and compare with experimental and observational constraints.  In Section \ref{sec:modNewt} we derive the modifications to the Newtonian potential induced by the KK tower of spin-2 modes, which can be parameterised by a Yukawa-type interaction with a single-massive field and parameters $(\alpha, \lambda)$.  In Section \ref{sec:paramgen}, we relate these phenomenological parameters to the parameters of the Klebanov-Strassler solution.  By fixing the warped-down scale at the brane to be $\sim$ TeV scale, and bearing in mind the need for D3-tadpole cancellation, we explore the predictions for $(\alpha, \lambda)$ for a range of string parameters and compare with current constraints.   Then, in Sections \ref{sec:fullywarped} and \ref{sec:fullyunwarped}, we restrict to the fully warped and unwarped cases, respectively, where we can obtain clear theoretical bounds on the $(\alpha, \lambda)$ parameter space depending on where the brane lies along the conifold. 

%%%%%%%%%%%%%%%%%%%%%%%%%%%%%%%%%%%%%%%%%%%%%%%%%%%%%%%%%%%%%%%%%%%%%%%%
%   4.1 MODIFIED NEWTONIAN POTENTIAL
%%%%%%%%%%%%%%%%%%%%%%%%%%%%%%%%%%%%%%%%%%%%%%%%%%%%%%%%%%%%%%%%%%%%%%%%

\subsection{Modified Newtonian potential}
\label{sec:modNewt}

In order to study the effects of the KK tower of spin-2 modes obtained in Section \ref{sec:conifoldKKtower}, we picture a braneworld scenario, with the Standard Model fields localised on a (3+1)-dimensional brane (stack) at $y_{brane}$ in the compact space. We want to consider the corrections to the Newtonian potential between masses living on the brane due to the presence of the infinite tower of massive KK gravitons. The (Fourier transformed) potential is obtained by looking at a scattering diagram where two particles interact through the exchange of a virtual graviton, in the limit where the energy of the graviton goes to zero \cite{Callin:2004detail},
\begin{figure}[h!]
	\centering
	\includegraphics[width=0.75\textwidth]{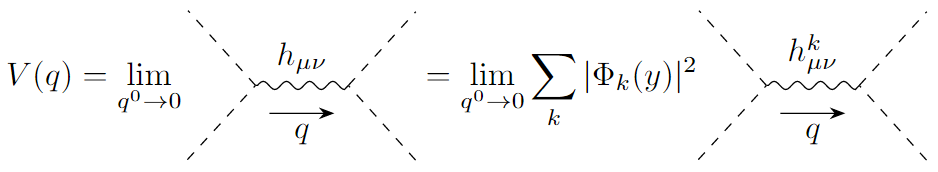}
\end{figure}

For this we need both the graviton propagator and the interactions with matter on the brane. Using the decomposition (\ref{eq:decomposition_munu}) and normalising the wavefunctions such that (\ref{eq:orthogonal_modes}) is satisfied, each propagator is simply the 4d propagator for a spin-2 field. Assuming the spacetime is approximatly flat Minkowski, we have \cite{Rattazzi}
\begin{equation}
    D_{\mu\nu\alpha\beta}^{(4,m_k)}(x,x') = \int \frac{d^4q}{(2\pi)^4}\frac{P_{\mu\nu\alpha\beta}^{(m_k)}(q)}{q^2 - m_q^2 + i\varepsilon}e^{-iq\cdot (x-x')} \,,
\end{equation}
where $D_{\mu\nu\alpha\beta}^{(4,m_k)}(x,x')$ is the 4d propagator of the $k^{th}$ mode (with mass $m_k$) and with the polarisation tensor 
\begin{align}
    P_{\mu\nu\alpha\beta}^{(m=0)}(q) &= \frac{1}{2}(\eta_{\mu\alpha}\eta_{\nu\beta} + \eta_{\mu\beta}\eta_{\nu\alpha}-\eta_{\mu\nu}\eta_{\alpha\beta})\,, \\
    P_{\mu\nu\alpha\beta}^{(m>0)}(q) &= \frac{1}{2}(\eta_{\mu\alpha}\eta_{\nu\beta} + \eta_{\mu\beta}\eta_{\nu\alpha}-\eta_{\mu\nu}\eta_{\alpha\beta}) 
    -\frac{1}{2m^2}(\eta_{\mu\alpha}q_\nu q_\beta + \eta_{\mu\beta}q_\nu q_\alpha +\eta_{\mu\beta}q_\mu q_\alpha) \nonumber \\
    &+\frac{1}{6}\left(\eta_{\mu\nu} + \frac{2}{m^2}q_\mu q_\nu\right)\left(\eta_{\alpha\beta} + \frac{2}{m^2}q_\alpha q_\beta\right) \,.
\end{align}

%In this context, when we are treating the perturbations $h_{MN}$ as particle fields (corresponding to the graviton), these cannot be dimensionless (in GR $\delta g_{MN} = h_{MN}$ is dimensionless). In fact, in order to have canonical kinetic terms, $h_{MN}$ must have dimension $(D-2)/2$ in $D$ dimensions (i.e. $4$ in $D=10$ and $1$ in $D=4$). This means that the eigenfunctions $\Phi_k(y)$ have dimension $3$. This is consistent with $\delta g_{MN} =\frac{\sqrt{V_w}}{M_P}h_{MN}$, as it was introduced below (\ref{eq:graviton_kinetic_terms}) and which indeed gives dimension 4 to the $h_{MN}$.

\newcommand{\gbrane}{\mathfrak{g}}
We now consider the brane action, which contains the fields living in the worldvolume of the brane at $y_{brane}$ and their interactions with the graviton,
\begin{equation}
	S_{brane} = \int d^4x \sqrt{-\gbrane} ~ \mathcal{L}_M \,,
\end{equation}
where $\gbrane_{\mu\nu}$ is the pullback of the metric
%\footnote{Here we denote the 10d metric $G_{MN}$ rather than $g_{MN}$ for convenience of notation.} 
$G_{MN}$ onto the (3+1)-dimensional brane
\begin{equation}
	\gbrane_{\mu\nu} = G_{MN}\frac{\partial X^M}{\partial x^{\mu}}\frac{\partial X^N}{\partial x^{\nu}}\,,
	\label{eq:metric_pullback_def}
\end{equation} 
for 10d coordinates $X^M$ and 4d coordinates on the brane $x^\mu$, which we align with the 4d coordinates of (\ref{eq:10dmetric}). Hence $\partial_\mu X^M = \delta^M_\mu + \partial_\mu\xi^M$, with brane fluctuations $\xi^M(x^{\mu})$, which gives for the pulled-back metric
\begin{align}
	\gbrane_{\mu\nu} = G_{\mu\nu} + 2G_{M(\mu}\partial_{\nu)}\xi^M + G_{MN}\partial_\mu\xi^M\partial_\nu\xi^N \,.
	\label{eq:metric_pullback}
\end{align}
Since the brane perturbations are assumed to be small, we may expand $\sqrt{-\gbrane}$ in the action\footnote{We use the known result $\det(1+\epsilon A) = 1+\epsilon \mathrm{Tr}(A) + \Op(\epsilon^2)$, when $\epsilon\ll 1$.} and express it in terms of $G_{MN}$,
\begin{equation}
	S_{brane} = \int d^4x \sqrt{-\det G_{\mu\nu}} ~ \mathcal{L}_M \Big(1 + G_{M\mu}G^{\mu\nu}\partial_\nu\xi^M + \frac{1}{2}G_{MN}G^{\mu\nu}\partial_\mu\xi^M\partial_\nu\xi^N\Big) \,.
\end{equation}
We will neglect the fluctuations of the brane and study the interactions between graviton KK modes and matter fields only, so that the action is simply
\begin{equation}
	S_{brane} = \int d^4x \sqrt{-\det G_{\mu\nu}} ~ \mathcal{L}_M \,.
\end{equation}

\noindent When we perturb the bulk metric $G_{MN}\to G_{MN}^{0} + \delta G_{MN}$, the action becomes
\begin{equation}
	S_{brane} = S_{brane}^0 - \frac{1}{2} \int d^4x \sqrt{-\det G_{\mu\nu}^0} ~ \tilde{T}^{\mu\nu} \delta G_{\mu\nu} \,,
\end{equation}
where $\tilde{T}_{\mu\nu}$ is the energy-momentum tensor with respect to $G_{\mu\nu}$. Using the background metric (\ref{eq:10dmetric}) and the fluctuations in terms of (canonically normalised) perturbations of $g_{\mu\nu}$, with $\delta G_{\mu\nu} = \kappa ~ H^{-1/2} ~ h_{\mu\nu}$, 
\begin{equation}
	S_{brane} = S_{brane}^0 - \frac{1}{2} \int d^4x \sqrt{-\det g_{\mu\nu}} ~ \sum_k \big(\kappa~\uPhi_k(y_b)\big) T^{\mu\nu} h_{\mu\nu}^k \,,
\end{equation}
where $T_{\mu\nu}$ is defined with respect to $g_{\mu\nu}$,
\begin{align}
	%-\frac{2}{\sqrt{-\det G_{\mu\nu}}}\frac{\delta(\sqrt{-G_{\mu\nu}}\mathcal{L}_M)}{\delta G^{\mu\nu}}
	\tilde{T}_{\mu\nu}=  -\frac{2}{\sqrt{-\det g_{\mu\nu}}}\frac{\delta(\sqrt{-\det G_{\mu\nu}}\mathcal{L}_M)}{\delta g^{\mu\nu}} H(y_b)^{3/2} 
	= T_{\mu\nu} H(y_b)^{3/2} \,.
\end{align}
From this action we conclude that the $k^{th}$ mode couples to matter with a coupling $(\kappa ~ \uPhi_k(y_b))$ that depends on the 10d gravitational coupling $\kappa = \sqrt{V_w}/M_P$ and its wavefunction evaluated at the position of the brane $y_b$. For the zero mode, with constant wavefunction (\ref{eq:graviton_zeromode_wavefunction}), the coupling is therefore the usual $1/M_P$.

Using $S_{brane}^0$, we can also see how the warping affects the energy scales on the braneworld theory. Taking $\mathcal{L}_M$ to include a single scalar field $\varphi$ with a Higgs-like potential \cite{Randall:1999ee}, 
\begin{align}
	S_{brane}^0 &= \int d^4x \sqrt{-\det G_{\mu\nu}} ~ \{G^{\mu\nu}(D_\mu\varphi)^\dagger(D_{\nu}\varphi) - \lambda(|\varphi|^2 - v_0^2)^2\} \nonumber \\ 
	%
%	&= \int d^4x \sqrt{-\det g_{\mu\nu}} ~ H(y_b)^{-1} \{H(y_b)^{1/2}g^{\mu\nu}(D_\mu\varphi)^\dagger(D_{\nu}\varphi) - \lambda(|\varphi|^2 - v_0^2)^2\} \nonumber \\
	%
	&= \int d^4x \sqrt{-\det g_{\mu\nu}} ~ \{H(y_b)^{-1/2}g^{\mu\nu}(D_\mu\varphi)^\dagger(D_{\nu}\varphi) - \lambda(H(y_b)^{-1/2}|\varphi|^2 - H(y_b)^{-1/2}v_0^2)^2\} \nonumber \\
	&\to \int d^4x \sqrt{-\det g_{\mu\nu}} ~ \{g^{\mu\nu}(D_\mu\varphi)^\dagger(D_{\nu}\varphi) - \lambda(|\varphi|^2 - v^2)^2\} \,,
	\label{eq:brane_hierarchy}
\end{align} 
with the field redefinition $\varphi\to H(y_b)^{1/4}\varphi$, so that the field is canonically normalised. The mass scales are then warped down as $v=H(y_b)^{-1/4}v_0$, which depends on the position of the brane in the compact space --- the biggest hierarchy is achieved by placing the brane at the tip of the throat, i.e. $y_b=0$.

Putting everything together, the gravitational potential in momentum space is given by 
\begin{equation}
    V(q) = \frac{V_w}{M_P^2}\sum_k |\uPhi_k(y_b)|^2\frac{T_1^{\mu\nu}P^{(m_k)}_{\mu\nu\alpha\beta}T_2^{\alpha\beta}}{|q^2 - m^2|}\Bigg |_{q^0\to 0}\,,
\end{equation}
%where $T_1^{\mu\nu}(x)=m_1\delta(x)u^\mu u^\nu$ and $T_2^{\alpha\beta}(x)=m_2\delta(x)u^\alpha u^\beta$ are the energy momentum tensors of two point particles of masses $m_1$ and $m_2$ at rest.
where $T_1^{\mu\nu} = m_1\delta^\mu_0\delta^\nu_0$ and $T_2^{\alpha\beta} = m_2\delta^\alpha_0\delta^\beta_0$ are the energy-momentum tensors of two point particles of masses $m_1$ and $m_2$ at rest, so that only $P_{0000}^{(m)}$ is relevant, which in the $q^0\to 0$ limit simply gives
\begin{equation}
    P_{0000}^{(m)}(q)
    \begin{cases}
    \frac{1}{2} & m=0\,, \\
    \frac{2}{3} & m>0\,. 
    \end{cases}
\end{equation}
Inserting this in the potential, we obtain
%\footnote{Recall that the wavefunctions $\Phi^k(y)$ have mass dimension 3, so that $h_{\mu\nu}(x,y)=h_{\mu\nu}^k(x)\Phi^k(y)$ has dimension 4, even though $h_{\mu\nu}(x)$ has dimension 1. This includes the factor $V_w/M_P^2$ which appears in the normalisation. This cancels the dimensions coming from $V_6$ in front of the potential, making the result dimensionally consistent.}
\begin{equation}
    V(q) = \frac{m_1m_2V_w}{M_P^2}\left\{
    \frac{1}{2}\frac{|\uPhi_0(y_b)|^2}{\textbf{q}^2} + \frac{2}{3}\sum_{k>0}\frac{|\uPhi_k(y_b)|^2}{\textbf{q}^2+m_k^2}
    \right\}\,,
\end{equation}
or in position space
\begin{equation}
    V(r) = G_N\frac{m_1m_2}{r}V_w\left\{
        |\uPhi_0(y_b)|^2 + \frac{4}{3}\sum_{k>0}|\uPhi_k(y_b)|^2e^{-m_k r}
    \right\}\,,
\end{equation}
where we used $M_P^{-2} = 8\pi G_N$, together with the factor of $4\pi$ coming from the Fourier transform, to write the potential in terms of Newton's constant $G_N$. The first term gives the contribution of the massless graviton, which is independent of the position of the brane due to (\ref{eq:graviton_zeromode_wavefunction}) and reproduces the Newtonian gravitational potential. The second term contains the contribution from the tower of massive KK-modes, weighed by their respective wavefunctions and suppressed by the exponential $e^{-m_kr}$, which corrects the Newtonian potential and becomes negligible for large distances and larger KK masses. 
%We need the wavefunctions $\uPhi_k(\tau)$ evaluated at $\tau=0$ with their correct normalizations, $N_{(k)}$, as defined in (\ref{eq:eigenmodeNormalisation}). 

\begin{figure}
    \centering
    \includegraphics[width=0.7\textwidth]{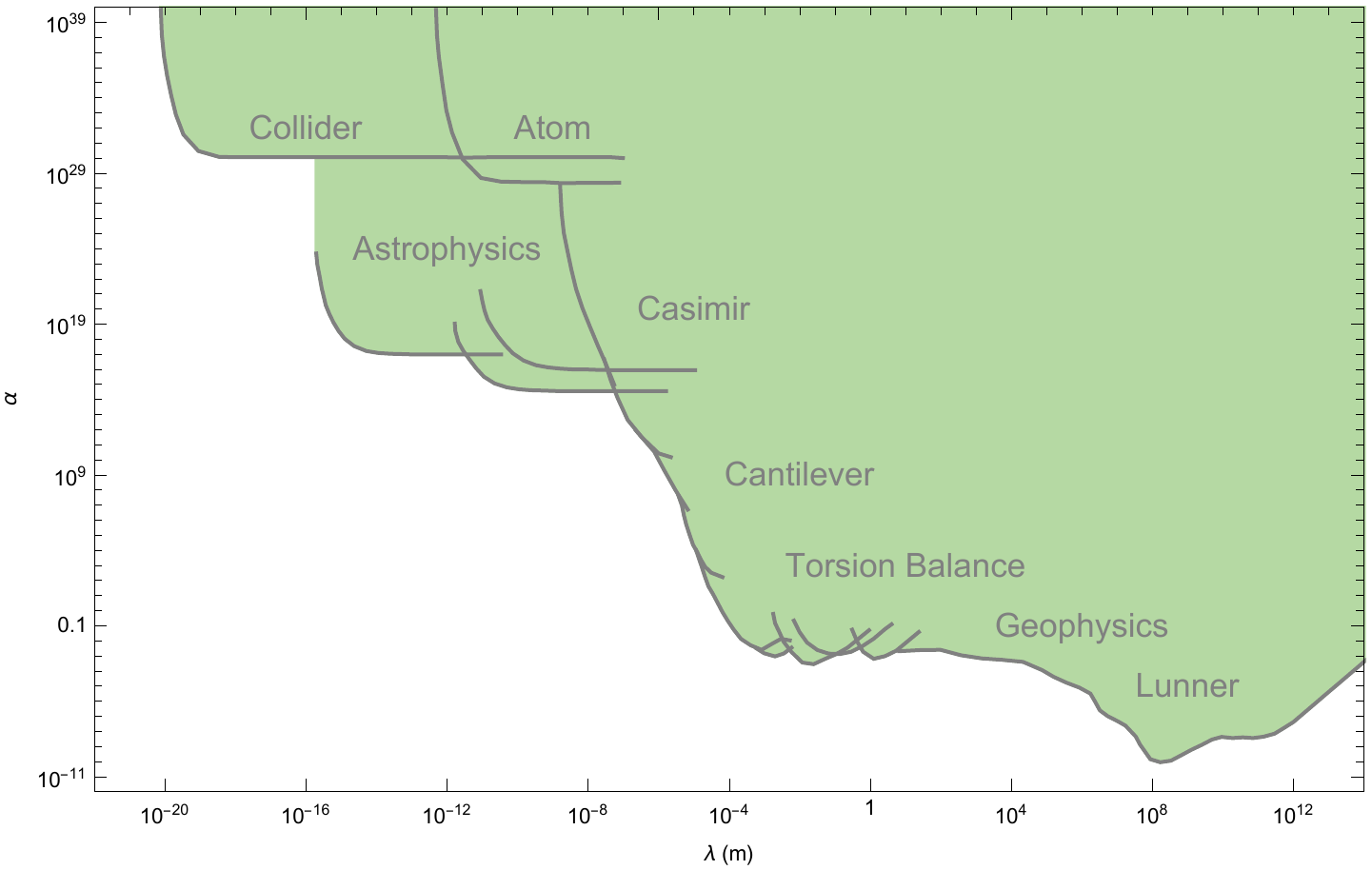}
    \caption{Experimental constraints on the parameters $\alpha$ (coupling strength) and $\lambda$ (range) of a Yukawa-type interaction, with the shaded area indicating the excluded region of parameter space at 95\% confidence level. Figure adapted from \cite{Murata:2014nra,Cembranos:2017vgi}. See \cite{Murata:2014nra,Cembranos:2017vgi} for details and further references.} 
    \label{fig:greenplot}
\end{figure}

Using the zero mode (\ref{eq:graviton_zeromode_wavefunction}) we have 
\begin{equation}
    V(r) = G_N\frac{m_1m_2}{r}\left\{
        1 + \frac{4}{3}V_w\sum_{k>0}|\uPhi_k(y_b)|^2e^{-m_k r}
    \right\}\,.
    \label{eq:gravitationalPotential}
\end{equation}
Figure \ref{fig:greenplot} shows the experimental bounds on deviations from the Newtonian gravitational potential --- from laboratory, geophysical, astrophysical and collider constraints\footnote{See \cite{Hall:1999mk,ParticleDataGroup:2020ssz} for cosmological constraints from overclosure and the diffuse cosmic gamma ray background, which however assume Planckian couplings with the KK tower.}  \cite{Murata:2014nra,Cembranos:2017vgi} --- parametrised as $V(r) = G_N \frac{m_1m_2}{r}(1+\delta V)$ with
\begin{equation}
    \delta V = \alpha \,e^{-r/\lambda}\,,
    \label{eq:YukawaCorrection}
\end{equation}
where $\alpha$ is a dimensionless parameter describing the strength of the interaction and $\lambda$ has dimensions of length and is given in meters (m). This parametrisation arises from considering the correction coming from a Yukawa-type interaction involving a single massive field, which takes the exponential form above. Since we have an infinite tower of massive scalars, we will have an infinite sum of such Yukawa terms. We must, however, either keep only the first mode, which is the dominant contribution due to the exponential suppression for larger masses, or rewrite the sum of Yukawa terms in the form of $\delta V$ given above if we want to compare our predictions with the experimental constraints. 

After exploring the general case, we will consider in detail the warped $(\tau_c\to T)$ and unwarped $(\tau_c\to 0)$ limits, with the brane located at different points in the compact space --- either at the tip $(\tau=0)$ or away from the tip $(\tau\gg 1)$. In these limits we are able to exclude large portions of parameter space from the usual control requirements in string compactifications; $g_s<1$, $g_s M>1$, and well-motivated upper bounds on the flux number $M$ coming from the flux contribution to the D3 tadpole.

%%%%%%%%%%%%%%%%%%%%%%%%%%%%%%%%%%%%%%%%%%%%%%%%%%%%%%%%%%%%%%%%%%%%%%%%
%   4.2 GENERAL CASE
%%%%%%%%%%%%%%%%%%%%%%%%%%%%%%%%%%%%%%%%%%%%%%%%%%%%%%%%%%%%%%%%%%%%%%%%

\subsection{General case}
\label{sec:paramgen}

As discussed in Section \ref{sec:numericalSolutions}, solving numerically the Schr\"odinger equation (\ref{eq:Bk_schrodinger}) with the potential\footnote{With trivial angular dependence, i.e. s-orbital.} (\ref{eq:Veff}) for a given choice of $(\tau_c,T)$ we find a set $(\hat{E}_k,B_k(\tau))$ of eigenvalues and eigenfunctions. We then substitute
\begin{align}
    \uPhi_k(\tau) &= N_{(k)}\tilde{\uPhi}_k(\tau) \,, \\
    N_{(k)}^{-2} &= \frac{2\pi^3}{3}c^{3/2}\epsilon^4 \mathcal{N}_{(k)}^{-2}(\tau_c,T) \,, \\
    G(\tau) &= (\sinh(2\tau) - 2\tau)^{2/3} \,, \\
    m_k &= \hat{E}_k/(\epsilon^{2/3}c^{1/4}) \,,
\end{align}
with $\mathcal{N}_{(k)}^{-2}(\tau_c,T)$ as defined in (\ref{eq:Normalisation_general_numerical}) and $\tilde{\uPhi}_k(\tau) = G(\tau)^{-1/2}B_k(\tau)$, into (\ref{eq:gravitationalPotential}) to obtain
\begin{equation}
    \delta V = \frac{2}{\pi^3} \frac{V_w}{c^{3/2}\epsilon^4} \sum_{k>0} \mathcal{N}_{(k)}^2(\tau_c,T)   |\tilde{\uPhi}_k(\tau_b)|^2 e^{- \hat{E}_k \frac{r}{ \epsilon^{2/3}c^{1/4}}} \,.   
\end{equation}
Defining $A(\tau_b,\tau_c,T)$ and $\mu(\tau_c,T)$ such that
%$\delta V_0 =  \frac{2}{\pi^3} \frac{V_w}{c^{3/2}\epsilon^4}$, we have
%\begin{equation}
%	\delta V_N = \frac{\delta V}{\delta V_0} =  \sum_{k>0}^N \mathcal{N}_{(k)}^2(\tau_c,T)   |\tilde{\uPhi}_k(\tau_b)|^2 e^{- \hat{E}_k \rho} 
%	\equiv A(\tau_b,\tau_c,T) e^{-\mu(\tau_c,T)\rho}\,,
%\end{equation}
%where we emphasize that $A$ and $\mu$ are functions of $(\tau_c,T)$. 

\begin{equation}
	 A(\tau_b,\tau_c,T) e^{-\mu(\tau_c,T)\frac{r}{ \epsilon^{2/3}c^{1/4}}} \equiv \sum_{k>0}^N \mathcal{N}_{(k)}^2(\tau_c,T)   |\tilde{\uPhi}_k(\tau_b)|^2 e^{- \hat{E}_k \frac{r}{ \epsilon^{2/3}c^{1/4}}} \,, 
\end{equation}
we can write the parameters $(\alpha,\lambda)$ in (\ref{eq:YukawaCorrection}) as
\begin{align}
	\alpha =  \frac{2}{\pi^3} \frac{V_w}{c^{3/2}\epsilon^4} A(\tau_b,\tau_c,T) 
	&& \lambda^{-1} =\frac{\mu(\tau_c,T)}{\epsilon^{2/3}c^{1/4} } \,.
	\label{eq:parCorr_definition}
\end{align}

%We can write $V_w$ as
%\begin{align}
%	V_w &= \left\{V_{CY} + \frac{2\pi^3}{3}c^{3/2}\epsilon^4 \int_0^{T} \sinh^2(\tau) H(\tau)\right\} \nonumber \\
%	&= \left\{V_{bulk} + \frac{2\pi^3}{3}c^{3/2}\epsilon^4 \int_0^{\tau_c} \sinh^2(\tau) H(\tau)\right\} \nonumber \\ 
%	&= c^{3/2} l_s^6 \left\{1 + \frac{2\pi^3}{3}|z|^2 \int_0^{\tau_c} \sinh^2(\tau) H(\tau)\right\} \equiv \V_w l_s^6 \,,
%	\label{eq:Vw_explicit}
%\end{align}
%after defining the coordinates in the CY$_3$ such that $V_{bulk} = V_{CY} + V_{unwarped} = c^{3/2}l_s^6$ and using $\epsilon^2=|z|/l_s^3$. Note that if the bulk (the CY$_3$ plus the unwarped piece of the conifold) dominates the volume, $\V_w \approx c^{3/2}$. However, 

It will be useful to replace the volume\footnote{The total volume includes both the bulk and the throat, $V_w = V_{bulk} + V_{th}$, with the bulk being a combination of the CY$_3$ and the unwarped region of the deformed conifold, $V_{bulk} = V_{CY} + V_{con} = c^{3/2}l_s^6$.} with the hierarchy between the tip of the throat and the bulk, which has a direct physical interpretation. From (\ref{eq:brane_hierarchy}) we know that the hierarchy between the fundamental scale and the brane is given by the warp factor (\ref{eq:warp_factor_tauc}) on the brane,
\begin{equation}
     H_{\tau_c}(\tau_b) 
     = 1 + \frac{I(\tau_b)}{I(\tau_c)} \,.
    \label{eq:hierarchy_general}
\end{equation}
For string theory, the natural scale in the UV is the string scale $m_s$. 
Let $M_{b}$ be the scale on the brane. Using (\ref{eq:msMp}), we can write it in terms of the known Planck scale $M_P$,
    \begin{equation}
        H_{\tau_c}(\tau_b)^{-1/4} = \frac{v}{v_0} 
        = \frac{M_{b}}{m_s} 
        = \frac{M_{b}}{M_P}\frac{\sqrt{4\pi\V_w}}{g_s} \,,
    \end{equation}
where $M_P = 2.14\times 10^{18}$ GeV. This implies that the hierarchy between the known scales $M_{b}$ and $M_{P}$, 
\begin{equation}
	\mathcal{H}\equiv H_{\tau_c}(\tau_b)^{-1/4} \frac{g_s}{\sqrt{4\pi\V_w}} \,,
	\label{eq:hierarchy_definition}
\end{equation}
depends on the volume and string coupling, as well as the warp factor $H_{\tau_c}(\tau_b)$. If we choose $M_{b}=1~$TeV, trying to solve the hierarchy problem, the hierarchy takes the value $\mathcal{H}\sim 10^{-15}$.

Using this in (\ref{eq:parCorr_definition}), we find
\begin{subequations}
\begin{align}
	\alpha &= \frac{(2\pi)^2}{(g_sM)^3} \frac{2A(\tau_b,\tau_c,T)}{I(\tau_c)^{3/2}}H_{\tau_c}(\tau_b)^{-1/2}\frac{g_s^2}{\mathcal{H}^2} \label{eq:parCorr_alpha}\,, \\
	\lambda^{-1} &= \frac{\mathcal{H} }{2^{1/6}}\frac{2\pi}{\sqrt{g_sM}} \frac{H_{\tau_c}(\tau_b)^{1/4}}{I(\tau_c)^{1/4}} \frac{\mu(\tau_c,T)}{l_p} \,.
	\label{eq:parCorr_lambda}
\end{align}
	\label{eq:parCorr}
\end{subequations}
The free parameters in (\ref{eq:parCorr}) are $(\tau_c,T,g_s,M,\mathcal{H},\tau_b)$, six in total. We should remember that $\mathcal{H}$ is keeping the dependence on $\V_w$, which is fully determined by the choice of $(\tau_c,\tau_b,g_s,\mathcal{H})$ through (\ref{eq:hierarchy_definition}) --- one should check that a choice of these parameters is consistent with the supergravity requirement $\V_w\gg 1$. Note also that $\tau_c$ is determined by the deformation modulus $|z|=\epsilon^2/l_s^3$, which means we can think of a choice of $\tau_c$ as representing a choice of $|z|$ (which in turn depends on the flux parameter $K$ via $z \sim e^{-\frac{2\pi K}{g_s M}}$ \cite{GKP,Douglas:2007tu,upliftingrunaways2019,Blumenhagen:2019qcg,LVSdS:2010.15903,Bento:2021nbb}). We will fix the $\mathcal{H}$ and $\tau_b$, the position of the brane, which leaves four free parameters.  In principle, the position of the brane should also be determined dynamically, since it becomes a modulus that experiences a potential due to several different ingredients \cite{Baumann:2006th,Baumann:2007ah,Baumann:2010sx}. In this work we will assume that the position can be fixed to a certain value due to the balance between these ingredients, without addressing the issue explicitly.

\begin{figure}[t]
     \centering
         \includegraphics[width=0.7\textwidth]{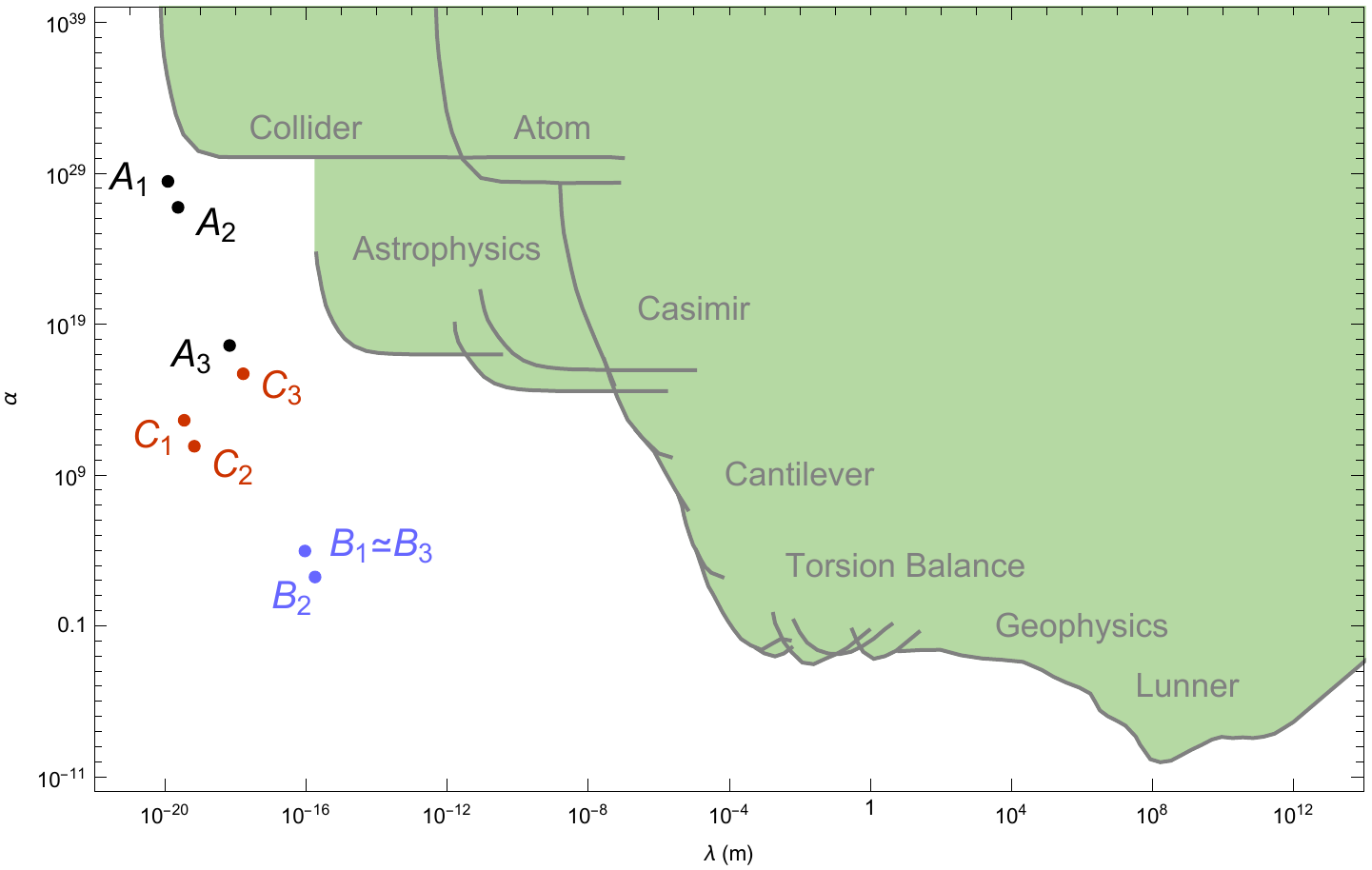}
     \caption{A few examples of the corrections $\delta V$ (\ref{eq:YukawaCorrection}) to the Newtonian potential for different choices of the free parameters in (\ref{eq:parCorr}). The hierarchy is fixed to $\mathcal{H}=10^{-15}$ and $g_s=0.2$ in all parameter sets. All points $A_i$ have $\tau_c=T=40$, corresponding to the fully warped conifold limit. All points $B_i$ have $\tau_c=0$ and $T=30$, corresponding to the unwarped conifold limit. All points $C_i$ have $\tau_c = T/2 = 15$, corresponding to a mid-regime with a warped throat and a piece of the bulk described by an unwarped conifold. We also choose $M=20$ and $75$ for $i=1,3$ and $i=2$, respectively, and $\tau_b=0$ and $15$ for $i=1,2$ and $i=3$, respectively. We notice in particular that all predictions are outside the excluded region, with the couplings being too small to be probed at colliders and the length scales too small to be probed by large scale experiments. See main text for detailed discussion. Figure adapted from \cite{Murata:2014nra,Cembranos:2017vgi}, with the shaded area indicating the excluded region of parameter space at 95\% confidence level.}
    \label{fig:greenplot_samplepoints}
\end{figure}

In Fig.~\ref{fig:greenplot_samplepoints} we show a sample of predictions $(\lambda_i,\alpha_i)$ for different choices of these parameters, divided in three main groups: the fully warped limit, with $\tau_c = T$; the unwarped conifold limit, with $\tau_c=0$; and a mid-regime with $\tau_c = T/2$ --- in this regime we see the competition between the throat trying to localise the modes and the bulk trying to spread them evenly throughout the compact space, since they are not forced to vanish in the bulk as in the fully warped case (see discussion on boundary conditions in Section \ref{sec:KKtower}). The hierarchy is fixed to $\mathcal{H}=10^{-15}$ and $g_s=0.2$ in all parameter sets. Tables \ref{tb:sample_points_A}-\ref{tb:sample_points_C} summarize the parameter choices and the relevant quantities for each set of examples. 

\begin{table}
	\centering
	\begin{tabular}{|c|c|c|c|c|c|c|c|c|}
		\hline
			\rowcolor{black!10!white!90}  & $M$ & $\tau_b$ & $\V_w$ & $M_{KK}$ & $H(\tau_b)^{-1/4}M_{KK}$  & $M_{KK}^w$ & $\V_{th}$ & $MK$ \\
		\hline
		\cellcolor{black!10!white!90} $A_1$ & $20$ & \multirow{2}{*}{$0$}  & \multirow{2}{*}{$9.6\times 10^{16}$} & \multirow{2}{*}{$600$} & \multirow{2}{*}{$3\times 10^{-3}$} & $14$ & $19.2$ & $755$ \\ \cline{1-2}\cline{7-9}
		\cellcolor{black!10!white!90} $A_2$ & $75$   &   & & & & $7.3$ & $1.0\times 10^3$ & $10382$ \\ \hline
		\cellcolor{black!10!white!90} $A_3$ & $20$ & $15$ & $3.0\times 10^{20}$ & $3$ & $8\times 10^{-4}$ & $0.253$ & $19.2$ & $806$ \\
		\hline
	\end{tabular}
	\caption{All points $A_i$ have $\tau_c=T=40$ (corresponding to the fully warped conifold) and fixed $\mathcal{H}=10^{-15}$, $g_s=0.2$. Masses are given in TeV and volumes are in string units. See main text for discussion.}
	\label{tb:sample_points_A}
\end{table}

\begin{table}
	\centering
	\begin{tabular}{|c|c|c|c|c|c|c|c|c|}
		\hline
			\rowcolor{black!10!white!90}  & $M$ & $\tau_b$ & $\V_w$ & $M_{KK}$ & $M_{KK}^{con}$ & $R$ & $\V_{con}$ & $MK$ \\
		\hline
		\cellcolor{black!10!white!90} $B_1$ & $20$ & \multirow{2}{*}{$0$}  & \multirow{2}{*}{$2.3\times 10^{27}$} & \multirow{2}{*}{$7\times 10^{-5}$} & $0.275$ & $7.0\times 10^{3}$ & $3.7\times 10^{23}$ & $556$ \\ \cline{1-2}\cline{6-9}
		\cellcolor{black!10!white!90} $B_2$ & $75$   &   & & & $0.142$ & $1.4\times 10^4$ & $2.0\times 10^{25}$ & $7585$ \\ \hline
		\cellcolor{black!10!white!90} $B_3$ & $20$ & $15$ & $3.2\times 10^{27}$ & $6\times 10^{-5}$ & $0.232$ & $7.0\times 10^3$ & $3.7\times 10^{23}$ & $558$ \\
		\hline
	\end{tabular}
	\caption{All points $B_i$ have $\tau_c=0, T=30$ (corresponding to the unwarped conifold) and fixed $\mathcal{H}=10^{-15}$, $g_s=0.2$. Masses are given in TeV and volumes are in string units. See main text for discussion.}
	\label{tb:sample_points_B}
\end{table}

\begin{table}
	\centering
	\begin{tabular}{|c|c|c|c|c|c|c|c|c|c|}
		\hline
			\rowcolor{black!10!white!90}  & $M$ & $\tau_b$ & $\V_w$ & $M_{KK}$ & $H(\tau_b)^{-1/4}M_{KK}$  & $M_{KK}^w$ & $R$ & $\V_{con}$ & $MK$ \\
		\hline
		\cellcolor{black!10!white!90} $C_1$ & $20$ & \multirow{2}{*}{$0$}  & \multirow{2}{*}{$1.0\times 10^{24}$} & \multirow{2}{*}{$1\times 10^{-2}$} & \multirow{2}{*}{$2\times 10^{-4}$} & $4.9$ & $125$ & $1.2\times 10^{13}$ & $546$ \\ \cline{1-2}\cline{7-9}
		\cellcolor{black!10!white!90} $C_2$ & $75$   &   & & & & $2.5$ & $243$ & $6.3\times 10^{14}$ & $7442$ \\ \hline
		\cellcolor{black!10!white!90} $C_3$ & $20$ & $15$ & $2.3\times 10^{24}$ & $7\times 10^{-5}$ & $6\times 10^{-5}$ & $0.104$ & $125$ & $1.2\times 10^{13}$ & $595$ \\
		\hline
	\end{tabular}
	\caption{All points $C_i$ have $\tau_c=15, T=30$ (corresponding a partially warped conifold) and fixed $\mathcal{H}=10^{-15}$, $g_s=0.2$. See main text for discussion.}
	\label{tb:sample_points_C}
\end{table}

The first thing to note is that none of these examples lies within the excluded region of parameter space, both due to the small couplings and small length scales --- this means that none of them can be excluded from this large set of gravitational experiments and observations. The most likely case to be probed in the near future is the fully warped limit, which is not far from the collider experiments --- these need to go slightly up in energy, but mostly be able to probe smaller couplings, which requires an increase in the statistics (i.e. higher luminosity). By contrast, the case that seems harder to probe is the unwarped conifold (representing an unwarped compactification\footnote{This can be compared with the ADD models \cite{ADD1,ADD2}, with $n=6$ large extra dimensions.  Note that we cannot model $n=1,2$ ADD models with the conifold.}), which must have all its scale suppression coming from a large compactification volume, $\V$. 

We can see that in all cases, increasing the flux number $M$ slightly lowers both the masses and the couplings of the graviton KK modes. It also leads to a larger tadpole charge $MK$, which is determined through the choice of $(\tau_c,g_s,M) \to c |z|^{4/3} \to MK$ from (\ref{eq:tauc_parameters}) and $z\sim e^{-\frac{2\pi K}{g_sM}}$, with $\V_w$ being determined through (\ref{eq:hierarchy_definition}) by further fixing $(\mathcal{H},\tau_b)$. In fact, we can put this together to find
\begin{equation}
	MK \sim - \frac{3(g_sM^2)}{2\pi} \log\left\{
		\frac{2^{1/3}}{(2\pi)^{5/6}}\frac{\sqrt{g_sM}}{g_s^{1/3}} I(\tau_c)^{1/6}I(\tau_b)^{1/12} \mathcal{H}^{1/3}
	\right\} \,.
	\label{eq:tadpoleMK}
\end{equation}
The tadpole cancellation condition is an important constraint in the context of moduli stabilisation, since one typically uses fluxes to stabilise complex structure moduli and the axio-dilaton --- in doing this, one needs to make sure that the D3-tadpole is cancelled once all the ingredients are taken into account. This interplay was recently given a lot of attention in the context of the Tapole Conjecture\footnote{\textbf{Tadpole Conjecture:} The fluxes that stabilise a large number of moduli at a generic point in moduli space have a positive contribution to the tadpole-cancelation condition that grows at least linearly with the number of moduli $n$,
\begin{equation}
	Q^{stabilisation} > \Op(1)\times n \,.
\end{equation}} \cite{tadpoleProblem,Bena:2021tadpole}, where it was argued that even with O3-plane, O7-plane and D7-brane negative contributions to the D3-tadpole, the presence of these ingredients will actually bring a \textit{larger} positive contribution to the tadpole once the stabilisation of all the moduli is taken into account. If we restrict to O3-planes, all examples in the literature have the number of O3-planes less than or equal to $64$, which gives a tadpole charge contribution $|Q_{O3}|\leq 32$. With no other ingredients, tadpole cancelation would require $MK\leq 32$. Note however that, even staying close to the boundary of control (e.g. $g_s=0.5, M=4 \implies g_sM = 2$) and with very small warping $\tau_c=1$, (\ref{eq:tadpoleMK}) gives $MK\sim 47 > 32$. Relaxing the hierarchy such that it connects the string scale with e.g. $10^3$ TeV rather than $1$ TeV still gives $MK\sim 38 > 32$.  

If one doubles $g_sM$ for better control of the supergravity approximation (e.g. $g_s=0.25, M=16 \implies g_sM=4$), (\ref{eq:tadpoleMK}) gives $MK\sim 360$.  For the choice of parameters in $A_1$ we find $MK\sim 755$ as reported in Table \ref{tb:sample_points_A} (in choosing the parameters for the sample points $A_i$ we also require $M_{KK}^w<M_{KK}$ and $r_{UV} > l_s$, which is why we actually use different values of $g_s$ and $M$ giving the same $g_sM=4$). Just asking for the hierarchy to come (at least partially) from the warping requires a large tadpole contribution $MK$, which leads us to consider the Tadpole Conjecture.

In these setups, we also see that moving the brane away from the tip of the throat has a big effect on the gravitational corrections. For $A_3$ the effect is more intuitive, being due to the localisation of the modes, since the couplings depend explicitly on the wavefunction profile (Fig.\ref{fig:NumericalSolutions}). Graviton KK modes will therefore have much weaker couplings with modes living away from the tip of the throat. However, (\ref{eq:parCorr_alpha}) tells us that the coupling strength also depends on the warp factor, with weaker warping giving stronger couplings, which explains why the coupling actually \textit{increases} for $C_3$, rather than decrease --- in this mid-regime the wavefunctions reach a constant plateau rather than quickly decaying towards zero (see Fig.~\ref{fig:NumericalSolutions}), so that the biggest effect will come from the decrease in the warp factor and the coupling increases.

Note also that in $C_i$ a big part of the hierarchy is coming from the volume rather than the warp factor, which is why the volume is much larger than in $A_i$ and why it does not change a lot from $C_1$ to $C_3$ when moving the brane away from the tip. This leads to smaller $M_{KK}$ and $H(\tau_b)^{-1/4}M_{KK}$ and does not require $MK$ as large as in $A_i$. 

%\red{For unwarped setups instead, $\tau_c$ is not well defined since a solution to $(\ref{eq:tauc_parameters})$ most likely does not exist. More precisely, a truly unwarped case would have $H\to 1$ everywhere in the compact space --- one can see that this is not the case for $\tau_c=0$, since $H(0) = 2$, representing instead a formal boundary between warped and unwarped compactifications. This is why $M_{KK}^{con}$ (which is the mass scale associated with the tower of gravitons) changes slightly when the brane is away from $\tau_b=0$, although the truly unwarped conifold gives constant wavefunctions throughout (most of) the compact space. We still choose an example with $\tau_c$ since this permits us to stay within the framework which is useful for warped compactifications and serves as a formal limit. 

%In this example, we are also not taking the conifold to be the whole bulk, with most of the compactification volume being due to the compact CY$_3$ --- this shows us clearly that $M_{KK}^{con}$ is tied to the region of the compact space over which the modes have support (i.e. where the wavefunctions do not vanish) rather that the full bulk. One may argue that it would be more natural to take the unwarped conifold to represent the whole bulk (in other words, not require the modes to vanish in most of the space), in which case one needs only to adjust the size of the conifold (e.g. for the parameters in $B_1$ we would have to choose $T\approx 34.5$).}

In all cases, the scale associated with the graviton tower ($M_{KK}^w$ for $A_i,C_i$ and $M_{KK}^{con}$ for $B_i$) is around the TeV scale, which is reflecting our choice of hierarchy. One might expect that this alone would be enough to make these modes detectable since we are able to access these energies at colliders like the LHC. However, as is well-known, the energy scale (masses of the modes) alone is not enough to determine whether new effects are detectable. The way these extra modes couple with Standard Model particles (which in our setup are confined on the brane) is extremely important --- the KK gravitons might be extremely light and yet couple so weakly to Standard Model particles that they are still undetectable with current experiments and observations. This makes the details of the compactification crucial when studying these effects, not only because they determine the masses of the modes, but also because they will affect the profile of the wavefunctions over the extra dimensions and consequently the couplings to other modes. 

When looking for examples to show in Fig.~\ref{fig:greenplot_samplepoints} we are not able to find any points within the excluded region, despite apparently having a lot of freedom.\footnote{The conifold background we are considering does not allow for an anisotropic compactification, which would be required to realise an unwarped ADD with less than 6 large extra dimensions.} In the next two sections, we study the fully warped and unwarped limits, which provide more insight into the allowed region of parameter space. 

%%%%%%%%%%%%%%%%%%%%%%%%%%%%%%%%%%%%%%%%%%%%%%%%%%%%%%%%%%%%%%%%%%%%%%%%
%   4.3 FULLY WARPED DEFORMED CONIFOLD
%%%%%%%%%%%%%%%%%%%%%%%%%%%%%%%%%%%%%%%%%%%%%%%%%%%%%%%%%%%%%%%%%%%%%%%%

\subsection{Fully warped deformed conifold}
\label{sec:fullywarped}

The fully warped deformed conifold corresponds to the limit $\tau_c\to T$. In this limit, the solution pair $(\hat{E}_k,\Phi_k)$ only depends on $\tau_c$, so that
\iffalse
and the normalisation of the massive modes is well approximated by 
\begin{equation}
    \mathcal{N}_{(k)}^{-2} 
    \approx \frac{1}{I(\tau_c)} \int_0^{\tau_c} d\tau \frac{I(\tau)}{\K(\tau)^2}|B_k(\tau)|^2 \,.
    \label{eq:Normalisation_fullywarped}
\end{equation}
\fi
$A(\tau_b,\tau_c,T) = A(\tau_b,\tau_c)$ and $\mu(\tau_c,T)=\mu(\tau_c)$. In particular, we know from the analytical approximations (confirmed using the numerical solutions) that in this limit
\begin{equation}
	\hat{E}_k \approx I(\tau_c)^{1/2} e_k \,, \quad\quad e_k = \frac{2\pi}{a\nu} k + \left(\nu - \frac{1}{2}\right)\frac{\pi}{a\nu}\,,
\end{equation}
with $\nu\sim 2.45, a\sim 1.96$. Notice that we must always have $\tau_b < \tau_c$ since our boundary condition $\Phi_k(T=\tau_c)=0$ would imply vanishing countributions to a brane at $\tau_b \geq \tau_c$, and hence 
%If we further assume $I(\tau_b) \gg I(\tau_c)$ (which is necessary if we want to have a hierarchy of scales between the brane and the UV), then 
\begin{equation} 
	H_{\tau_c}(\tau_b) \approx \frac{I(\tau_b)}{I(\tau_c)}\,.
	\label{eq:hierarchy_fullywarped}
\end{equation}

\begin{figure}
     \centering
     \includegraphics[width=0.7\textwidth]{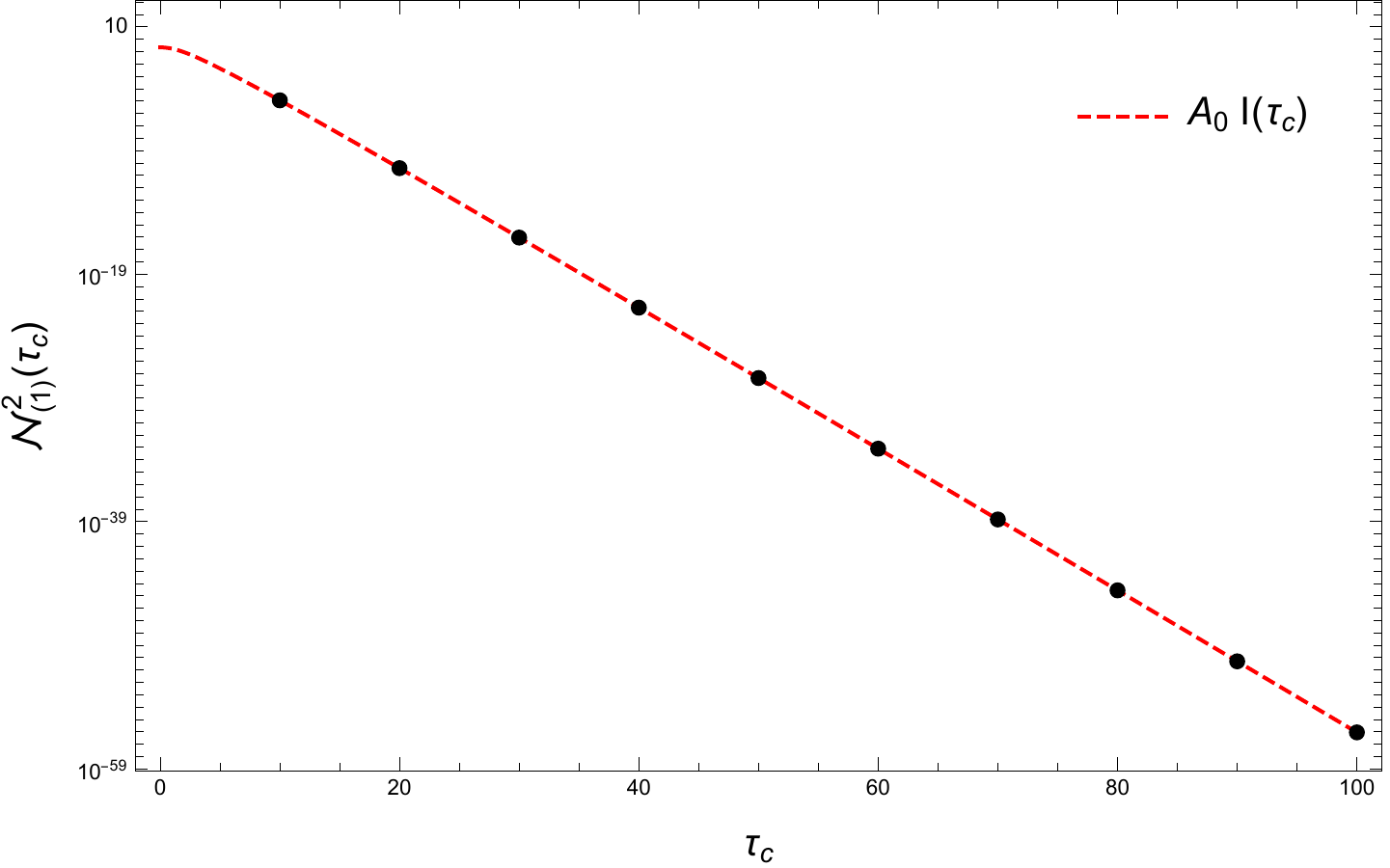}
    \caption{Plot of $\mathcal{N}_{(1)}^2(\tau_c)$ in the fully warped limit $\tau_c\to T$. We see that the normalisations fall with $A_0 I(\tau_c)$, with $A_0\approx 0.3$.}
	\label{fig:Afunction_fullywarped}
\end{figure}

Assuming that only the first massive mode has a relevant contribution\footnote{Although higher modes have larger couplings at the tip, as emphasized in \cite{Shiu:2007tn}, the exponential suppression from higher masses will dominate, so that the net result is an exponential suppression of the contributions from higher modes to the gravitational potential compared to the first mode in the tower.}, we find $A(\tau_b,\tau_c) \approx \mathcal{N}_{(1)}^2(\tau_c) |\tilde{\uPhi}_1(\tau_b)|^2$ and $\mu(\tau_c)\approx I(\tau_c)^{1/2}e_1$, where $\mathcal{N}_{(1)}^2(\tau_c) \approx A_0 I(\tau_c)$ and $A_0\approx 0.3$ (see Fig.\ref{fig:Afunction_fullywarped}) --- physically this implies weaker couplings for modes living in longer throats (i.e. with stronger warping), as well as smaller masses. The parameters $(\alpha,\lambda)$ become
\begin{subequations}
\label{eq:alphalambda_warped}
	\begin{align}
		\alpha &\approx \frac{(2\pi)^2}{(g_sM)^3} \frac{2|\tilde{\uPhi}_1(\tau_b)|^2}{I(\tau_b)^{1/2}}\frac{g_s^2}{\mathcal{H}^2} \,, \\
		\lambda^{-1} &\approx \frac{\mathcal{H} }{2^{1/6}}\frac{2\pi}{\sqrt{g_sM}} I(\tau_b)^{1/4}\frac{e_1}{l_p} \,.
		\label{eq:lambda_warped}
	\end{align}
\end{subequations}

\begin{figure}
	\centering
        \includegraphics[width=0.9\textwidth]{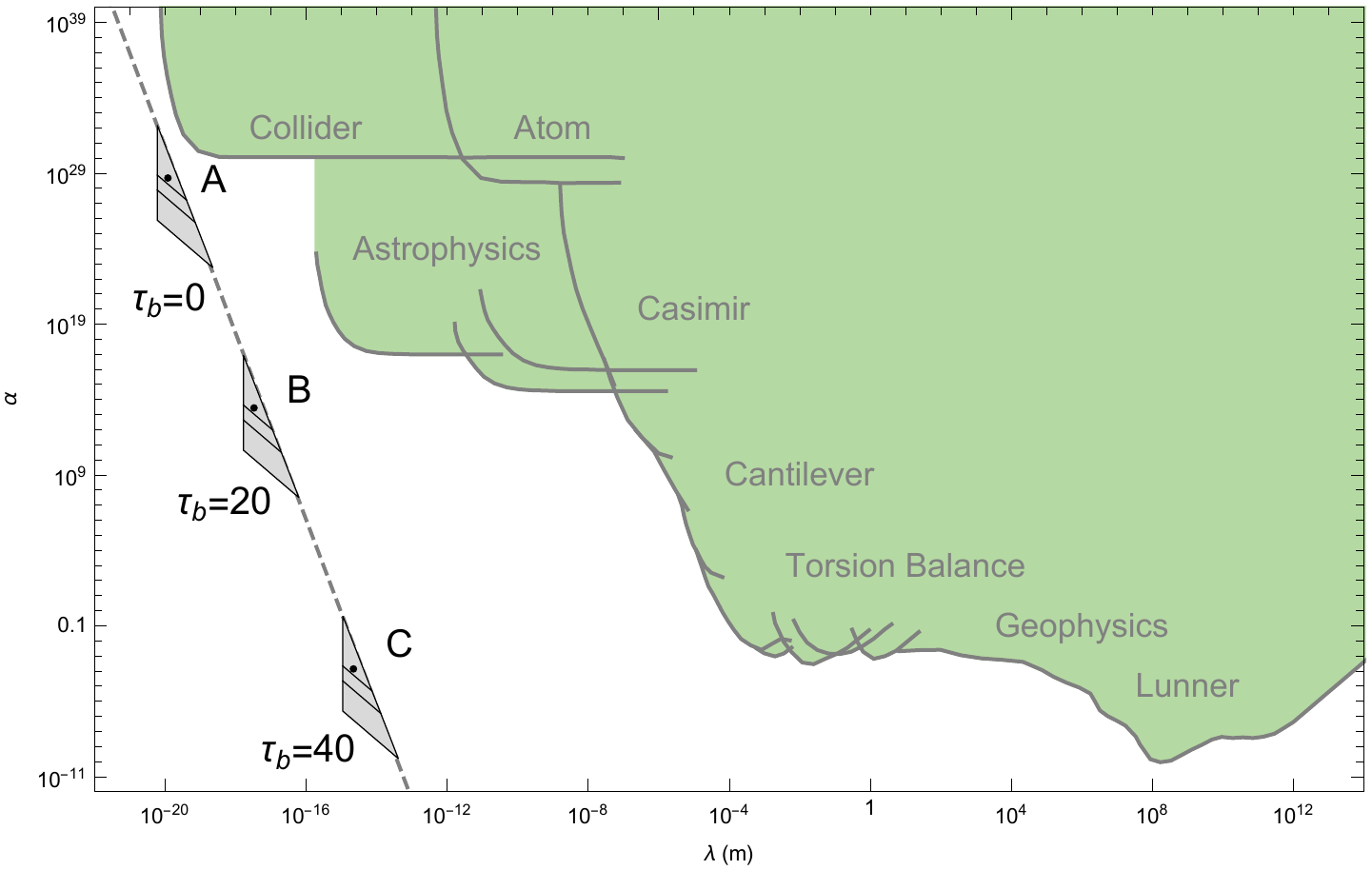}
	\caption{The shaded triangles correspond to the allowed regions of parameter space for the fully warped limit $\tau_c\to T$, for different choices of $\tau_b$ and fixed $\mathcal{H}=10^{-15}$. All points $(A,B,C)$ have $g_s=0.2, M=20$.
The upper right bounds on each region follow from (\ref{eq:fullywarped_alpha1lambda6}) and $g_s<1$, while the left bounds follow from (\ref{eq:lambda_warped}) and $g_sM>1$. The lower lines in each triangle represent the lower bound on (\ref{eq:fullywarped_alpha1lambda2}) with $M<32,100,1000$ (with larger values giving weaker bounds, i.e.~lower lines). Figure adapted from \cite{Murata:2014nra,Cembranos:2017vgi}, with the shaded area indicating the excluded region of parameter space at 95\% confidence level.}
	\label{fig:triangle_plots_warped}
\end{figure}

%From the definition of $\tau_c$ (\ref{eq:tauc_parameters}), we see that $\tau_c$ increases as $g_sM$ decreases, and since $\tilde{\uPhi}_1(\tau_b)$ slighly increases with $\tau_c$ (recall that the solutions $\tilde{\uPhi}_k$ depend on $\tau_c$), the lower bound $g_sM>1$ translates into an upper bound $\tilde{\uPhi}_1(\tau_b) < \tilde{\uPhi}_1^{max}(\tau_b)$. Therefore, 
For fixed $(\tau_b,\mathcal{H})$, the bounds $g_s<1$ and $g_sM>1$, required for control of the string loop expansion and supergravity approximation, respectively, translate into direct bounds on $(\alpha,\lambda)$, 
\begin{subequations}
	\begin{align}
		\alpha &< \frac{(2\pi)^2}{\mathcal{H}^2} \frac{4|\tilde{\uPhi}_1(\tau_b)|^2}{I(\tau_b)^{1/2}} \,, \\
		\lambda^{-1} &<\frac{\mathcal{H} }{2^{1/6}}(2\pi) I(\tau_b)^{1/4}\frac{e_1}{l_p}\,.
	\end{align}
\end{subequations}
However, there is a stronger bound on $\alpha$ given by the combination
\begin{equation}
	\alpha \lambda^6 = \frac{1}{(2\pi)^4}\frac{4|\tilde{\uPhi}_1(\tau_b)|^2}{I(\tau_b)^2} \frac{g_s^2}{\mathcal{H}^8}\frac{l_p^6}{e_1^6} \,,
	\label{eq:fullywarped_alpha1lambda6}
\end{equation}
which is bounded from above and is a diagonal line with fixed negative slope in Fig.\ref{fig:triangle_plots_warped}. 

Finally, one can find a lower bound on the combination 
\begin{equation}
	\alpha \lambda^2 = \frac{2^{1/3}}{\mathcal{H}^4} \frac{2|\tilde{\uPhi}_1(\tau_b)|^2}{I(\tau_b)}\frac{l_p^2}{e_1^2}\frac{1}{M^2} \,,
	\label{eq:fullywarped_alpha1lambda2}
\end{equation}
by putting an upper bound on the flux number $M$. Thinking in terms of the tadpole contribution $MK$ and since the flux numbers are both positive, we have $MK < (MK)_{max} \implies M<(MK)_{max}$. Imposing the bound $M<(MK)_{max}$ allows us to connect with the tapole cancellation discussion in the previous section. Inspired by the examples given below (\ref{eq:tadpoleMK}) we choose three reference bounds, $M < (MK)_{max} = 32,100,1000$. Note that all of these will include point $A_1$ in Fig.~\ref{fig:greenplot_samplepoints} ($M=20$), although $MK\sim 755$ --- for a given choice of bound $M<(MK)_{max}$, all points with $MK < (MK)_{max}$ clearly lie within the allowed region $M<(MK)_{max}$, even though there are also points inside that region with $MK > (MK)_{max}$, such as $A_1$. 

The allowed regions of parameter space are shown in Fig.~\ref{fig:triangle_plots_warped}. 
%In this fully warped limit, the parameters $\tau_c$ and $\V_w$ become redundant, with only a combination of $\tau_c$ (which could be thought of in terms of $|z|$) and $\V_w$ being fixed by the hierarchy --- this case actually represents a line in the $(|z|,\V_w)$ plane along which the hierarchy and corrections $\delta V$ are the same (these must satisfy the control conditions $|z|\ll 1$, $\V_w \gg 1$ and also $\V_{th}<\V_w$ (\ref{eq:V_throat})). 
The points $(A,B,C)$ correspond to the specific choice $g_s=0.2$ and $M=20$, for each choice of $\tau_b=0,20,40$ (note that $\alpha$ and $\lambda$ in (\ref{eq:alphalambda_warped}) are independent of $\tau_c$, though recall that a combination of $\tau_c$ and $\mathcal{V}_w$ is fixed by our choice of hierarchy via (\ref{eq:hierarchy_definition})). 
%Importantly, the allowed region of parameter space never intersects the excluded region. 
Putting the brane away from the tip allows the KK graviton modes to have lower masses due to stronger warping without affecting the hierarchy between the bulk and the brane (which we keep fixed), while at the same time suppressing their couplings to matter on the brane, which depends on their wavefunctions (see Fig.~\ref{fig:NumericalSolutions}). This means that we can move to regions of larger $\lambda$ but at the expense of also moving to lower $\alpha$ --- this gives rise to the diagonal dashed line in Fig.\ref{fig:triangle_plots_warped}, which is an upper bound for the allowed regions which never crosses the excluded region.

The upper right bounds on each region follow from (\ref{eq:fullywarped_alpha1lambda6}) and $g_s<1$, while the left bounds follow from (\ref{eq:lambda_warped}) and $g_sM>1$. The lower lines in each triangle represent the lower bound on (\ref{eq:fullywarped_alpha1lambda2}) with $M<32,100,1000$ (with larger values giving weaker bounds, i.e.~lower lines). 

%%%%%%%%%%%%%%%%%%%%%%%%%%%%%%%%%%%%%%%%%%%%%%%%%%%%%%%%%%%%%%%%%%%%%%%%
%   4.4 UNWARPED DEFORMED CONIFOLD
%%%%%%%%%%%%%%%%%%%%%%%%%%%%%%%%%%%%%%%%%%%%%%%%%%%%%%%%%%%%%%%%%%%%%%%%

\subsection{Unwarped deformed conifold}
\label{sec:fullyunwarped}

The unwarped deformed conifold corresponds to the limit $\tau_c\to 0$. In this limit, the solution pair $(\hat{E}_k,\Phi_k)$ only depends on $T$, so that
$A(\tau_b,\tau_c,T) = A(\tau_b,T)$ and $\mu(\tau_c,T)=\mu(T)$. In particular, we know from the analytical approximations (confirmed using the numerical solutions) that 
\begin{equation}
	\hat{E}_k \approx e^{-T/3} e_k \,, \quad\quad e_k = \frac{2^{5/6}}{3^{1/2}}\Big\{\pi k + \frac{3\pi}{4}\Big\} \,.
\end{equation}
Notice that we must always have $\tau_b < T$ since our boundary condition $\Phi_k(T)=0$ would imply vanishing countributions to a brane at $\tau_b \geq T$. In this limit
\begin{equation} 
	\mathcal{H} \approx \frac{g_s}{\sqrt{4\pi\V_w}} \,.
\end{equation}

\begin{figure}
     \centering
     \includegraphics[width=0.7\textwidth]{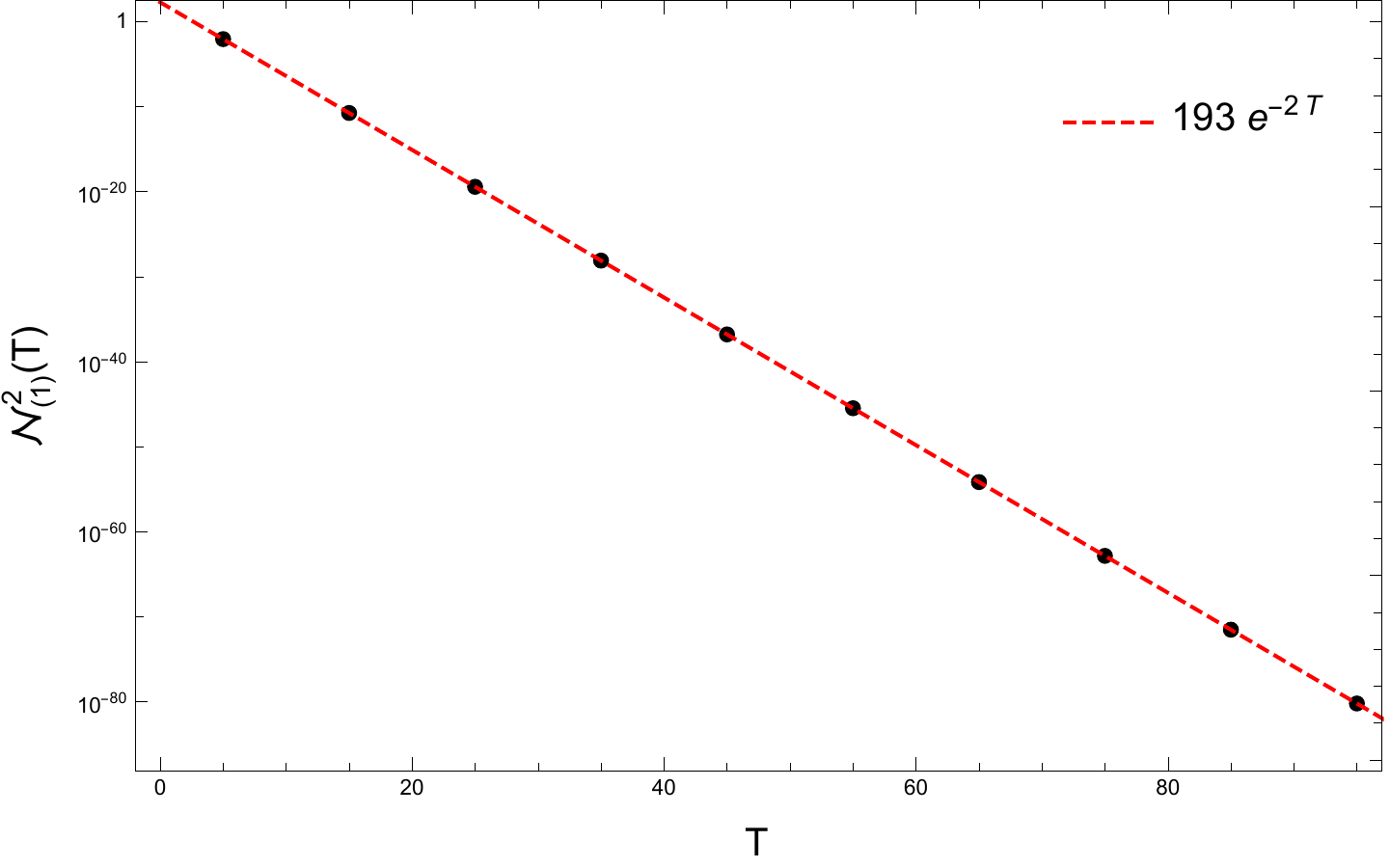}
    \caption{Plot of $\mathcal{N}_{(1)}^2(T)$ in the unwarped limit $\tau_c\to 0$. We see that the normalisations fall with $T$ as $\chi e^{-a T}$, with $\chi\approx 193$ and $a\approx 2$. See main text for further discussion.}
	\label{fig:normalisation_unwarped}
\end{figure}

The general result (\ref{eq:parCorr}) is expressed in terms of $\tau_c$, whose (implicit) definition is (\ref{eq:tauc_parameters}). However, this follows from the condition $H(\tau_c) - 1 = 1$, whereas a fully unwarped conifold would have $H(\tau)=1$ for all $\tau$. This is not possible for the deformed conifold, since the presence of fluxes, necessary to deform the conifold, will automatically source some warping, i.e. $H(\tau)\neq 1$ for any finite $\tau$. Nevertheless, we could still have a deformed conifold for which
\begin{equation}
	H(\tau) - 1 = 2^{2/3}\frac{(\alpha' g_s M)^2}{c ~ \epsilon^{8/3}}I(\tau) \ll 1 \quad \forall \,\,\,\tau \,,
\end{equation}
which gives $H(\tau)\approx 1$ as we would expect from an unwarped compactification. In this case, the definition of $\tau_c$ (\ref{eq:tauc_parameters}) is no longer useful --- one should instead substitute back in (\ref{eq:parCorr}) the parameters of the conifold and remove $\tau_c$ completely. This gives 
\begin{subequations}
	\begin{align}
		\alpha &= \frac{1}{(2\pi)^4} \frac{8A(\tau_b,T)}{c^{3/2}|z|^2}\frac{g_s^2}{\mathcal{H}^2} \,, \\
		\lambda^{-1} &= \frac{\mathcal{H} }{c^{1/4}|z|^{1/3}} \frac{\mu(T)}{l_p} \,.
	\end{align}
\end{subequations}
One can now relate $c^{1/4}|z|^{1/3}$ to the physical size of the conifold,
\begin{equation}
	R_{con} = \frac{3^{1/2}}{2^{5/6}}(c^{1/4}|z|^{1/3}) e^{T/3} l_s\,,
	\label{eq:conifoldR}
\end{equation}
such that the parameters become
\begin{subequations}
	\begin{align}
		\alpha &= \frac{1}{(2\pi)^4} \frac{27A(\tau_b,T)}{8} e^{2T}\left(\frac{l_s}{R_{con}}\right)^6\frac{g_s^2}{\mathcal{H}^2} \,, \\
		\lambda^{-1} &= \frac{3^{1/2}}{2^{5/6}} \mathcal{H}  \left(\frac{l_s}{R_{con}}\right) e^{T/3} \frac{\mu(T)}{l_p} \,.
	\end{align}
\end{subequations}

Assuming that only the first massive mode has a relevant contribution (as all other modes will be exponentially suppressed), we find $A(\tau_b,T) \approx  \mathcal{N}_{(1)}^2(T)   |\tilde{\uPhi}_1(\tau_b)|^2$ and $\mathcal{N}_{(1)}^2 \approx \chi e^{-2T}$, with $\chi\approx 193$ (see Fig.\ref{fig:normalisation_unwarped}), and $\mu(\tau_c)\approx e^{-T/3}e_1$ --- we can interpret this physically by noting that larger conifolds (larger $R_{con}\sim e^{T/3} \implies V_{con}\sim R_{con}^6 \sim e^{2T}$) will lead to mode wavefunctions spreading over a larger volume, and since they spread evenly through the internal space, they consequently have a smaller amplitude at each point.
This means that modes living on larger unwarped conifolds have weaker couplings to fields living on the brane. 
With these simplifcations we find
\begin{subequations}
	\begin{align}
		\alpha &\approx \frac{1}{(2\pi)^4} \frac{27\chi |\tilde{\uPhi}(\tau_b)|^2}{8} \left(\frac{l_s}{R_{con}}\right)^6\frac{g_s^2}{\mathcal{H}^2} \,, \\
		\lambda^{-1} &\approx \frac{3^{1/2}}{2^{5/6}} \mathcal{H}  \left(\frac{l_s}{R_{con}}\right) \frac{e_1}{l_p} \,.
	\end{align}
\end{subequations}

From the condition $R_{con} > R_{min} > l_s$, one can immediately put a bound on $\lambda$ for each choice of $R_{min}$, while the strongest bound on $\alpha$ comes from eliminating $R_{con}$ through the combination
\begin{equation}
	\alpha\lambda^6 =  \frac{4\chi |\tilde{\uPhi}(\tau_b)|^2}{(2\pi)^4} \left(\frac{l_p}{e_1}\right)^6 \frac{g_s^2}{\mathcal{H}^8}\,,
	\label{eq:unwarped_plots_upperlimit}
\end{equation}
which is bounded from above by $g_s<1$. 

\begin{figure}
	\centering
	\includegraphics[width=0.9\textwidth]{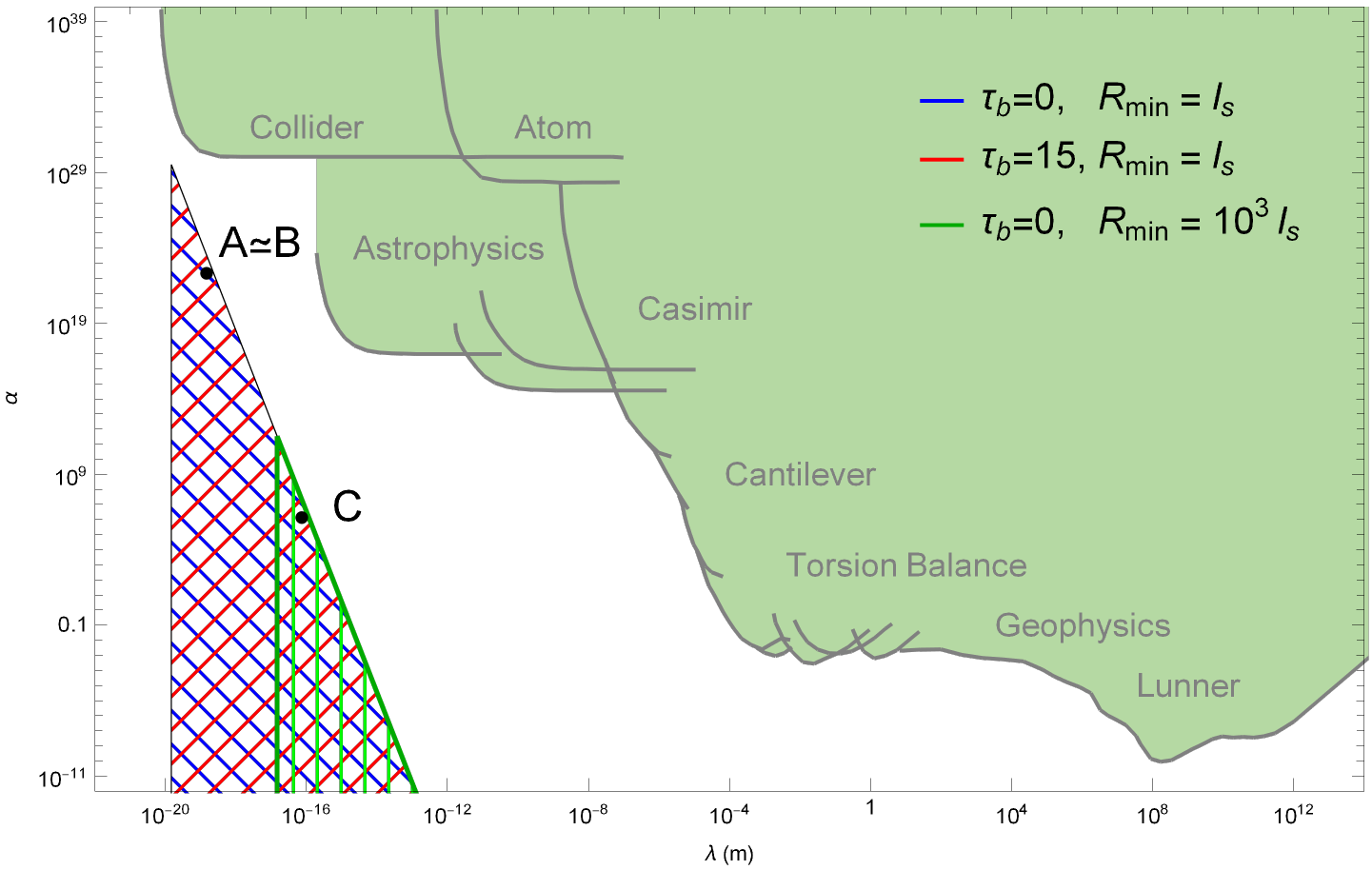}
	\caption{Allowed regions of parameter space for the unwarped limit $\tau_c\to 0$, for different choices of the parameters $(\tau_b,R_{min}/l_s)$ and fixed $\mathcal{H}=10^{-15}$.  See main text for further discussion.
	Figure adapted from \cite{Murata:2014nra,Cembranos:2017vgi}, with the shaded area indicating the excluded region of parameter space at 95\% confidence level. 
}
	\label{fig:triangle_plots_unwarped}
\end{figure}

With these bounds we can plot the allowed region of parameter space over the experimental and observational constraints (see Fig.\ref{fig:triangle_plots_unwarped}). The points $(A,B,C)$ correspond to the specific choice $g_s=0.2, M=20$, and with $R_{con}=10 l_s$ for $A$ and $B$, and $R_{con}=10^4 l_s$ for $C$. Notice that the upper limit following from (\ref{eq:unwarped_plots_upperlimit}) is only a function of $\tau_b$ and in the unwarped limit the wavefunctions are constant for most values of $\tau$ (see Fig.\ref{fig:NumericalSolutions}).
Therefore the allowed regions in this case are only functions of $R_{min}$, larger values of which decrease their size within the region with $R_{min}=l_s$ (bigger triangle). One should remember that the example points in Fig.~\ref{fig:greenplot_samplepoints} have $c^{1/4}|z|^{1/3}$ fixed by $(g_s,M,\tau_c=0)$ through (\ref{eq:tauc_parameters}) --- for $T=30$ as in Fig.~\ref{fig:greenplot_samplepoints}, this implies $R_{con}\approx 5\times 10^3 ~ l_s$. There is again no overlap between the theoretically allowed regions of parameter space and the experimentally excluded regions.

\section{Towards Gravitational Wave Observations}
\label{sec:GW}

All the available and planned GW experiments give us the opportunity to probe regimes of strong gravity, which could potentially give interesting constraints to explicit models \cite{Seahra:2004fg,Clarkson:2005eb,Chakraborty:2017qve,Andriot:2017oaz,Du:2020rlx}. Although a more careful study is left for future research, we end this paper by working out how a warped throat affects the scales involved in GW detection.   In particular, we will identify points in the Klebanov-Strassler parameter space where the hierarchy problem is solved by warping and the gravitational wave frequencies corresponding to the graviton KK tower reach LISA, LIGO-Virgo/ET and UHF windows.

From (\ref{eq:waveEq_KKmodes}) we obtain a wave equation for each mode $k$ in an infinite tower, whose frequency is given by\footnote{This follows from
\begin{equation}
	(p_k)^\mu (p_k)_\mu = - m_k^2 \implies - E_k^2 + |\vec{p}_k|^2 = - m_k^2 \,, 
\end{equation}
and by noting that for a wave $E_k = \hbar \omega_k$ (with $\hbar = 1$ and $c=1$ in all equations). 
} 
\begin{equation}
	\omega_k^2 = m_k^2 + |\vec{p}_k|^2\,,
\end{equation}
where $\vec{p}_k$ is the momentum of the wave. When an event (e.g. a black hole merger) generates gravitational waves it does so by exciting $h_{\mu\nu}(x^{\mu},y^p)$, which is decomposed into an infinite tower of modes in the 4d EFT (\ref{eq:decomposition_munu}), all of which are excited. 
For the zero mode we have $m_0=0$, $h^0_{\mu\nu}$ corresponds to the massless graviton and the gravitational wave will have a frequency associated to the source event. Higher modes $h_{\mu\nu}^{k>0}$ belong to the tower of massive spin-2 modes, which have $m_k > 0$. If $m_k \gg |\vec{p}_k|$, the frequency of the wave is given by the mass of the respective mode,
\begin{equation}
	\omega_k \approx m_k \,.
\end{equation}
When these frequencies are much higher than the frequency range covered by GW experiments, we can only probe the zero mode, while the massive tower remains out of reach together with the extra dimensions that it encodes. If on the other hand the masses of the KK modes are low enough to give frequencies covered by current or future experiments, then one might hope to detect a characteristic signature of the extra dimensions --- a tower of signals whose frequencies are separated by a constant gap $\Delta\omega = m_1 \equiv M_{KK}^w$. For reference, we give in Table \ref{tb:fGW} the frequency ranges and corresponding $M_{KK}^w$ for some current and future GW detectors.\footnote{We obtain $f_{GW}$ by reintroducing factors of $c$ and $\hbar$, which were set to $1$, such that the result is expressed in Hz,
\begin{equation}
	f_{GW} = \frac{M_{KK}^w c^2}{2\pi\hbar}\,,
\end{equation}
where $M_{KK}^w$ must be in kg. 
}

\begin{table}[h!]
	\centering
	\begin{tabular}{|c|c|c|}
		\hline
		\rowcolor{black!10!white!90} & $f_{GW}$ (Hz) & $M_{KK}^w$ (eV) \\
		\hline
		\cellcolor{black!10!white!90} LISA & $10^{-4} - 10^0$ & $10^{-31} - 10^{-27}$\\
		\cellcolor{black!10!white!90} LIGO-Virgo/ET & $10^1 - 10^4$ & $10^{-26} - 10^{-23}$ \\
		\cellcolor{black!10!white!90} UHF & $10^6 - 10^9$ & $10^{-21} - 10^{-18}$ \\
		\hline
	\end{tabular}
	\caption{Frequency range of different gravitational wave experiments and the associated mass ranges for the case in which the wave corresponds to the excitation of a massive spin-2 mode of mass $M_{KK}^w$.}
	\label{tb:fGW}
\end{table}

We focus on the fully warped case studied in Section \ref{sec:fullywarped}. Since $m_k = \lambda^{-1}$, (\ref{eq:parCorr}) is fixed by choosing a value for $m_k$ that could potentially be detected by one of these GW experiments. In this limit $m_k$ is given by (\ref{eq:lambda_warped}) and we can fix the ratio
\begin{equation}
	\frac{M_{KK}^w}{\mathcal{H}} = \frac{I(\tau_b)^{1/4}}{2^{1/6}}\frac{2\pi}{\sqrt{g_sM}} \frac{e_1}{l_p} \,,
\end{equation} 
by fixing both $M_{KK}^w \equiv m_1$ and $\mathcal{H}$, and choosing the values of the parameters $g_s$ and $M$. In what follows we choose $g_s=0.2$ and $M=20$ as before.
This determines the position of the brane $\tau_b$. One can then use (\ref{eq:hierarchy_definition}) to determine $|z|/\V_w$, which emphasises the fact that only a combination of these parameters is fixed by our choices. To have a concrete example, we fix the volume to $\V_w\sim 10^{15}$, which gives $|z|$ and therefore $MK$. In Table \ref{tb:fGW_parameters} we give three examples by taking the upper bounds on the frequency ranges for LISA, LIGO-Virgo/ET and UHF.
\begin{table}
	\centering
	\begin{tabular}{|c|c|c|c|c|c|c|c|c|}
		\hline
		\rowcolor{black!10!white!90} & $f_{GW}$ (Hz) & $M_{KK}^w$ (eV) & $\tau_b$ & $\tau_c$ & $z^{1/3}$ & $r_{UV}$ & $\V_{th}$ & $MK$  \\
		\hline
		\cellcolor{black!10!white!90} LISA & $10^0$ & $10^{-27}$ & 195 & 239 & $1.51\times 10^{-47}$  & $1.70$ & $290$ & $3259$ \\
		\cellcolor{black!10!white!90} LIGO-Virgo/ET & $10^4$ & $10^{-23}$ & 168 & 211 & $1.51\times 10^{-43}$   & $1.64$ & $240$ & $2906$ \\
		\cellcolor{black!10!white!90} UHF & $10^9$ & $10^{-18}$ & 133 & 176 & $1.51\times 10^{-38}$   & 1.57 & 183 & 2464 \\
		\hline
	\end{tabular}
	\caption{Explicit examples of parameters and scales associated with GW signals covered by LISA, LIGO-Virgo/ET and UHF. We fix $\mathcal{H}=10^{-15}$ and $\V_w = 10^{15}$ in all cases. Lengths and volumes are given in string units.}
	\label{tb:fGW_parameters}
\end{table}
We see that all these examples require tadpole contributions of $\Op(10^3)$, which brings us back to considering the Tadpole Conjecture \cite{tadpoleProblem}. The lower the frequency, the larger the warping must be to suppress $M_{KK}^w$, which we can see by the larger $\tau_c$ that results in longer throats. Since we are fixing the volume to $\V_w\sim 10^{15}$ in all cases, this also requires a larger $\tau_b$ in order to keep the hierarchy fixed --- if the warping is stronger, the brane must be farther away from the tip. Different values of $\V_w$ will require different values of $MK$, since the strength of the warping depends on the combination $\V_w^{1/6}|z|^{1/3}$ (\ref{eq:tauc_parameters}). For the UHF case, choosing $\V_w\sim 10^2$ would require $MK\sim 2021$, while $\V_w\sim 10^{30}$ would require $MK\sim 2847$, which is always $\Op (10^3)$. 

It is important to note that even if the warping is such that $M_{KK}^w$ is low enough to give frequencies that lie in the ranges of any of these GW experiments, one is not guaranteed to make a detection --- we must also take into account the amplitude of the waves. This requires a more careful analysis of the wave equations, which includes in particular the source of the GWs. One would expect a source localised on the brane (e.g.~a neutron star merger) to have very small couplings, due to the large suppression of the KK mode wavefunctions away from the tip of the throat. On the other hand, a source which is higher dimensional (e.g.~a higher dimensional black hole) might be able to couple strongly to the KK modes, not being confined to the brane far away from the tip. Bulk moduli (e.g.~complex structure and K\"ahler moduli) might also source the wave equation and these could have wavefunctions with big overlaps with the KK modes, and may therefore provide signals with higher amplitudes. It will also be interesting to understand if any resonance effects can enhance the wave amplitude. We leave a careful treatment of these questions for future work.

\section{Conclusions}
\label{S:Conclusions}

Gravity is currently being probed with unprecedent precision, with experiments and observations covering strong and weak field regimes, and providing crucial information that can be used to constrain modifications to GR, including those descending from specific UV completions such as string theory. It is therefore important to derive concrete predictions from phenomenologically interesting models that can be compared with observations. In the context of string theory, warped throats naturally arise due to the fluxes used to stabilise some of the many moduli present in the low-energy 4d theory, they provide a possible explanation of the hierarchy problem in the Standard Model, and they are crucial ingredients of de Sitter constructions such as KKLT \cite{KKLT} and LVS \cite{originalLVS}. In this work we explored the gravitational effects of a warped flux compactification of Type IIB supergravity on matter confined to a (3+1)-dimensional brane. More precisely, we derived the tower of KK graviton modes arising from a compactification in which the internal space contains a warped throat described by the KS solution and studied how this tower corrects the Newtonian potential felt by matter confined to the brane.  We derived the relations between the KS string solution parameters and the phenomenological parameters associated with a modified Newtonian potential, and thus found some tight bounds on the latter, as illustrated e.g.~by Fig.~\ref{fig:triangle_plots_warped}.

After going through the dimensional reduction of the 10d gravitational wave equation for Type IIB supergravity in Section \ref{sec:4dEFT}, taking into account the warp factor given by the KS solution and keeping track of the contribution of the Type IIB bosonic sector, we focused on the simple case of 4d Minkowski spacetime with the fluctuations of all fields apart from the 10d graviton set to zero. This allowed us to explicitly derive the KK tower of 4d spin-2 fields in Section \ref{sec:conifoldKKtower}, obtaining both their masses and wavefunctions, and studying how it is affected by the warping. We recovered the results in \cite{Tye:2005qs} for the mass spectrum, and found larger wavefunction amplitudes for higher KK modes, which lead to larger couplings as emphasised in \cite{Shiu:2007tn}. We then discussed how the results depend on the balance between the strengh of the warping and the size of the bulk --- if the warping is strong enough, it dictates the suppression of the KK masses, but if the bulk is so large that it dominates over the effect of the warping, the masses are instead suppressed by the volume of the compact space. We studied this effect by describing part of the bulk with the unwarped deformed conifold, 
%(which was also done in \cite{Tye:2005qs}, where there is no CY$_3$ and an orbifold of a finite deformed conifold is considered)
 since the unknown metric of the CY$_3$ does not allow us to find the mode wavefunctions explicity. Interestingly, the warping becomes sufficiently strong to suppress the masses and localise the modes when roughly half of the conifold is warped, in terms of the radial coordinate $\tau$ i.e. $\tau_c/T\sim 1/2$ (see Fig.~\ref{fig:masses_taucTratio}).  Since the more intuitive radial coordinate $\rl\sim e^{\tau/3}$ behaves exponentially in $\tau$, this means that the throat can be an exponentially small portion of the internal space and still provide the dominant effect. 

In Section \ref{sec:NewtPotCorrections}, we showed how the masses and wavefunction profiles of the KK modes manifest in the low-energy theory, through the range and strength of the corrections to the gravitational interaction, respectively. After reviewing how gravity manifests on a brane, especially when the extra dimensions are warped, and how the warping affects the scales of the braneworld theory, we gave the general form of corrections to the gravitational potential 
%felt by masses confined to the brane due to the tower of spin-2 modes 
expressed as a single Yukawa-type interaction, which allowed us to compare with experimental and observational constraints \cite{Murata:2014nra}. We gave three sets of examples, covering the fully warped, unwarped and partially warped regimes.  We also considered placing the brane at the tip of the throat or higher up along the throat, always fixing the hierarchy such that the string scale is lowered by warping and/or volume dilution to $\sim 1$ TeV in the braneworld theory.
%, which we express in terms of the known Planck scale through $\mathcal{H} \equiv 1\text{ TeV}/M_P \sim 10^{-15}$. The three examples give us some insight on how these regimes manifest on the corrections and how they compare with the available experimental data --- we conclude that the 
The presence of warping brings the corrections closer to collider constraints so that we might be able to probe warped models with future collider experiments (this requires an increase in luminosity in order to probe smaller couplings). However, strong warping typically comes with a large tadpole contribution $MK$, which has been given special attention in recent works \cite{tadpoleProblem,Bena:2021tadpole}. Isotropic unwarped models on the other hand are much harder to probe, staying much further from the excluded region of parameter space.\footnote{Models with a smaller number of large extra dimensions \cite{ADD1} can be excluded, but using the deformed conifold as a background we can only consider the isotropic case with 6 large extra dimensions \cite{Murata:2014nra}.} This can be understood by noting that unwarped scenarios must provide the hierarchy through a much larger volume when compared to the warped case, which, whilst bringing down the masses, also suppresses the couplings of the KK modes since their wavefunctions will spread evenly through the internal space.

We then focused on the fully warped and unwarped limits, in which we were able to reduce the number of free parameters and study the parameter space more carefully. In particular, there exist combinations of the Yukawa parameters $\alpha$ and $\lambda$ which depend only on the hierarchy, the position of the brane and on either $g_s$, $M$ or $g_sM$ (see e.g.~eqs.~(\ref{eq:lambda_warped}), (\ref{eq:fullywarped_alpha1lambda6}) and (\ref{eq:fullywarped_alpha1lambda2})). Therefore, using consistency conditions required by the string loop expansion ($g_s<1$) and the supergravity approximation ($g_sM>1$, $R_{con}>1$), and some well-motivated constraints on the flux number $M$ based on the recent discussion of the D3-tadpole cancellation condition within flux compactifications, we were able to exclude large portions of parameter space. In the fully warped scenario, the tadpole cancellation considerations reduce the allowed region considerably, with smaller upper bounds on $M$ excluding larger regions. In both the fully warped and fully unwarped cases, the theoretically allowed regions never overlap with the experimentally excluded region.  It would be interesting to consider how these constraints could improve with future experiments and new technologies.  

The position of the brane along the warped throat drastically changes the predictions, lowering the masses of the graviton KK modes, as well as their couplings to matter on the brane, as the brane moves up the throat.  As we have seen in Section \ref{sec:GW}, this could be especially interesting in the context of GWs, since the masses of the KK modes give the frequency of the corresponding waves. With KK masses close to the UV scale, these frequencies are too high for us to probe, but if they are suppressed by the warping one might hope to detect them in future experiments. Our results show that this lowering of the KK masses does not necessarily create any tension with current experimental data, since the allowed regions of parameter space stay away from the experimentally excluded regions. %However, one might still worry about the relation between a strong warping and a large tadpole contribution $MK$, which has been given special attention in recent works \cite{tadpoleProblem,Bena:2021tadpole}. 
In Section \ref{sec:GW} we considered the required warping of the KK mode masses to bring gravitational wave frequencies within the ranges of current and future GW experiments, giving explicit examples of KS background parameters that achieve this. All examples required a tadpole contribution of $\Op(10^3)$; it would be important to understand how much warping one can have in a consistent flux compactification with all the moduli stabilised and compatible with tadpole cancellation.

There are several further questions which naturally arise from this work that we would like to address in the future. Firstly, it will be very interesting to complete our exploration of the effects of a warped throat described by the KS solution within this Type IIB setup on gravitational wave signals.  We have shown that the gravitational wave frequencies can be brought down to observable windows, but it will be important to work out the signal amplitudes.   Not only do we expect extra higher frequency signals associated to the massive spin-2 modes, but there could also be extra polarisations and modified dispersion relations, which one might use to constrain these models with current and upcoming gravitational wave experiments \cite{Andriot:2017oaz,Andriot:2019hay,Andriot:2021gwv,Du:2020rlx,Clarkson:2006pq,Chakraborty:2017qve,Baker:2022rhh}. Tightly related to these questions is the effect of the scalar field fluctuations we neglected in Section \ref{sec:4dEFT}. The zero modes of these fluctuations are the complex structure and K\"ahler moduli of the compact space --- even in this work we encounter at least two moduli, the deformation modulus $z$ and the volume modulus $c$. One might then expect them to affect the GW signals, in a similar way to the breathing mode identified in \cite{Andriot:2017oaz}. On the other hand, we should expect these scalar fluctuations (and possibly even fluctuations arising from the other bosonic fields in Type IIB) to also have their masses lowered by the warping and it would be important to check both their effects on the 4d theory and whether these would change our results. Finally it would be interesting to explore the connections between our results and the swampland program in the context of KK towers and/or spin-2 fields \cite{DeRham:2018bgz,Klaewer:2018yxi,Blumenhagen:2019qcg}.

\newpage
\section*{Acknowledgements}
We are grateful to Pedro Fernandes and Gianmassimo Tasinato for helpful discussions. D.C. is supported by Internacionalizaci\'on de la Investigaci\'on (BUAP) grant. IZ is partially supported by STFC, grant ST/T000813/1.

\appendix

\bibliographystyle{utphys}

\bibliography{references}

\providecommand{\href}[2]{#2}\begingroup\raggedright\begin{thebibliography}{10}

\bibitem{LIGO2016}
{\bf LIGO Scientific, Virgo} Collaboration, B.~P. Abbott {\em et al.},
  ``{Observation of Gravitational Waves from a Binary Black Hole Merger}'',
  \href{http://dx.doi.org/10.1103/PhysRevLett.116.061102}{{\em Phys. Rev.
  Lett.} {\bf 116} (2016) no.~6, 061102},
  \href{http://arxiv.org/abs/1602.03837}{{\tt arXiv:1602.03837 [gr-qc]}}.

\bibitem{LIGOScientific:2016sjg}
{\bf LIGO Scientific, Virgo} Collaboration, B.~P. Abbott {\em et al.},
  ``{GW151226: Observation of Gravitational Waves from a 22-Solar-Mass Binary
  Black Hole Coalescence}'',
  \href{http://dx.doi.org/10.1103/PhysRevLett.116.241103}{{\em Phys. Rev.
  Lett.} {\bf 116} (2016) no.~24, 241103},
  \href{http://arxiv.org/abs/1606.04855}{{\tt arXiv:1606.04855 [gr-qc]}}.

\bibitem{LIGOScientific:2017bnn}
{\bf LIGO Scientific, VIRGO} Collaboration, B.~P. Abbott {\em et al.},
  ``{GW170104: Observation of a 50-Solar-Mass Binary Black Hole Coalescence at
  Redshift 0.2}'', \href{http://dx.doi.org/10.1103/PhysRevLett.118.221101}{{\em
  Phys. Rev. Lett.} {\bf 118} (2017) no.~22, 221101},
  \href{http://arxiv.org/abs/1706.01812}{{\tt arXiv:1706.01812 [gr-qc]}}.
  [Erratum: Phys.Rev.Lett. 121, 129901 (2018)].

\bibitem{Marion:2017enj}
{\bf LIGO Scientific, VIRGO} Collaboration, F.~Marion, ``{GW150914: Observation
  of gravitational waves from a binary black hole merger}'',
  \href{http://dx.doi.org/10.1393/ncc/i2016-16310-2}{{\em Nuovo Cim. C} {\bf
  39} (2017) no.~4, 310}.

\bibitem{LIGOScientific:2017ycc}
{\bf LIGO Scientific, Virgo} Collaboration, B.~P. Abbott {\em et al.},
  ``{GW170814: A Three-Detector Observation of Gravitational Waves from a
  Binary Black Hole Coalescence}'',
  \href{http://dx.doi.org/10.1103/PhysRevLett.119.141101}{{\em Phys. Rev.
  Lett.} {\bf 119} (2017) no.~14, 141101},
  \href{http://arxiv.org/abs/1709.09660}{{\tt arXiv:1709.09660 [gr-qc]}}.

\bibitem{LIGOScientific:2021qlt}
{\bf LIGO Scientific, KAGRA, VIRGO} Collaboration, R.~Abbott {\em et al.},
  ``{Observation of Gravitational Waves from Two Neutron Star\textendash{}Black
  Hole Coalescences}'', \href{http://dx.doi.org/10.3847/2041-8213/ac082e}{{\em
  Astrophys. J. Lett.} {\bf 915} (2021) no.~1, L5},
  \href{http://arxiv.org/abs/2106.15163}{{\tt arXiv:2106.15163 [astro-ph.HE]}}.

\bibitem{LIGOScientific:2017vwq}
{\bf LIGO Scientific, Virgo} Collaboration, B.~P. Abbott {\em et al.},
  ``{GW170817: Observation of Gravitational Waves from a Binary Neutron Star
  Inspiral}'', \href{http://dx.doi.org/10.1103/PhysRevLett.119.161101}{{\em
  Phys. Rev. Lett.} {\bf 119} (2017) no.~16, 161101},
  \href{http://arxiv.org/abs/1710.05832}{{\tt arXiv:1710.05832 [gr-qc]}}.

\bibitem{Berti:2015itd}
E.~Berti {\em et al.}, ``{Testing General Relativity with Present and Future
  Astrophysical Observations}'',
  \href{http://dx.doi.org/10.1088/0264-9381/32/24/243001}{{\em Class. Quant.
  Grav.} {\bf 32} (2015)  243001}, \href{http://arxiv.org/abs/1501.07274}{{\tt
  arXiv:1501.07274 [gr-qc]}}.

\bibitem{LIGOScientific:2016lio}
{\bf LIGO Scientific, Virgo} Collaboration, B.~P. Abbott {\em et al.}, ``{Tests
  of general relativity with GW150914}'',
  \href{http://dx.doi.org/10.1103/PhysRevLett.116.221101}{{\em Phys. Rev.
  Lett.} {\bf 116} (2016) no.~22, 221101},
  \href{http://arxiv.org/abs/1602.03841}{{\tt arXiv:1602.03841 [gr-qc]}}.
  [Erratum: Phys.Rev.Lett. 121, 129902 (2018)].

\bibitem{LIGOScientific:2018dkp}
{\bf LIGO Scientific, Virgo} Collaboration, B.~P. Abbott {\em et al.}, ``{Tests
  of General Relativity with GW170817}'',
  \href{http://dx.doi.org/10.1103/PhysRevLett.123.011102}{{\em Phys. Rev.
  Lett.} {\bf 123} (2019) no.~1, 011102},
  \href{http://arxiv.org/abs/1811.00364}{{\tt arXiv:1811.00364 [gr-qc]}}.

\bibitem{LIGOScientific:2019fpa}
{\bf LIGO Scientific, Virgo} Collaboration, B.~P. Abbott {\em et al.}, ``{Tests
  of General Relativity with the Binary Black Hole Signals from the LIGO-Virgo
  Catalog GWTC-1}'', \href{http://dx.doi.org/10.1103/PhysRevD.100.104036}{{\em
  Phys. Rev. D} {\bf 100} (2019) no.~10, 104036},
  \href{http://arxiv.org/abs/1903.04467}{{\tt arXiv:1903.04467 [gr-qc]}}.

\bibitem{Johnson-McDaniel:2019zkl}
{\bf LIGO Scientific, Virgo} Collaboration, N.~K. Johnson-McDaniel, ``{Summary
  of Tests of General Relativity with the Binary Black Hole Signals from the
  LIGO-Virgo Catalog GWTC-1}'', in {\em {54th Rencontres de Moriond on
  Gravitation}}.
\newblock 5, 2019.
\newblock \href{http://arxiv.org/abs/1905.05565}{{\tt arXiv:1905.05565
  [gr-qc]}}.

\bibitem{LIGOScientific:2020tif}
{\bf LIGO Scientific, Virgo} Collaboration, R.~Abbott {\em et al.}, ``{Tests of
  general relativity with binary black holes from the second LIGO-Virgo
  gravitational-wave transient catalog}'',
  \href{http://dx.doi.org/10.1103/PhysRevD.103.122002}{{\em Phys. Rev. D} {\bf
  103} (2021) no.~12, 122002}, \href{http://arxiv.org/abs/2010.14529}{{\tt
  arXiv:2010.14529 [gr-qc]}}.

\bibitem{LIGOScientific:2021sio}
{\bf LIGO Scientific, VIRGO, KAGRA} Collaboration, R.~Abbott {\em et al.},
  ``{Tests of General Relativity with GWTC-3}'',
  \href{http://arxiv.org/abs/2112.06861}{{\tt arXiv:2112.06861 [gr-qc]}}.

\bibitem{Punturo:2010zz}
M.~Punturo {\em et al.}, ``{The Einstein Telescope: A third-generation
  gravitational wave observatory}'',
  \href{http://dx.doi.org/10.1088/0264-9381/27/19/194002}{{\em Class. Quant.
  Grav.} {\bf 27} (2010)  194002}.

\bibitem{eLISA:2013xep}
{\bf eLISA} Collaboration, P.~A. Seoane {\em et al.}, ``{The Gravitational
  Universe}'', \href{http://arxiv.org/abs/1305.5720}{{\tt arXiv:1305.5720
  [astro-ph.CO]}}.

\bibitem{LISA:2017pwj}
{\bf LISA} Collaboration, P.~Amaro-Seoane {\em et al.}, ``{Laser Interferometer
  Space Antenna}'', \href{http://arxiv.org/abs/1702.00786}{{\tt
  arXiv:1702.00786 [astro-ph.IM]}}.

\bibitem{Aggarwal:2020olq}
N.~Aggarwal {\em et al.}, ``{Challenges and opportunities of gravitational-wave
  searches at MHz to GHz frequencies}'',
  \href{http://dx.doi.org/10.1007/s41114-021-00032-5}{{\em Living Rev. Rel.}
  {\bf 24} (2021) no.~1, 4}, \href{http://arxiv.org/abs/2011.12414}{{\tt
  arXiv:2011.12414 [gr-qc]}}.

\bibitem{Adelberger:2003zx}
E.~G. Adelberger, B.~R. Heckel, and A.~E. Nelson, ``{Tests of the gravitational
  inverse square law}'',
  \href{http://dx.doi.org/10.1146/annurev.nucl.53.041002.110503}{{\em Ann. Rev.
  Nucl. Part. Sci.} {\bf 53} (2003)  77--121},
  \href{http://arxiv.org/abs/hep-ph/0307284}{{\tt arXiv:hep-ph/0307284}}.

\bibitem{Dimopoulos:2006nk}
S.~Dimopoulos, P.~W. Graham, J.~M. Hogan, and M.~A. Kasevich, ``{Testing
  general relativity with atom interferometry}'',
  \href{http://dx.doi.org/10.1103/PhysRevLett.98.111102}{{\em Phys. Rev. Lett.}
  {\bf 98} (2007)  111102}, \href{http://arxiv.org/abs/gr-qc/0610047}{{\tt
  arXiv:gr-qc/0610047}}.

\bibitem{EventHorizonTelescope:2019dse}
{\bf Event Horizon Telescope} Collaboration, K.~Akiyama {\em et al.}, ``{First
  M87 Event Horizon Telescope Results. I. The Shadow of the Supermassive Black
  Hole}'', \href{http://dx.doi.org/10.3847/2041-8213/ab0ec7}{{\em Astrophys. J.
  Lett.} {\bf 875} (2019)  L1}, \href{http://arxiv.org/abs/1906.11238}{{\tt
  arXiv:1906.11238 [astro-ph.GA]}}.

\bibitem{Psaltis:2018xkc}
D.~Psaltis, ``{Testing General Relativity with the Event Horizon Telescope}'',
  \href{http://dx.doi.org/10.1007/s10714-019-2611-5}{{\em Gen. Rel. Grav.} {\bf
  51} (2019) no.~10, 137}, \href{http://arxiv.org/abs/1806.09740}{{\tt
  arXiv:1806.09740 [astro-ph.HE]}}.

\bibitem{Murata:2014nra}
J.~Murata and S.~Tanaka, ``{A review of short-range gravity experiments in the
  LHC era}'', \href{http://dx.doi.org/10.1088/0264-9381/32/3/033001}{{\em
  Class. Quant. Grav.} {\bf 32} (2015) no.~3, 033001},
  \href{http://arxiv.org/abs/1408.3588}{{\tt arXiv:1408.3588 [hep-ex]}}.

\bibitem{Baker:2014zba}
T.~Baker, D.~Psaltis, and C.~Skordis, ``{Linking Tests of Gravity On All
  Scales: from the Strong-Field Regime to Cosmology}'',
  \href{http://dx.doi.org/10.1088/0004-637X/802/1/63}{{\em Astrophys. J.} {\bf
  802} (2015)  63}, \href{http://arxiv.org/abs/1412.3455}{{\tt arXiv:1412.3455
  [astro-ph.CO]}}.

\bibitem{Randall:1999vf}
L.~Randall and R.~Sundrum, ``{An Alternative to compactification}'',
  \href{http://dx.doi.org/10.1103/PhysRevLett.83.4690}{{\em Phys. Rev. Lett.}
  {\bf 83} (1999)  4690--4693}, \href{http://arxiv.org/abs/hep-th/9906064}{{\tt
  arXiv:hep-th/9906064}}.

\bibitem{Randall:1999ee}
L.~Randall and R.~Sundrum, ``{A Large mass hierarchy from a small extra
  dimension}'', \href{http://dx.doi.org/10.1103/PhysRevLett.83.3370}{{\em Phys.
  Rev. Lett.} {\bf 83} (1999)  3370--3373},
  \href{http://arxiv.org/abs/hep-ph/9905221}{{\tt arXiv:hep-ph/9905221}}.

\bibitem{Seahra:2004fg}
S.~S. Seahra, C.~Clarkson, and R.~Maartens, ``{Detecting extra dimensions with
  gravity wave spectroscopy: the black string brane-world}'',
  \href{http://dx.doi.org/10.1103/PhysRevLett.94.121302}{{\em Phys. Rev. Lett.}
  {\bf 94} (2005)  121302}, \href{http://arxiv.org/abs/gr-qc/0408032}{{\tt
  arXiv:gr-qc/0408032}}.

\bibitem{Clarkson:2005eb}
C.~Clarkson and R.~Maartens, ``{Gravity-wave detectors as probes of extra
  dimensions}'', \href{http://dx.doi.org/10.1007/s10714-005-0150-8}{{\em Gen.
  Rel. Grav.} {\bf 37} (2005)  1681--1687},
  \href{http://arxiv.org/abs/astro-ph/0505277}{{\tt arXiv:astro-ph/0505277}}.

\bibitem{Clarkson:2006pq}
C.~Clarkson and S.~S. Seahra, ``{A gravitational wave window on extra
  dimensions}'', \href{http://dx.doi.org/10.1088/0264-9381/24/9/F01}{{\em
  Class. Quant. Grav.} {\bf 24} (2007)  F33--F40},
  \href{http://arxiv.org/abs/astro-ph/0610470}{{\tt arXiv:astro-ph/0610470}}.

\bibitem{Chakraborty:2017qve}
S.~Chakraborty, K.~Chakravarti, S.~Bose, and S.~SenGupta, ``{Signatures of
  extra dimensions in gravitational waves from black hole quasinormal modes}'',
  \href{http://dx.doi.org/10.1103/PhysRevD.97.104053}{{\em Phys. Rev. D} {\bf
  97} (2018) no.~10, 104053}, \href{http://arxiv.org/abs/1710.05188}{{\tt
  arXiv:1710.05188 [gr-qc]}}.

\bibitem{Andriot:2017oaz}
D.~Andriot and G.~Lucena~G\'omez, ``{Signatures of extra dimensions in
  gravitational waves}'',
  \href{http://dx.doi.org/10.1088/1475-7516/2017/06/048}{{\em JCAP} {\bf 06}
  (2017)  048}, \href{http://arxiv.org/abs/1704.07392}{{\tt arXiv:1704.07392
  [hep-th]}}. [Erratum: JCAP 05, E01 (2019)].

\bibitem{Andriot:2019hay}
D.~Andriot and D.~Tsimpis, ``{Gravitational waves in warped
  compactifications}'', \href{http://dx.doi.org/10.1007/JHEP06(2020)100}{{\em
  JHEP} {\bf 06} (2020)  100}, \href{http://arxiv.org/abs/1911.01444}{{\tt
  arXiv:1911.01444 [hep-th]}}.

\bibitem{Andriot:2021gwv}
D.~Andriot, P.~Marconnet, and D.~Tsimpis, ``{Warp factor and the gravitational
  wave spectrum}'', \href{http://dx.doi.org/10.1088/1475-7516/2021/07/040}{{\em
  JCAP} {\bf 07} (2021)  040}, \href{http://arxiv.org/abs/2103.09240}{{\tt
  arXiv:2103.09240 [hep-th]}}.

\bibitem{Du:2020rlx}
Y.~Du, S.~Tahura, D.~Vaman, and K.~Yagi, ``{Probing Compactified Extra
  Dimensions with Gravitational Waves}'',
  \href{http://dx.doi.org/10.1103/PhysRevD.103.044031}{{\em Phys. Rev. D} {\bf
  103} (2021) no.~4, 044031}, \href{http://arxiv.org/abs/2004.03051}{{\tt
  arXiv:2004.03051 [gr-qc]}}.

\bibitem{KKLT}
S.~Kachru, R.~Kallosh, A.~D. Linde, and S.~P. Trivedi, ``{De Sitter vacua in
  string theory}'', \href{http://dx.doi.org/10.1103/PhysRevD.68.046005}{{\em
  Phys. Rev. D} {\bf 68} (2003)  046005},
  \href{http://arxiv.org/abs/hep-th/0301240}{{\tt arXiv:hep-th/0301240}}.

\bibitem{originalLVS}
J.~P. Conlon, F.~Quevedo, and K.~Suruliz, ``{Large-volume flux
  compactifications: Moduli spectrum and D3/D7 soft supersymmetry breaking}'',
  \href{http://dx.doi.org/10.1088/1126-6708/2005/08/007}{{\em JHEP} {\bf 08}
  (2005)  007}, \href{http://arxiv.org/abs/hep-th/0505076}{{\tt
  arXiv:hep-th/0505076}}.

\bibitem{KS2000supergravity}
I.~R. Klebanov and M.~J. Strassler, ``{Supergravity and a confining gauge
  theory: Duality cascades and $\chi$SB resolution of naked singularities}'',
  \href{http://dx.doi.org/10.1088/1126-6708/2000/08/052}{{\em JHEP} {\bf 08}
  (2000)  052}, \href{http://arxiv.org/abs/hep-th/0007191}{{\tt
  arXiv:hep-th/0007191}}.

\bibitem{Tye:2005qs}
H.~Firouzjahi and S.~H.~H. Tye, ``{The Shape of gravity in a warped deformed
  conifold}'', \href{http://dx.doi.org/10.1088/1126-6708/2006/01/136}{{\em
  JHEP} {\bf 01} (2006)  136}, \href{http://arxiv.org/abs/hep-th/0512076}{{\tt
  arXiv:hep-th/0512076}}.

\bibitem{Shiu:2007tn}
G.~Shiu, B.~Underwood, K.~M. Zurek, and D.~G.~E. Walker, ``{Probing the
  geometry of warped string compactifications at the LHC}'',
  \href{http://dx.doi.org/10.1103/PhysRevLett.100.031601}{{\em Phys. Rev.
  Lett.} {\bf 100} (2008)  031601}, \href{http://arxiv.org/abs/0705.4097}{{\tt
  arXiv:0705.4097 [hep-ph]}}.

\bibitem{Callin:2004detail}
P.~Callin and F.~Ravndal, ``{Higher order corrections to the Newtonian
  potential in the Randall-Sundrum model}'',
  \href{http://dx.doi.org/10.1103/PhysRevD.70.104009}{{\em Phys. Rev. D} {\bf
  70} (2004)  104009}, \href{http://arxiv.org/abs/hep-ph/0403302}{{\tt
  arXiv:hep-ph/0403302}}.

\bibitem{Callin:2004short}
P.~Callin, ``{Corrections to the Newtonian potential in the two-brane
  Randall-Sundrum model}'', \href{http://arxiv.org/abs/hep-ph/0407054}{{\tt
  arXiv:hep-ph/0407054}}.

\bibitem{PTconifold}
G.~Papadopoulos and A.~A. Tseytlin, ``{Complex geometry of conifolds and
  five-brane wrapped on two sphere}'',
  \href{http://dx.doi.org/10.1088/0264-9381/18/7/315}{{\em Class. Quant. Grav.}
  {\bf 18} (2001)  1333--1354}, \href{http://arxiv.org/abs/hep-th/0012034}{{\tt
  arXiv:hep-th/0012034}}.

\bibitem{candelas1990conifolds}
P.~Candelas and X.~C. de~la Ossa, ``{Comments on Conifolds}'',
\href{http://dx.doi.org/10.1016/0550-3213(90)90577-Z}{{\em Nucl. Phys.} {\bf
  B342} (1990)  246--268}.
%%CITATION = NUPHA,B342,246;%%.

\bibitem{Minasian:1999tt}
R.~Minasian and D.~Tsimpis, ``{On the geometry of nontrivially embedded
  branes}'', \href{http://dx.doi.org/10.1016/S0550-3213(00)00035-3}{{\em Nucl.
  Phys. B} {\bf 572} (2000)  499--513},
  \href{http://arxiv.org/abs/hep-th/9911042}{{\tt arXiv:hep-th/9911042}}.

\bibitem{Aganagic:1999fe}
M.~Aganagic, A.~Karch, D.~Lust, and A.~Miemiec, ``{Mirror symmetries for brane
  configurations and branes at singularities}'',
  \href{http://dx.doi.org/10.1016/S0550-3213(99)00608-2}{{\em Nucl. Phys. B}
  {\bf 569} (2000)  277--302}, \href{http://arxiv.org/abs/hep-th/9903093}{{\tt
  arXiv:hep-th/9903093}}.

\bibitem{KT}
I.~R. Klebanov and A.~A. Tseytlin, ``{Gravity duals of supersymmetric SU(N) x
  SU(N+M) gauge theories}'',
  \href{http://dx.doi.org/10.1016/S0550-3213(00)00206-6}{{\em Nucl. Phys. B}
  {\bf 578} (2000)  123--138}, \href{http://arxiv.org/abs/hep-th/0002159}{{\tt
  arXiv:hep-th/0002159}}.

\bibitem{frey2009universal}
A.~R. Frey, G.~Torroba, B.~Underwood, and M.~R. Douglas, ``{The Universal
  K{\"a}hler Modulus in Warped Compactifications}'',
  \href{http://dx.doi.org/10.1088/1126-6708/2009/01/036}{{\em JHEP} {\bf 01}
  (2009)  036}, \href{http://arxiv.org/abs/0810.5768}{{\tt arXiv:0810.5768
  [hep-th]}}.

\bibitem{Giddings:2005ff}
S.~B. Giddings and A.~Maharana, ``{Dynamics of warped compactifications and the
  shape of the warped landscape}'',
  \href{http://dx.doi.org/10.1103/PhysRevD.73.126003}{{\em Phys. Rev. D} {\bf
  73} (2006)  126003}, \href{http://arxiv.org/abs/hep-th/0507158}{{\tt
  arXiv:hep-th/0507158}}.

\bibitem{Aparicio:2015psl}
L.~Aparicio, F.~Quevedo, and R.~Valandro, ``{Moduli Stabilisation with
  Nilpotent Goldstino: Vacuum Structure and SUSY Breaking}'',
  \href{http://dx.doi.org/10.1007/JHEP03(2016)036}{{\em JHEP} {\bf 03} (2016)
  036}, \href{http://arxiv.org/abs/1511.08105}{{\tt arXiv:1511.08105
  [hep-th]}}.

\bibitem{Bento:2021nbb}
B.~V. Bento, D.~Chakraborty, S.~L. Parameswaran, and I.~Zavala, ``{A New de
  Sitter Solution with a Weakly Warped Deformed Conifold}'',
  \href{http://arxiv.org/abs/2105.03370}{{\tt arXiv:2105.03370 [hep-th]}}.

\bibitem{GKP}
S.~B. Giddings, S.~Kachru, and J.~Polchinski, ``{Hierarchies from fluxes in
  string compactifications}'',
  \href{http://dx.doi.org/10.1103/PhysRevD.66.106006}{{\em Phys. Rev. D} {\bf
  66} (2002)  106006}, \href{http://arxiv.org/abs/hep-th/0105097}{{\tt
  arXiv:hep-th/0105097}}.

\bibitem{Higuchi}
A.~Higuchi, ``{Forbidden Mass Range for Spin-2 Field Theory in De Sitter
  Space-time}'', \href{http://dx.doi.org/10.1016/0550-3213(87)90691-2}{{\em
  Nucl. Phys. B} {\bf 282} (1987)  397--436}.

\bibitem{Higuchi:1989gz}
A.~Higuchi, ``{Massive Symmetric Tensor Field in Space-times With a Positive
  Cosmological Constant}'',
  \href{http://dx.doi.org/10.1016/0550-3213(89)90507-5}{{\em Nucl. Phys. B}
  {\bf 325} (1989)  745--765}.

\bibitem{deRham:2016nuf}
C.~de~Rham, J.~T. Deskins, A.~J. Tolley, and S.-Y. Zhou, ``{Graviton Mass
  Bounds}'', \href{http://dx.doi.org/10.1103/RevModPhys.89.025004}{{\em Rev.
  Mod. Phys.} {\bf 89} (2017) no.~2, 025004},
  \href{http://arxiv.org/abs/1606.08462}{{\tt arXiv:1606.08462 [astro-ph.CO]}}.

\bibitem{Rattazzi}
G.~F. Giudice, R.~Rattazzi, and J.~D. Wells, ``{Quantum gravity and extra
  dimensions at high-energy colliders}'',
  \href{http://dx.doi.org/10.1016/S0550-3213(99)00044-9}{{\em Nucl. Phys. B}
  {\bf 544} (1999)  3--38}, \href{http://arxiv.org/abs/hep-ph/9811291}{{\tt
  arXiv:hep-ph/9811291}}.

\bibitem{Hall:1999mk}
L.~J. Hall and D.~Tucker-Smith, ``{Cosmological constraints on theories with
  large extra dimensions}'',
  \href{http://dx.doi.org/10.1103/PhysRevD.60.085008}{{\em Phys. Rev. D} {\bf
  60} (1999)  085008}, \href{http://arxiv.org/abs/hep-ph/9904267}{{\tt
  arXiv:hep-ph/9904267}}.

\bibitem{ParticleDataGroup:2020ssz}
{\bf Particle Data Group} Collaboration, P.~A. Zyla {\em et al.}, ``{Review of
  Particle Physics}'', \href{http://dx.doi.org/10.1093/ptep/ptaa104}{{\em PTEP}
  {\bf 2020} (2020) no.~8, 083C01}.

\bibitem{Cembranos:2017vgi}
J.~A.~R. Cembranos, A.~L. Maroto, and H.~Villarrubia-Rojo, ``{Constraints on
  hidden gravitons from fifth-force experiments and stellar energy loss}'',
  \href{http://dx.doi.org/10.1007/JHEP09(2017)104}{{\em JHEP} {\bf 09} (2017)
  104}, \href{http://arxiv.org/abs/1706.07818}{{\tt arXiv:1706.07818
  [hep-ph]}}.

\bibitem{Douglas:2007tu}
M.~R. Douglas, J.~Shelton, and G.~Torroba, ``{Warping and supersymmetry
  breaking}'', \href{http://arxiv.org/abs/0704.4001}{{\tt arXiv:0704.4001
  [hep-th]}}.

\bibitem{upliftingrunaways2019}
I.~Bena, E.~Dudas, M.~Gra\~na, and S.~L\"ust, ``{Uplifting Runaways}'',
  \href{http://dx.doi.org/10.1002/prop.201800100}{{\em Fortsch. Phys.} {\bf 67}
  (2019) no.~1-2, 1800100}, \href{http://arxiv.org/abs/1809.06861}{{\tt
  arXiv:1809.06861 [hep-th]}}.

\bibitem{Blumenhagen:2019qcg}
R.~Blumenhagen, D.~Kläwer, and L.~Schlechter, ``{Swampland Variations on a
  Theme by KKLT}'', \href{http://dx.doi.org/10.1007/JHEP05(2019)152}{{\em JHEP}
  {\bf 05} (2019)  152}, \href{http://arxiv.org/abs/1902.07724}{{\tt
  arXiv:1902.07724 [hep-th]}}.

\bibitem{LVSdS:2010.15903}
C.~Crin\`o, F.~Quevedo, and R.~Valandro, ``{On de Sitter String Vacua from
  Anti-D3-Branes in the Large Volume Scenario}'',
  \href{http://dx.doi.org/10.1007/JHEP03(2021)258}{{\em JHEP} {\bf 03} (2021)
  258}, \href{http://arxiv.org/abs/2010.15903}{{\tt arXiv:2010.15903
  [hep-th]}}.

\bibitem{Baumann:2006th}
D.~Baumann, A.~Dymarsky, I.~R. Klebanov, J.~M. Maldacena, L.~P. McAllister, and
  A.~Murugan, ``{On D3-brane Potentials in Compactifications with Fluxes and
  Wrapped D-branes}'',
  \href{http://dx.doi.org/10.1088/1126-6708/2006/11/031}{{\em JHEP} {\bf 11}
  (2006)  031}, \href{http://arxiv.org/abs/hep-th/0607050}{{\tt
  arXiv:hep-th/0607050}}.

\bibitem{Baumann:2007ah}
D.~Baumann, A.~Dymarsky, I.~R. Klebanov, and L.~McAllister, ``{Towards an
  Explicit Model of D-brane Inflation}'',
  \href{http://dx.doi.org/10.1088/1475-7516/2008/01/024}{{\em JCAP} {\bf 01}
  (2008)  024}, \href{http://arxiv.org/abs/0706.0360}{{\tt arXiv:0706.0360
  [hep-th]}}.

\bibitem{Baumann:2010sx}
D.~Baumann, A.~Dymarsky, S.~Kachru, I.~R. Klebanov, and L.~McAllister,
  ``{D3-brane Potentials from Fluxes in AdS/CFT}'',
  \href{http://dx.doi.org/10.1007/JHEP06(2010)072}{{\em JHEP} {\bf 06} (2010)
  072}, \href{http://arxiv.org/abs/1001.5028}{{\tt arXiv:1001.5028 [hep-th]}}.

\bibitem{ADD1}
N.~Arkani-Hamed, S.~Dimopoulos, and G.~R. Dvali, ``{The Hierarchy problem and
  new dimensions at a millimeter}'',
  \href{http://dx.doi.org/10.1016/S0370-2693(98)00466-3}{{\em Phys. Lett. B}
  {\bf 429} (1998)  263--272}, \href{http://arxiv.org/abs/hep-ph/9803315}{{\tt
  arXiv:hep-ph/9803315}}.

\bibitem{ADD2}
I.~Antoniadis, N.~Arkani-Hamed, S.~Dimopoulos, and G.~R. Dvali, ``{New
  dimensions at a millimeter to a Fermi and superstrings at a TeV}'',
  \href{http://dx.doi.org/10.1016/S0370-2693(98)00860-0}{{\em Phys. Lett. B}
  {\bf 436} (1998)  257--263}, \href{http://arxiv.org/abs/hep-ph/9804398}{{\tt
  arXiv:hep-ph/9804398}}.

\bibitem{tadpoleProblem}
I.~Bena, J.~Bl\r{a}b\"ack, M.~Gra\~na, and S.~L\"ust, ``{The Tadpole
  Problem}'', \href{http://arxiv.org/abs/2010.10519}{{\tt arXiv:2010.10519
  [hep-th]}}.

\bibitem{Bena:2021tadpole}
I.~Bena, J.~Bl\r{a}b\"ack, M.~Gra\~na, and S.~L\"ust, ``{Algorithmically
  solving the Tadpole Problem}'', \href{http://arxiv.org/abs/2103.03250}{{\tt
  arXiv:2103.03250 [hep-th]}}.

\bibitem{Baker:2022rhh}
T.~Baker {\em et al.}, ``{Measuring the propagation speed of gravitational
  waves with LISA}'', \href{http://arxiv.org/abs/2203.00566}{{\tt
  arXiv:2203.00566 [gr-qc]}}.

\bibitem{DeRham:2018bgz}
C.~De~Rham, L.~Heisenberg, and A.~J. Tolley, ``{Spin-2 fields and the weak
  gravity conjecture}'',
  \href{http://dx.doi.org/10.1103/PhysRevD.100.104033}{{\em Phys. Rev. D} {\bf
  100} (2019) no.~10, 104033}, \href{http://arxiv.org/abs/1812.01012}{{\tt
  arXiv:1812.01012 [hep-th]}}.

\bibitem{Klaewer:2018yxi}
D.~Klaewer, D.~L\"ust, and E.~Palti, ``{A Spin-2 Conjecture on the
  Swampland}'', \href{http://dx.doi.org/10.1002/prop.201800102}{{\em Fortsch.
  Phys.} {\bf 67} (2019) no.~1-2, 1800102},
  \href{http://arxiv.org/abs/1811.07908}{{\tt arXiv:1811.07908 [hep-th]}}.

\end{thebibliography}\endgroup

\end{document}